\documentclass[aps,prx,showpacs,showkeys,amssymb,amsmath,superscriptaddress,reprint,maxbibnames=10]{revtex4-2}

\usepackage{graphicx,bm,amsmath}
\usepackage{overpic}
\usepackage[usenames,dvipsnames]{xcolor}
\usepackage[colorlinks,bookmarks=false,citecolor=NavyBlue,linkcolor=Red,urlcolor=blue,
]{hyperref}
\usepackage{xr-hyper}      
\usepackage{appendix}

\usepackage[bbgreekl]{mathbbol}
\usepackage[normalem]{ulem}
\usepackage{braket}
\usepackage{comment}
\usepackage{amsthm}
\usepackage{enumerate}
\usepackage{diagbox}
\usepackage{tabularx}
\usepackage{booktabs}
\usepackage{mdframed}
\usepackage[table]{xcolor}
\usepackage{colortbl}
\def\doi{http://dx.doi.org/}
\newcommand{\be}{\begin{equation}}
\newcommand{\ee}{\end{equation}}
\newcommand{\bec}{\begin{equation*}}
\newcommand{\eec}{\end{equation*}}
\newcommand{\bea}{\begin{eqnarray}}
\newcommand{\eea}{\end{eqnarray}}

\usepackage{xcolor}

\definecolor{mypurple}{RGB}{150,50,200}

\begin{document}
\title{Pulling strings in real time: flux tube  dynamics in (2+1)-d $\mathbb{Z}_2$-Higgs Gauge Theories}

\author{Zeno Bacciconi}
\thanks{These authors contribute equally}
\affiliation{SISSA --- International School for Advanced Studies, via Bonomea 265, 34136 Trieste, Italy}
\affiliation{ICTP --- The Abdus Salam International Centre for Theoretical Physics, Strada Costiera 11, 34151 Trieste, Italy}
\author{Martina Frau}
\thanks{These authors contribute equally}
\affiliation{SISSA --- International School for Advances Studies, via Bonomea 265, 34136 Trieste, Italy}
\affiliation{ICTP --- The Abdus Salam International Centre for Theoretical Physics, Strada Costiera 11, 34151 Trieste, Italy}

\author{Luca Tagliacozzo}
\affiliation{Institute of Fundamental Physics IFF-CSIC, C/ Serrano 113b, Madrid 28006, Spain}
\affiliation{Quantum Advanced Research Center (QuARC), CSIC, Calle Serrano 113b, 28006 Madrid, Spain}
\author{Michele Caselle}
\affiliation{Università degli studi di Torino and INFN – sezione di Torino ,
Via Pietro Giuria 1, Torino, Italy}
\author{Marcello Dalmonte}
\affiliation{ICTP --- The Abdus Salam International Centre for Theoretical Physics, Strada Costiera 11, 34151 Trieste, Italy}
\affiliation{Dipartimento di Fisica e Astronomia, Università di Bologna, and INFN, Sezione di Bologna, via Irnerio 46, I-40126 Bologna, Italy}

\begin{abstract}
Understanding real-time flux-tube dynamics in more than one spatial dimension is key to unlocking the non-perturbative physics of confinement, and is now actively pursued by quantum computing and simulation experiments. However, describing such dynamics has proven to be extremely challenging with both experiments and state-of-the-art numerical simulations limited to small volumes and short timescales. Here we investigate flux tube statics and real-time evolution in a genuine two-dimensional $\mathbb{Z}_2$ Higgs gauge theory at system sizes and timescales orders of magnitude beyond present experiments and numerics. The key enabling element is the recently introduced Clifford-augmented matrix product states (CAMPS) framework, which we demonstrate to parametrically reduce the entanglement that must be represented in the matrix product state;  both in the pure-gauge limit and in the presence of dynamical matter. 
We benchmark this capability through stringent tests of effective string theory, including universal spectral features and flux tube roughening properties in the presence of matter. 
We then introduce a string-pull protocol that selectively excites transverse modes and reconstructs their finite-size spectrum in real time. In the rough regime, the response is collective, and our simulations show that this is also well captured by universal effective string theory predictions. Strong confinement instead produces long-lived, lattice-locked local dynamics persisting to times $tJ \gtrsim 100$. These results provide ab initio evidence that effective string theory captures nonequilibrium string dynamics and reveal a hitherto unexplored long-lived prethermal regime of strongly confined flux tubes, providing a novel angle on how confinement dictates dynamics in more than one spatial dimension.

\end{abstract}


\maketitle
\section{Introduction}

Lattice gauge theories (LGTs) \cite{Wilson1974,Kogut1979,montvay1994quantum} represent one of the most successful frameworks to understand fundamental phenomena, ranging from quark confinement in quantum chromodynamics (QCD) \cite{Wilson1974}, to topological matter in electronic and synthetic quantum systems \cite{Wen_rmp2017,Kitaev2003,moessner1945topological,savary2017quantum}. The ubiquitous non-perturbative nature of these theories makes them an ideal testing ground for quantum many-body techniques. Monte Carlo simulations have had tremendous success at targeting equilibrium properties \cite{Creutz_prd1980_montecarlo,Kronfeld2012_montecarlorev} - with archetypal predictions such as the low-lying QCD spectrum \cite{fodor_rmp2012_qcd}. However, they do suffer from a severe complex action problem when applied to out-of-equilibrium dynamics \cite{wiese2013ultracold,Gattringer2016_signproblem}, rendering most of the latter phenomena inaccessible to sampling methods. 

The growing interface of quantum information science and high-energy physics has opened an entirely new perspective on the problem~\cite{wiese2013ultracold,DalmonteMontangero2016_rev,banuls2020simulating,bauer2023quantum,di2024quantum,davoudi2026quantum,halimeh2025_quantumsimulationoutofequilibriumdynamics}, along two complementary routes. First, quantum computation and simulation experiments, from trapped ions ~\cite{Martinez_nat2016_quantumsimulationgauge,deMonroe_2024_1Dobservationstringbreakingdynamicsquantum,meth2025simulating,joshiHalimeh_2026observationgenuine21dstring} to Rydberg atom arrays \cite{bernien2017probing,GonzlezCuadra2025_nat2025_observationstringbreaking,xiang2025real,mark2025observation,liang2025observation}, neutral atoms in optical lattices~\cite{aidelsburger2021cold,mil2020scalable,zhu2024probing,karch2026dynamical}, and superconducting circuits~\cite{Cochran2025_visualizingstring_google,schuhmacher2025observation,cobosRico_2025realtimedynamics21dgauge}, are beginning to realize and control LGTs in the laboratory. Second, variational ansatze based on tensor network (TN) states \cite{montangero2018introduction,Silvi2019tn,schollwock2011density,verstraete2008matrix,carmen2020review,emonts2020gauss} continue to push the boundaries of what is computationally tractable in out-of-equilibrium dynamics \cite{carmen2020review,montangero2022loop}, offering a robust and scalable alternative for benchmarking and discovery.

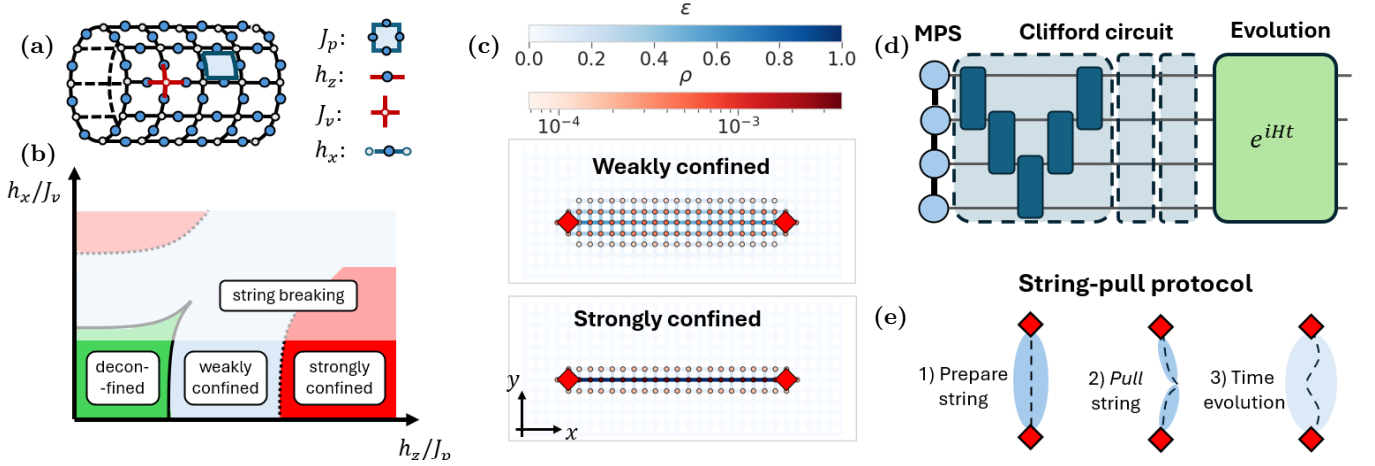
\begin{figure*}
 \begin{overpic}[width=\linewidth]{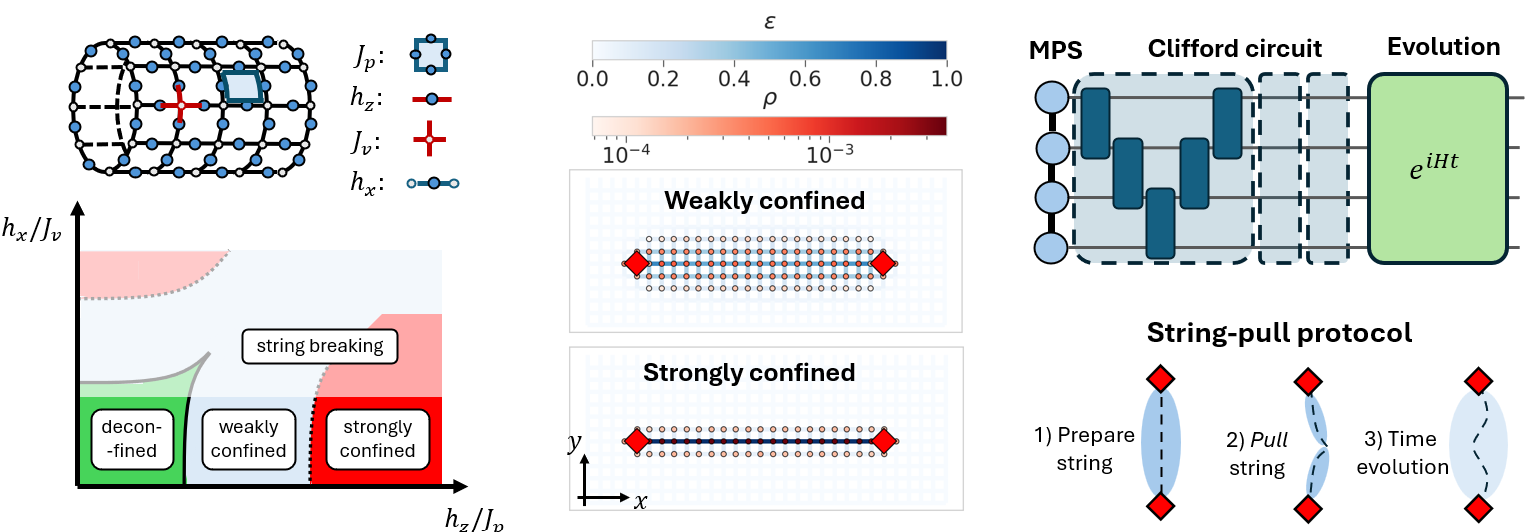}
        \put(1, 30){\textbf{(a)}}
        \put(1, 22){\textbf{(b)}}
        \put(34,30){\textbf{(c)}}
        \put(64,30){\textbf{(d)}}
        \put(64,10){\textbf{(e)}}
    \end{overpic}

    \caption{\textbf{Framework for large-scale string simulations}. \textbf{(a)} Schematic representation of the $\mathbb{Z}_2$-Higgs gauge theory Hamiltonian introduced in Eq.~\ref{eq:TC_hamiltonian}, comprising plaquette ($J_p$), vertex ($J_v$), confining field ($h_z$), and gauge-matter coupling ($h_x$) terms.
    \textbf{(b)} Phase diagram as a function of $h_x/J_v$ and $h_z/J_p$, showing the deconfined, weakly confined, and strongly confined regimes, as discussed in Sec.~\ref{subsec:phasediag}. Owing to electric--magnetic duality, the phase diagram is symmetric with respect to the diagonal; hence, we restrict our analysis to $h_x<0.2$, indicated by the unshaded region. String breaking can occur for any finite gauge--matter coupling, $h_x>0$. \textbf{(c)} Representative snapshots of strings connecting two charges, prepared using the sequential protocol described in Sec.~\ref{subsec:stringstateprep}. Shades of blue encode the electric-field strength $\epsilon_l$, defined in Eq.~\ref{eq:electric_field}, while shades of red encode the charge density $\rho_r$, defined as 
$\rho_r=\langle A_r\rangle_{\mathrm{str}}-\langle A_r\rangle_{\mathrm{vac}}$,
with $A_r$ the vertex operator centered at site $r$. The red diamonds mark the string endpoints. The parameters are $h_z=0.5$ and $h_z=1.3$ for the weakly and strongly confined cases, respectively, with $h_x=0.2$, $L_x=30$, $L_y=12$, and bond dimension $\chi=600$ in both cases. \textbf{(d)} Clifford-augmented matrix-product-state (CAMPS) ansatz used to represent the initial state, and subsequent quench dynamics. Crucially, Clifford disentangling is applied only during the preparation of the initial state, and not during time evolution. \textbf{(e)} String-pull protocol: starting from an equilibrium string connecting two fixed charges, the string is projected to pass through a point above its midpoint, preparing a stretched configuration that is subsequently evolved in real time.}
    \label{fig:fig1}
\end{figure*}

A particularly intriguing example of real-time processes that is of paramount importance in field theory is the formation and dynamics of a chromoelectric flux tube between two static colour sources - a hallmark of confinement in gauge theories. Real-time string dynamics has been widely studied in 1D experiments \cite{ZhangPan_natphys2024_1Dlgtexp,MildenbergerHauke_natphys2025_1Dlgtexp,deMonroe_2024_1Dobservationstringbreakingdynamicsquantum,luo2025quantumsimulationbubblenucleation,LiuPan_prl2025_stringbreaking,Tan2021_dwconf} and simulations \cite{carmen2020review,buyens2017real,gupta2026stringbreakingstaticsdynamics11d,pichlermontangero_prx2016_U1TNdynamics,VerdelHeyl_prb2020_stringbreaking,surace_prxquantum2026_stringbreaking1d,shirali2025break,florio2024quantum,Florio_prl2024_jetproduction,grieninger2026quantumcomplexitystringbreaking,magnifico2020real}, where the qualitative string behavior is different from that of higher dimensions due to the lack of a transverse direction. Progress beyond 1D has been the focus of recent quantum simulation experiments, targeting instances of string breaking and far-off-equilibrium dynamics \cite{Cochran2025_visualizingstring_google,GonzlezCuadra2025_nat2025_observationstringbreaking,xuHalimeh_2026observationglueballexcitationsstring,cobosRico_2025realtimedynamics21dgauge}, and classical TN numerical methods on both static \cite{Pollman2025roughening,uedatagliacozzo_2026interfaceroughening3dising,negro2026finitevolumeschemecontinuumextrapolation} and dynamical properties \cite{wu2025_pepsdynamics_purelgt,DiMarcantonioRico_2026commphys_roughening,bombieri2026u1latticegaugetheory,KrinitsinSchmitt_prl2025_rougheningising,xu2025string}. 
However, so far, despite remarkable experimental advances and computational efforts, progress has been limited in volumes and time, offering limited insight on the flux tube away from perturbative regimes. A natural question then follows, to which extent can one explore aspects of flux tube dynamics with presently available computational methods, that are informative about the underlying physics, and at the same time, provide fundamental new insights on how to probe such phenomena in experiments - as well as reset the target for true quantum advantage in quantum computing and simulation platforms.

In this work, we present a full-fledged investigation of flux tube statics and dynamics in two-dimensional $\mathbb{Z}_2$- Higgs gauge theories (Fig.~\ref{fig:fig1}a-b), in regimes so far inaccessible to both quantum and classical simulations. The enabling element of our investigation is the use of a hybrid algorithm, that combines tensor network structures \cite{white1993density,white1992_prl,schollwock2011density} with stabilizer formalism \cite{gottesman1997stabilizer}. Building on recent advances, we utilize Clifford-augmented matrix product states ~\cite{qian2024augmenting,qian2024clifford,mello2024clifford,masotllima2024,AzarQuella_prl2025_stabilizer} as the backbone of our study, and then tailor dynamics evolutions with time-dependent variational principle (TDVP) \cite{Haegeman_prl2011_tdvp}. The success of this combination lies within the unique structure of gauge theories: Clifford-augmentation is able to incorporate successfully information due to gauge invariance (local) as well as approximate duality (non-local), so that the tensor network basis has to cope with a limited amount of entanglement. This allows us to treat systems of several hundred gauge variables with very modest bond dimensions: for instance, in case of statics, we utilize $\chi=600$, instead of the approximate $\chi=12000$ needed for traditional matrix product states that would make the simulation prohibitively expensive. As we elaborate upon below, this computational advantage (roughly corresponding to a memory saving of order $O(20^2)$, and a computational time save of order 1000) will be key for both qualitative and quantitative characterization of flux tube statics and dynamics.

Based on this enabling intuition, to set up a strong basis before investigating real time processes, we test a series of predictions about the Effective String Theory (EST) description of confinement~\cite{Luscher:1980ac, Luscher:1980fr} - a universal framework that applies to a wide range of confining gauge theories, ranging from the
(2+1)-dimensional gauge Ising model to the physically relevant (3 + 1)-dimensional SU (3) gauge theory. By accessing cylinders as large as 30x12, and string distances of order of 20 sites, we perform a systematic characterization of the confining potential: first, we verify the universal prediction of the L\"uscher term up to accuracy of 10\%, i.e. much higher precision with respect to previous tensor network simulations \cite{bombieri2026u1latticegaugetheory,DiMarcantonioRico_2026commphys_roughening} with considerably smaller finite volume effects, and verify its stability upon inclusion of matter dynamics. 
Then, we compute the mass of the lightest glueball in the continuum limit, a considerably more challenging task, for which we observe results compatible with Monte Carlo simulations. Finally, we compute the string roughening transition point (Fig.~\ref{fig:fig1}c), and, exploiting volumes as wide as 12 sites, we verify its finite-size convergence below percent level. Remarkably, our tensor network simulations offer a quantitative characterization of almost all static properties, something that, in the context of (2+1)-d gauge theories, could only be demonstrated so far with Monte Carlo simulations (in regimes free of the sign problem).

Against this systematic benchmarking for static properties, we explore real-time dynamics. In order to probe flux tube properties, we formulate a "string-pull" protocol, that is amenable to both numerical as well as real experiments \cite{Cochran2025_visualizingstring_google,GonzlezCuadra2025_nat2025_observationstringbreaking}. In simulations, as a pivotal difference from previously proposed algorithms \cite{qian2024clifford,mello2024clifford}, we do not time-adapt our Clifford augmentation (Fig.~\ref{fig:fig1}d) - based on the fact that both gauge invariance and duality are not affected by unitaries. This allows us to keep our simulations feasible despite the large volumes of over 700 gauge variables. We observe a string whose edges evolve almost perfectly following a light-cone pattern in a wide parameter range of microscopic couplings. At long time, strong confinement leads to the string being unable to relax on timescales of over 100 inverse coupling, signalling the emergence of prethermal dynamics that, to the best of our knowledge, has never been observed in such large 2D gauge theories.

We then utilize the string-pulling protocol (Fig.~\ref{fig:fig1}e) to perform flux-tube spectroscopy. We confirm that the energy spectrum can be well captured by a compactified boson field theory up to momenta comparable with the inverse string size, and extract the corresponding speed of light within the rough phase focusing on the lowest energy modes. The latter departs strongly from purely static estimates, and agrees well with expectations from Lorentz invariance. To the best of our knowledge, this represents the first ab-initio demonstration that effective string theory is not only able to capture statics, but also finite-energy string properties with remarkable accuracy. 

The results of our simulations show that gauge theories are by far, at present, the most promising system where Clifford-augmentation can extend the power of tensor network states. In particular, it allows us to push considerably upward the quantum advantage threshold for investigating flux tube statics and dynamics, as the systems we are able to treat with modest computational effort are an order of magnitude larger than recent, very remarkable experiments, and extend up to timescales over 100 inverse couplings.

The paper is organized as follows. In Sec. \ref{sec:model} we introduce the model and the effective theory that describes confinement properties. Then in Sec. \ref{sec:method} we first review the numerical framework used in this work, that is CAMPS, and compare its performance for the model under study against standard MPS simulations. In Sec. \ref{sec:statics} we show static results for the confining string and in Sec. \ref{sec:dynamics} we introduce and study the new string-pull protocol which allows us to reconstruct the spectrum of the string. 

\section{Model}\label{sec:model}

In this section, we present an extended account of the $\mathbb{Z}_2$-Higgs lattice gauge theory and its basic properties, including both phase diagram and effective string theory description. On purpose, we opted for a propaedeutical, overcomplete summary, for two reasons: first, to guarantee that the material presented here is self-contained and notation is consistent; second, to allow readers with different backgrounds to cast the problem in a language that is more familiar to their expertise, and, at the same time, to emphasize points that might have been extensively discussed within a given community, but less in others. For these reasons, we have included derivations of the effective spin model description we use in our numerical experiments, of duality properties, and on the connection between Hamiltonian and Lagrangian formulations of the problem. Readers already familiar with those aspects can proceed to Sec. \ref{sec:statics} and consult the first subsection below to establish the notation.

\subsection{Hamiltonian}
We consider the $\mathbb{Z}_2$ Higgs model defined on a square lattice (Fig. \ref{fig:fig1}\textbf{a}), with matter fields $\tau$ residing on vertices and gauge fields $\sigma$ defined on links.

We first introduce our notation and conventions. We work in units where the lattice constant is $a=1$ and also $\hbar=1$. Lattice sites are labeled by $r$, while links are denoted by $\ell=(r, r+\mathbf{e}_\mu)$, where $\mu \in \{x,y\}$ and $\mathbf{e}_x=(1,0)$, $\mathbf{e}_y=(0,1)$.

In this convention, a plaquette rooted at site $r$ (i.e., with $r$ as its bottom-left corner) is uniquely specified by the four links:
\[
(r,r+\mathbf{e}_x), \,
(r+\mathbf{e}_x,r+\mathbf{e}_x+\mathbf{e}_y), \,
(r,r+\mathbf{e}_y), \,
(r+\mathbf{e}_y,r+\mathbf{e}_x+\mathbf{e}_y).
\]

We then define the standard $\mathbb{Z}_2$ gauge-invariant operators. The plaquette operator rooted at site $r$ is
\[
B_{r} =
\sigma^x_{(r,r+\mathbf{e}_x)}
\sigma^x_{(r+\mathbf{e}_x,r+\mathbf{e}_x+\mathbf{e}_y)}
\sigma^x_{(r+\mathbf{e}_y,r+\mathbf{e}_x+\mathbf{e}_y)}
\sigma^x_{(r,r+\mathbf{e}_y)} ,
\]
while the vertex operator is defined as
\[
A_{r} =
\sigma^z_{(r-\mathbf{e}_x,r)}
\sigma^z_{(r,r+\mathbf{e}_x)}
\sigma^z_{(r-\mathbf{e}_y,r)}
\sigma^z_{(r,r+\mathbf{e}_y)} ,
\]
where the product runs over the four links adjacent to the vertex $r$.

The Hamiltonian of the $\mathbb{Z}_2$-gauge Ising model - schematically depicted in (Fig.~\ref{fig:fig1}\textbf{a}) - is given by:
\begin{align}\label{eq:LGT_hamiltonian}
H_{\mathrm{LGT}} =&
- J_v \sum_{r} \tau^z_{r}
- J_p \sum_{r} B_{r}
- h_z \sum_{r}\sum_{a \in {x, y}} \sigma^z_{(r, r+\textbf{e}_a)}\nonumber \\
&- h_x \sum_{r}\sum_{a \in {x, y}} \tau^x_{r} \sigma^x_{(r, r+\textbf{e}_a)}\,\tau^x_{r+\mathbf{e}_a}\nonumber\\
&- h_y\sum_{r}\sum_{a \in {x, y}} \tau^x_{r} \sigma^y_{(r, r+\textbf{e}_a)}\,\tau^x_{r+\mathbf{e}_a}.
\end{align}
where $\tau^\alpha_r$ and $\sigma^\alpha_l$ are pauli matrices respectively representing matter and gauge variables. 

The coupling $J_v$ is the mass term for the Higgs field, energetically favoring configurations polarized in the $\tau^z$ basis. The plaquette coupling $J_p$ corresponds to the magnetic flux energy of the $\mathbb{Z}_2$ gauge field, penalizing flux excitations through each plaquette. The $h_z$ term is the electric-field energy: it assigns an energy cost to electric strings and hence generates the confining potential between static charges.
The couplings $h_x$ and $h_y$ describe kinetic gauge-matter interactions. The term $h_x$ corresponds to the standard gauge invariant hopping obtained from the minimal coupling prescription, in which the matter tunneling is dressed by a $\sigma^x$ operator on the intermediating link. The $h_y$ term introduces an additional kinetic contribution, corresponding to a rotation in the local gauge-field basis from $\sigma^x$ to $\sigma^y$, thereby inducing a complex mixing of electric and magnetic operator components on each link. While we have also carried out extensive simulations with $h_y>0$ (see App. \ref{app:sign_problem_gain}), for the sake of clarity, below we focus on the case $h_y=0$.

Although Eq.~\eqref{eq:LGT_hamiltonian} is written in terms of both matter and gauge fields, gauge invariance can be used to eliminate the matter degrees of freedom and reformulate the theory entirely in terms of gauge variables. In this representation, the Hamiltonian reduces to the toric code in a magnetic field
\begin{align}\label{eq:TC_hamiltonian}
H
=&
- J_v \sum_r A_r
- J_p \sum_r B_r
- \vec{h}\sum_r\sum_{a \in {x, y}}\vec{\sigma}_{(r, r+\textbf{e}_{a})}
\end{align}
which we employ throughout this work. Its different terms are illustrated schematically in Fig.~\ref{fig:fig1}a.

\subsubsection{Eliminating the matter fields} 
We now show how the gauge-only description of Eq.\eqref{eq:TC_hamiltonian} can be recovered from Eq.~\eqref{eq:LGT_hamiltonian} in terms of a unitary operator $\mathcal{U}$ that can be represented by a finite-depth Clifford circuit. This is a reformulation in modern language of well known dualities see e.g. \cite{TupitsynKitaevProkofevStamp2010}.

The local gauge symmetry is generated by the Gauss-law operators 
\begin{equation}\label{eq:gauss_law}
G_r=\tau_r^z A_r 
\end{equation}
where $A_r$ is the vertex operator introduced above. 
These operators satisfy 
\begin{equation} G_r^2=\mathbb{I}, \qquad [G_r,G_{r'}]=0, \qquad [G_r,H_{\mathrm{LGT}}]=0.
\end{equation} 
In absence of external static charges,  the physical Hilbert space is the simultaneous $+1$ eigenspace of all Gauss-law operators, 
\begin{equation} G_r\ket{\Psi_{\mathrm{phys}}} = \ket{\Psi_{\mathrm{phys}}}, \qquad \forall r. \label{eq:Gauss_law_constraint} \end{equation}.

The elimination of the matter fields can be formulated as an exact local unitary transformation. Specifically, the transformation maps the Gauss-law constraint entirely onto the matter qubits, leaving the gauge fields as the only dynamical degrees of freedom. To construct it, for each link $\ell=(r,r+\mathbf{e}_a)$ and one of its endpoints $r$, we define a controlled-NOT gate with the gauge qubit on the link as the control and the matter qubit at the vertex as the target, 
\begin{equation}
C_{\ell\rightarrow r}
=
\ket{0}\!\bra{0}_{\ell}\otimes\mathbb{I}_{r}
+
\ket{1}\!\bra{1}_{\ell}\otimes\tau_r^x,
\label{eq:gauge_matter_CNOT}
\end{equation}
where $\ket{0}_\ell$ and $\ket{1}_\ell$ denote the eigenstates of $\sigma_\ell^z$. 
The complete unitary transformation is 
\begin{equation} \mathcal{U} = \prod_r\prod_{\ell\ni r}C_{\ell\rightarrow r}. \label{eq:disentangling_unitary} \end{equation}
where the product runs over the four links incident on the vertex $r$.

All gates entering Eq.~\eqref{eq:disentangling_unitary} commute, since the gauge qubits always act as controls while the matter qubits always act as targets. However, each matter qubit participates in four CNOT gates, one for each adjacent link, so these gates cannot all be applied simultaneously. They can instead be grouped into four layers, such that no two gates within the same layer act on the same qubit. As a result, $\mathcal U$ is a geometrically local Clifford circuit of depth four, independent of the system size.
The action of a CNOT gate on Pauli operators is
\begin{align} C_{\ell\rightarrow r}\,\sigma_\ell^x\, C_{\ell\rightarrow r}^{\dagger} &= \sigma_\ell^x\tau_r^x, & C_{\ell\rightarrow r}\,\sigma_\ell^z\, C_{\ell\rightarrow r}^{\dagger} &= \sigma_\ell^z, \nonumber\\ C_{\ell\rightarrow r}\,\tau_r^x\, C_{\ell\rightarrow r}^{\dagger} &= \tau_r^x, & C_{\ell\rightarrow r}\,\tau_r^z\, C_{\ell\rightarrow r}^{\dagger} &= \sigma_\ell^z\tau_r^z . \label{eq:CNOT_Pauli_action} \end{align}

It follows that the full circuit transforms the matter and link operators according to 
\begin{align} \mathcal{U}\tau_r^z\mathcal{U}^{\dagger} &= \tau_r^z A_r = G_r, \label{eq:tauz_to_Gauss}\\ \mathcal{U}\sigma^x_{(r,r+\mathbf e_a)} \mathcal{U}^{\dagger} &= \tau_r^x \sigma^x_{(r,r+\mathbf e_a)} \tau_{r+\mathbf e_a}^x, \label{eq:sigmax_to_hopping}\\ 
\mathcal{U}\sigma^y_{(r,r+\mathbf e_a)} \mathcal{U}^{\dagger} &= \tau_r^x \sigma^y_{(r,r+\mathbf e_a)} \tau_{r+\mathbf e_a}^x, \label{eq:sigmay_to_hopping}\\ 
\mathcal{U}\sigma^z_{(r,r+\mathbf e_a)} \mathcal{U}^{\dagger} &= \sigma^z_{(r,r+\mathbf e_a)}. \label{eq:sigmaz_unchanged} \end{align}

In Eq.~\eqref{eq:sigmay_to_hopping} we have used $\sigma^y=i\sigma^x\sigma^z$. The plaquette operator is also left invariant by the transformation 
\begin{equation} \mathcal{U}B_r\mathcal{U}^{\dagger}=B_r. \end{equation} 
since each matter operator generated by the transformation appears twice around a plaquette and hence cancels. 

Moreover, because $\mathcal{U}^2=\mathbb{I}$, Eqs.~\eqref{eq:sigmax_to_hopping} and \eqref{eq:sigmay_to_hopping} can be inverted to show that the gauge-invariant hopping operators are mapped onto bare link operators,, 
\begin{align} \mathcal{U} \left( \tau_r^x\sigma^\alpha_{(r,r+\mathbf e_a)} \tau_{r+\mathbf e_a}^x \right) \mathcal{U}^{\dagger} = \sigma^\alpha_{(r,r+\mathbf e_a)}, \qquad \alpha=x,y. \label{eq:dressed_to_bare} \end{align}

At the same time, the original Gauss-law generators become single-site matter operators, 
\begin{equation} \mathcal{U}G_r\mathcal{U}^{\dagger}=\tau_r^z. \label{eq:Gauss_to_tauz} \end{equation} 

Consequently, the physical Hilbert space, defined by Eq.~\eqref{eq:Gauss_law_constraint}, is mapped onto the subspace
\begin{equation} \tau_r^z=+1, \qquad \forall r, \label{eq:frozen_matter_sector} \end{equation} 
in which all matter qubits are frozen in a product state and completely disentangled from the gauge degrees of freedom.

Applying $\mathcal{U}$ to Eq.~\eqref{eq:LGT_hamiltonian} and then restricting to the sector \eqref{eq:frozen_matter_sector} gives back Eq. \eqref{eq:TC_hamiltonian}.

Thus, within the physical Hilbert space, the $\mathbb{Z}_2$ Higgs model is unitarily equivalent to the toric code in a uniform magnetic field, together with a set of disentangled matter qubits fixed in the $\tau^z=+1$ state. Conversely, applying $\mathcal{U}$ to the link-only model couples the auxiliary matter qubits to the gauge field and reconstructs the gauge-invariant Higgs hopping terms. This correspondence is considerably stronger than a low-energy or perturbative mapping: it is an exact equivalence between the physical sectors of the two Hamiltonians, implemented by a finite-depth local Clifford circuit. As a consequence, the two models have identical spectra within the corresponding sectors, while their entanglement properties can differ only by boundary-law contributions.

\subsubsection{Electric--magnetic duality} 
\label{subsec:phasediag}
A key property of the toric-code Hamiltonian in Eq.~\eqref{eq:TC_hamiltonian}, which plays a central role in understanding its phase diagram, is its electric--magnetic duality. In this subsection we discuss this symmetry for the case $h_y=0$, where it takes its simplest form.

The vertex and plaquette operators introduced above have complementary physical interpretations. Violations of the condition
\begin{equation}
A_r=+1
\end{equation}
correspond to electric charges, conventionally denoted by $e$, whereas violations of
\begin{equation}
B_r=+1
\end{equation}
correspond to magnetic fluxes, or $m$ anyons. 
The Pauli operators acting on a link create and move these excitations. Since $\sigma_\ell^x$ anticommutes with the two vertex operators adjacent to the link $\ell$, it creates, moves, or annihilates pairs of electric charges. Similarly, $\sigma_\ell^z$ anticommutes with the two plaquette operators sharing the link $\ell$, and therefore creates, moves, or annihilates pairs of magnetic fluxes. 

On the square lattice, electric--magnetic duality exchanges vertices with plaquettes and maps
\[
A_r\longleftrightarrow B_{r^\star},
\qquad
\sigma_\ell^x\longleftrightarrow\sigma_{\ell^\star}^z,
\]
where $r^\star$ and $\ell^\star$ denote sites and links of the dual lattice.
Accordingly,
\begin{equation} (J_v,J_p,h_x,h_z) \longleftrightarrow (J_p,J_v,h_z,h_x). \label{eq:EM_duality} \end{equation} 
For equal vertex and plaquette couplings, \begin{equation} J_v=J_p\equiv J, \end{equation} the line 
\begin{equation} h_x=h_z \label{eq:self_dual_line} \end{equation} 
is therefore self-dual.

This electric--magnetic duality should be distinguished from the transformation \eqref{eq:disentangling_unitary}: the latter disentangles matter and gauge degrees of freedom, whereas the former exchanges the electric and magnetic sectors of the resulting link-only theory. 

\subsubsection{Phase diagram and roughening transition}

We now summarize the zero-temperature phase diagram of the toric-code Hamiltonian in Eq.~\eqref{eq:TC_hamiltonian}. The different terms entering the Hamiltonian are illustrated in Fig.~\ref{fig:fig1}\textbf{a}, while the corresponding phase diagram is shown in Fig.~\ref{fig:fig1}\textbf{b}. In what follows, we restrict our attention to the case $h_y=0$.

As shown in Fig.~\ref{fig:fig1}\textbf{b}, the phase diagram consists of a central topologically ordered region (green), corresponding to the deconfined phase of the original gauge theory, surrounded by a confined phase. The latter is further divided by the dashed roughening line into a weakly confined regime (blue) and a strongly confined regime (red). While the confinement transition separates distinct thermodynamic phases, the roughening transition instead distinguishes two regimes of the confining string with qualitatively different transverse fluctuations.

At zero field, $h_x=h_z=0$, the system realizes the $\mathbb Z_2$ topological phase, characterized by a fourfold ground-state degeneracy on the torus and deconfined electric and magnetic anyons. Along the coordinate axes, this phase is destroyed by the condensation of either electric charges ($h_z=0$) or magnetic fluxes ($h_x=0$), giving rise to the Higgs and confined phases, respectively. Both transitions belong to the three-dimensional Ising universality class and occur at \cite{DengProkofev2012}
\begin{equation}
\left.\frac{h_x}{J}\right|_{\rm c}
=
\left.\frac{h_z}{J}\right|_{\rm c}
\simeq0.32847,
\end{equation}
for $J_v=J_p\equiv J$. 

Away from the axes, the Higgs and confinement transition lines bend towards the self-dual line and meet at a self-dual multicritical point. Along the two branches, the transition out of the deconfined phase is driven respectively by the condensation of the electric $e$ and magnetic $m$ excitations and belongs to the three-dimensional Ising universality class. At the self-dual multicritical point, electric and magnetic excitations become critical simultaneously; numerical studies provide evidence that this transition is continuous and belongs to a distinct self-dual universality class~\cite{SomozaSernaNahum2021}. From this point, a first-order transition extends for a finite interval along the self-dual line, separating regimes predominantly associated with electric-charge and magnetic-flux condensation and spontaneously breaking the electric--magnetic duality. This first-order line terminates at a conventional critical endpoint, located for the toric-code Hamiltonian at
\begin{equation}
h_x=h_z\simeq 0.418 ,
\end{equation}
beyond which the Higgs and confined regimes are smoothly connected. More generally, these regimes do not correspond to distinct thermodynamic phases and can be adiabatically connected without crossing a phase transition, in accordance with the Fradkin--Shenker continuity~\cite{Fradkin:1978dv,DengProkofev2012}.

Within the confined phase, the electric flux connecting two static charges exhibits two qualitatively distinct behaviors, illustrated in Fig.~\ref{fig:fig1}\textbf{c}. In the weakly confined (or rough) regime, the flux tube broadens with increasing charge separation, reflecting strong transverse fluctuations of the confining string. In contrast, the strongly confined (or stiff) regime is characterized by a narrow flux tube whose width remains essentially independent of its length. These two regimes are separated by the roughening transition (represented with a dashed line in  Fig.\ref{fig:fig1}(c)), which belongs to the Berezinskii--Kosterlitz--Thouless universality class and is well captured by an Effective String theory which we review in ec.~\ref{sec:effective_stringtheory}. Note that at finite gauge-matter coupling $h_x>0$ long strings can eventually break when the energy stored in the string meets the cost of creating another pair of charges (see App.\ref{app:string_breaking}) and the interpretation of this line as a thermodynamic transition is more loose.


\subsubsection{Relation to the Euclidean lattice gauge-theory formulation} 

We here briefly review the relation between the Hamiltonian formulation (Eq. \eqref{eq:LGT_hamiltonian}) introduced above with the more standard Euclidean lattice field-theory of the $\mathbb{Z}_2$ gauge-Higgs model \cite{KogutSusskind1975}. In the Euclidean formulation one works on a three-dimensional cubic lattice whose vertices are located at $x=(\boldsymbol{r},\tau)$, with two spatial directions $\boldsymbol{r}$ and one imaginary-time direction $\tau$. Gauge fields are Ising variables 
\begin{equation} U_{\lambda_\mu}(x)=\pm 1, \end{equation} 
living on spacetime links $\lambda=(r,r+\boldsymbol{e}_\mu)$ - whose spacetime direction is denoted by $\mu=x,y,\tau$ - of the Euclidean  lattice,  and also  the Higgs field is an Ising variable 
\begin{equation} \phi(x)=\pm 1 \end{equation}
living on sites. 

The standard Euclidean action is 
\begin{equation} 
S_E[U,\phi] = -\beta_g \sum_{\square} \prod_{\lambda\in\square}U_{\lambda} -\beta_h \sum_{x,\mu} \phi(x)\, U_{\lambda_\mu}(x)\, \phi(x+\boldsymbol{e}_\mu). \label{eq:Euclidean_Z2_Higgs_action} \end{equation}

The first term is the Wilson plaquette action for the $\mathbb{Z}_2$ gauge field with $\square$ being either a space ($s$) or time ($\tau$) oriented plaquette, while the second term is the nearest-neighbor hopping term for the Higgs field.
The action is invariant under the local $\mathbb{Z}_2$ gauge transformation 
\begin{align} 
&\phi(x)\rightarrow \eta(x)\phi(x), \nonumber\\  
&U_{\lambda_{\mu}}(x)\rightarrow \eta(x)U_{\lambda_\mu}(x)\eta(x+e_\mu),. \label{eq:Euclidean_gauge_symmetry} 
\end{align}
parametrized by $\eta(x)=\pm1$.

The connection with the Hamiltonian formulation is obtained by fixing the temporal gauge and viewing the Euclidean time direction as generated by a transfer matrix \cite{Creutz1977}. The local gauge redundancy \eqref{eq:Euclidean_gauge_symmetry} can be partially fixed to the the temporal gauge, requiring all temporal gauge fields to be one
\begin{align}
    \textrm{temporal gauge:} \qquad U_{\lambda_\tau}(x) =1  
\end{align}
This then requires the residual gauge transformation to be time-independent:

\begin{equation}
\begin{aligned}
U_{\lambda_\mu}(\mathbf r,\tau)
&\mapsto
\eta(\mathbf r)\,
U_{\lambda_\mu}(\mathbf r,\tau)\,
\eta(\mathbf r+\mathbf e_\mu),
\\
\phi(\mathbf r,\tau)
&\mapsto
\eta(\mathbf r)\phi(\mathbf r,\tau),
\qquad
\eta(\mathbf r)=\pm1,
\\
\end{aligned}
\end{equation}
The residual gauge degree of freedom $\eta(\boldsymbol{r})$ is then generated by the $G_r$ of the previous section
\begin{equation}
\begin{aligned}
\eta(\mathbf r)\text{ gauge flip}
&\longleftrightarrow
G_{\mathbf r}
=
\tau^z_{\mathbf r}
\prod_{\ell\ni\mathbf r}\sigma^z_\ell .
\end{aligned}
\label{eq:residual_gauge_Gauss_dictionary}
\end{equation}
with $\ell=\lambda_{s}$ a space only link.

In this gauge, the transfer matrix view of the partition function becomes more transparent. In particular, in an anisotropic discretization, spatial $\square_s$ and temporal $\square_\tau$ plaquettes are allowed to have different couplings and represent different terms in the final Hamiltonian Eq. \eqref{eq:LGT_hamiltonian}. The spatial plaquette term maps to the magnetic term in the Hamiltonian, 
\begin{equation} -\beta_g^{(s)} \sum_{\square_s} \prod_{\ell\in\square_s}U_l \quad \longleftrightarrow \quad -J_p\sum_p B_p . \end{equation} 

Temporal plaquettes generate the electric-field energy
\begin{equation}
-\beta_g^{(\tau)} \sum_{\square_\tau} \prod_{l\in\square_\tau}U_l \quad \longleftrightarrow \quad -h_z\sum_{\ell}\sigma_\ell^z ,
\end{equation}
where the two temporal links are omitted because of being gauge fixed to 1.

Similarly, spatial Higgs hopping terms in the Euclidean action map to the gauge-invariant matter hopping terms 
\begin{equation}
\phi(x)\, U_{l}(x)\, \phi(x+e_a) \quad \longleftrightarrow
-h_x \sum_{r,\mu} \tau_r^x \sigma^x_{(r,r+\mathbf e_\mu)} \tau_{r+\mathbf e_\mu}^x . \end{equation} 
The temporal Higgs hopping term gives rise to the Higgs mass term 
\begin{equation}\phi(x)\, U_{\lambda_\tau}(x)\, \phi(x+e_\tau) \quad \longleftrightarrow -J_v\sum_r\tau_r^z . 
\end{equation} 

The precise relation between the Euclidean couplings $(\beta_g^{(s)},\beta_g^{(\tau)},\beta_h^{(s)},\beta_h^{(\tau)})$ and the Hamiltonian couplings $(J_p,h_z,h_x,J_v)$ depends on the anisotropic transfer-matrix limit.
For the purposes of the present work, only the operator dictionary is needed. Note however that the Euclidean lattice gauge-theory literature often uses the $\mathbb{Z}_2$ Higgs action \eqref{eq:Euclidean_Z2_Higgs_action} to discuss the confinement-Higgs phase diagram. In that language, the large-$\beta_h$ region is the Higgs regime (large $h_x/J_v$ regime in Fig. \ref{fig:fig1}(b)), while the small-$\beta_g$ region is the confining regime (large $h_z/J_p$ regime in Fig. \ref{fig:fig1}(b)). For Higgs matter in the fundamental representation these two regimes are not necessarily separated by a thermodynamic phase transition: they can be analytically connected, as shown by Fradkin and Shenker \cite{Fradkin:1978dv}. 

The term proportional to $h_y$ deserves a separate comment. It is fully gauge invariant at the Hamiltonian level, since $\sigma_\ell^y$ anticommutes with the $\sigma_\ell^z$ factor appearing in the Gauss-law generators at the two endpoints of $\ell$, exactly as $\sigma_\ell^x$ does. 
Therefore
\begin{equation} \left[ G_r, \tau_r^x\sigma_\ell^y\tau_{r'}^x \right]=0 \end{equation} 
for all vertices $r$. Nevertheless, this term is not present in the standard Euclidean $\mathbb{Z}_2$ gauge-Higgs action. In that formulation the gauge and Higgs fields are classical Ising variables, and the elementary matter-gauge coupling is a real Boltzmann weight.
By contrast, $\sigma^y=i\sigma^x\sigma^z$ combines the link variable with its conjugate electric field. It is therefore an intrinsically quantum Hamiltonian perturbation, corresponding in a transfer-matrix description to complex or sign-changing off-diagonal matrix elements, rather than to an additional positive local Boltzmann weight of the standard Euclidean $\mathbb{Z}_2$ gauge-Higgs model. For this reason, when making contact with the Euclidean Fradkin--Shenker formulation, the plane $h_y=0$ should be regarded as the direct Hamiltonian counterpart of the conventional real $\mathbb{Z}_2$ gauge-Higgs lattice action. The $h_y$ direction is a gauge-invariant Hamiltonian extension of that model.

\subsection{Effective string theory}\label{sec:effective_stringtheory}

A hallmark of confinement in non-abelian gauge theories is the formation of a chromoelectric flux tube between two static colour sources. Within the framework of effective string theory, this object is described as a fluctuating string connecting the quark and the antiquark~\cite{Luscher:1980ac, Luscher:1980fr}. Although the underlying string is assumed to be infinitely thin, quantum fluctuations give rise to a finite transverse width~\cite{Luscher:1980iy}. The properties of this width can be investigated analytically within effective string theory (EST) and have been extensively tested in lattice gauge theory simulations.

The simplest realization of EST is provided by the Nambu--Goto string~\cite{Nambu:1978bd,Goto:1971ce}. One of its best-known predictions is that, at low temperatures, the squared width of the flux tube increases logarithmically with the interquark separation~\cite{Luscher:1980iy}. This logarithmic broadening has been confirmed by lattice simulations in several pure gauge theories~\cite{Caselle:1995fh,Gliozzi:2010zv}. Beyond this logarithmic growth, it is also well known that the flux profile is also characterized by an intrinsic width which give rise to a long distance exponential tail \cite{caselle2026intrinsicwidthfluxtube}.

A remarkable feature of the EST description of confinement is its high degree of universality. The same qualitative behaviour is observed in essentially all confining gauge theories, ranging from the $(2+1)$-dimensional gauge Ising model~\cite{Wegner:1971app} to the physically relevant $(3+1)$-dimensional $SU(3)$ gauge theory. For this reason, many of the characteristic properties of the confining string (the Lüscher term, the transverse broadening of the flux tube, the effective string spectrum, ...) were first investigated in the gauge Ising model (see, for instance,~\cite{Caselle:2002ah,Caselle:1995fh}), where its simplicity allows high-precision numerical simulations to be performed at a much lower computational cost, and were only later studied in the more complex non-Abelian gauge theories. In the present work we follow the same spirit: we apply the CAMPS algorithm to the gauge Ising model as a preliminary step towards the investigation of the static and dynamical properties of the confining string in non-Abelian gauge theories.

We now review the effective field theory for the roughening transition which will also be used to describe the weakly confining phase. We directly work in an open string geometry where two charges are pinned at a distance $R$ in units of the lattice spacing $a=1$. The transverse dynamics of the string can be written in terms of a bosonic scalar height field $\phi(x)$ and its conjugate field $\Pi(x)$  ($[\phi(x),\Pi(x')]=i\delta(x-x')$):
\begin{align}\label{eq:EST_hamiltonian}
    {H}_{str}= \int_0^R dx\, \frac{v_0}{2}\left[\frac{1}{K_0}(\partial_x\phi)^2 + K_0 \,\Pi^2\right] -g \cos(2\pi \phi)
\end{align}
where $v_0$ is a non-universal velocity, $K_0$ the Luttinger parameter and $g$ controls the pinning to the lattice. The sine-gordon model above features a BKT transition when the cosine perturbation (with scaling dimension $\Delta[\cos(2\pi \phi)]=\pi K_0$) becomes relevant at $K_c=2/\pi $ \cite{Giamarchi_1d} pinning the height field to the lattice. When $K_0>K_c$ the cosine perturbation is irrelevant and just produces shifts in the infrared ($R\gg1$) for the quadratic part:
\begin{align}
    H_{str}\to \int_0^R dx\, \frac{v}{2}\left[\frac{1}{K}(\partial_x\phi)^2 + K \,\Pi^2\right] 
\end{align}
where $v$ and $K$ are the renormalized velocity and Luttinger parameter $K$. Ultimately these are the values that will be observed from a long-wavelength analysis of the string properties at a particular value of the microscopic parameters.  

We now describe some predictions of the effective string model in Eq. \eqref{eq:EST_hamiltonian} in the rough phase where the cosine perturbation can be neglected. First of all, let us remark that the pinned charges translate in open boundary conditions for the string at position $x=0$ and $x=R+1$, just outside the string region determined by $x=1,\dots, R$. The expected height field fluctuations at finite sizes are then:
\begin{align}\label{eq:widths}
    \langle\phi^2(x)\rangle = \frac{K}{2\pi} \log\left[ \frac{R+1}{\pi}\sin\left( \frac{\pi x}{R+1}\right)\right] + const
\end{align}
where we are using $R+1$ as the effective string length as the distance between the points where the fixed boundaries are set is $R+1$ and not $R$. 
The dynamics of the height field $\phi(x)$ can be expanded on its normal modes $\hat{\phi}(x)=\sum_{n}\phi_n(x) \hat{b}_n$ where $\hat{b}_k$ is a set of bosonic modes and:
\begin{align}\label{eq:normal_modes}
    &\phi_n(x)= \sqrt{\frac{2}{R+1}}\sin(k_n x)\nonumber \\ & k_n=\frac{\pi}{R+1}(n+1) \;\;(n=0,..,R-1)
\end{align}
The spectrum at long-wavelength ($k_n R\ll 1$) is linear:
\begin{align}\label{eq:spectrum}
    \omega_n\simeq v\, k_n + \dots
\end{align}
with $v$ and effective velocity which can be renormalized with respect to the bare value $v_0$ in Eq. \eqref{eq:EST_hamiltonian} by the cosine perturbation $g$. We expect sub-leading contributions to the spectrum coming from lattice effects both in the longitudinal and transverse direction. A simple expectation for the former is to give rise to a periodic spectrum $\omega_n=2v\sin(k_n /2)$; while the latter can bring more subtle power law corrections as it enters through a cosine perturbation.

Another important prediction of the effective string model lies in a correction to the confining potential $V(R)$ felt by the charges known as Lüscher correction. This is nothing but the Casimir energy of the transverse modes which, for the gapless phase of the sine-gordon model in Eq. \eqref{eq:EST_hamiltonian}, is the one of a free boson in 1+1D:
\begin{align}
    E_{casimir}= -\frac{\pi}{24}\frac{v}{(R+1)}
\end{align}
where $R+1$ is the effective length of the string. The confining potential at large $R$ is then expected to behave as:
\begin{align}\label{eq:confining_potential}
    V(R)=\sigma R + E_0 - \frac{\gamma_0}{R} +O(R^{-2})
\end{align}
where $\gamma_0 = \gamma v$, with $\gamma=\pi/24$. The same Casimir correction in $1/R$ can be recovered by a lattice regularization of a free boson with the modes in Eq. \eqref{eq:normal_modes} and a dispersion $\omega_n=2v\sin(k_n/2)$.

The action discussed above is only the first order approximation of the actual EST action. In the continuum limit the transverse mode is described by the Nambu-Goto action~\cite{Nambu:1978bd,Goto:1971ce}, which includes all higher order derivatives of the field $\phi$. In particular, the Nambu-Goto prediction for the confining potential is:
\begin{align}\label{eq:confining_potential_NG}
    V_{NG}(R)=E_0 + \sigma R\sqrt{1-\frac{2\gamma v}{\sigma R^2}}
\end{align}
which recovers the leading order lattice expression in Eq. \eqref{eq:confining_potential}. Note that in these expression the velocity $v$ is explicitly appearing. For the values of $\sigma$ and $R$ that we studied in this paper (see the table in Fig.\ref{fig:stringproperties}) we can safely approximate the Nambu-Goto results to its first order approximation (the "L\"uscher term") .

We now also remark that, within the rough phase,  one can rewrite the velocity and Luttinger parameter in terms of the string tension as:
\begin{align}
    \sigma= \frac{v_0}{a^2K_0}
\end{align}
The most important microscopic parameter which tunes the transition is $h_z$, which controls the electric field 
\section{Clifford augmented matrix product states for $\mathbb{Z}_2$ LGT}\label{sec:method}
In this section, we describe our numerical algorithm and compare it with standard MPS approaches. 
Below we first review the Clifford Augmented MPS ansatz and how it is used to perform both ground state search and string states preparation (Sec.\ref{subsec:method_camps}). Then provide a systematic study of its performance (Sec. \ref{subsec:Performanceresults}) on the $\mathbb{Z}_2$-gauge ising model showing a parametric gain on the exponential growth of the computational cost with system size.
\subsection{Method}\label{subsec:method_camps}
The CAMPS ansatz consist of two parts, a Clifford unitary $C$ and an MPS $\ket{\psi(A)}$ which can be schematically written as:
\begin{align}
    \ket{\boldsymbol{\Psi}(A,C)}=C\ket{\psi(A)}
\end{align}
The MPS ansatz is generically written as:
\be\label{eq:MPS}
|\psi(A)\rangle=\sum_{s_1,s_2,\cdots,s_{N_s}} A^{s_1} A^{s_2} \cdots A^{s_{N_s}} |s_1,s_2,\cdots s_{N_s} \rangle
\ee
where $A^{s_i}$ are $\chi \times \chi$ matrices, except for the boundaries $i = 1$, $i =N_s$, where the corresponding tensors are, respectively, a row and a column vector of length $\chi$. The computational cost of handling CAMPS is typically similar of that of handling an MPS with the same \textit{bond dimension} $\chi$ ~\cite{qian2024augmenting,qian2024clifford,mello2024clifford,masotllima2024}. For example calculating the expectation value of an observable $O$ which can be written as a Pauli string can be done as:
\begin{equation}
    \bra{\Psi}O\ket{\Psi}= \bra{\psi}(C^\dagger O C)\ket{\psi}
\end{equation}
which has the same cost of standard MPS as the transformed operator can be efficiently calculated $\mathcal{O}(N_s^2)$ with the stabilizer formalism \cite{gottesman1997stabilizer} and is also a single Pauli string. 

The bond dimension of the MPS is, for standard MPS, directly related to the entanglement of the state. The scaling of the bond dimension with system size, and thus of entanglement, governs the efficiency of the MPS representation: it is efficient in one-dimensional systems in area-law entangled phases, as well as critical points where entanglement exhibits logarithmic scaling. It can also be extended to two-dimensional systems by mapping the lattice onto a one-dimensional chain using various strategies; in this work, we employ a \textit{snake} mapping, which is particularly well suited for cylindrical geometries. However, even for area-law entangled phases in two dimensions, the required bond dimension typically grows exponentially with \textit{linear} system size $\chi\propto e^{\alpha L}$, which severely limits simulations to relatively small lattices. 

The Clifford circuit can introduce entanglement. It is interesting to note that in model under study the exactly solvable point $\vec{h}=\vec{0}$ has area law entanglement and can be exactly represented by a CAMPS with $\chi=1$. Indeed the pure toric code Hamiltonian without $\vec{h}$ gives a full stabilizer ground state that can be generated with a finite depth Clifford circuit starting from a product state. We remark however that the actual Clifford circuit $C$ used in the ansatz is automatically found by the ground state search as described below. 

\subsubsection{Augmented Groundstate Search}
To obtain the ground state, we employ the Clifford-Augmented Density Matrix Renormalization Group (CA-DMRG) introduced in \cite{qian2024augmenting}, which proceeds as follows:
\begin{enumerate}
    \item[(i)] Starting from an initial MPS state $|\Psi_0\rangle$, we compute the effective Hamiltonian $H_{\mathrm{eff}}^{(1,2)}$ and diagonalize it using Lanczos methods, as in standard DMRG\cite{schollwock2011density}. This yields un updated pair of site-tensors $A^{s_1}, A^{s_2}$ that minimize the energy on the bond $(1, 2)$.
        
   \item[(ii)] We apply all two-qubit Clifford gates on the bond \( (1,2) \). In practice, it is sufficient to consider the 20 inequivalent representatives of the two-qubit Clifford group up to local Clifford transformations \cite{Frau_2025}, which generate the set of transformed tensors
    \[
        W[C_{1}] = C_{1,2}\,(A^{s_1} A^{s_2}) \, .
    \]

    \item[(iii)] For each candidate in \( W[C_{1}] \), we evaluate the bipartite entanglement entropy across the cut \((1,2)\) and select the state with minimal entanglement. We denote the resulting tensor by \( W[C^{m}_{1}] \), with corresponding optimal Clifford gate \( C^{m}_{1,2} \).
    
   \item[(iv)] We perform a singular value decomposition (SVD) of the tensor \( W[C^{m}_{1}] \), thereby obtaining a new pair of site tensors $A^{s_1}_m A^{s_2}_m$ that are reinserted into the MPS representation. The Hamiltonian and all the relevant observables $O$ are updated accordingly as,
\[
H, O \rightarrow C^{m}_{1,2} \, H \, {C^{m}_{1,2}}^\dagger,C^{m}_{1,2} \, O \, {C^{m}_{1,2}}^\dagger \, .
\]

\item[(v)] Repeat steps (i)–(iv) sweeping through the chain until convergence of the energy is achieved.
\end{enumerate}
In this way, we simultaneously optimize the state to minimize the energy while searching along the Clifford orbit of $|\psi\rangle$ for a representation that minimizes the entanglement.

A crucial feature underlying the efficiency of this procedure, rooted in the fact that Clifford operations are the normalizer of the Pauli group, is that they do not arbitrarily increase the complexity of evaluating expectation values. For a Hamiltonian expressed as $H = \sum_i \alpha_i P_i$, with $P_i$ a Pauli String of $N$ qubits, Clifford conjugation yields $H' = \sum_i \alpha_i P'_i$, where each term remains a Pauli string. As a consequence, the structure of the Pauli expansion is preserved, upper-bounding the growth in operator complexity. In practice, throughout each DMRG sweep we keep track of the Hamiltonian both in MPO form and as a sum of Pauli strings. The former is updated at step (iv) by applying the gate as a tensor and a subsequent SVD while the latter by updating the global Clifford $C$. This causes the bond dimension of the MPO to grow during the sweep and a truncation error in the MPO to accumulate. At the end of the sweep we then check wether the MPO form is still close to that given by the sum of Pauli strings; if not we recalculate the environments with the correct Hamiltonian obtained as the sum of Pauli strings. This produces a negligible overhead in the overall computational cost.

We further note that when pushing the bond dimension to large values performing an augmented search instead of normal DMRG sweeps does not necessarily help. Nevertheless for Sec. \ref{subsec:Performanceresults} we always perform augmented searches.

\subsubsection{Preparation of String State}\label{subsec:stringstateprep}
We here discuss how to adiabatically prepare string states. The core idea is to sequentially converge strings of length $1,...,R$. This is done schematically as follows:
\begin{enumerate}
    \item[(i)] Prepare the ground state $\ket{\Psi}$ of $H$;
    \item[(ii)] Apply a particle pair creation operator  and get $\ket{\psi^{2c}_{\{r,r+e_x\}}}=\sigma^x_{(r,r+e_x)}\ket{\Psi}$;
    \item[(iii)] Run DMRG with initial state $\ket{\psi^{2c}_{\{r,r+e_x\}}}$ and a new Hamiltonian $H_{\{r,r+e_x\}}=H-\mu_0 (V_r+V_{r+e_x})$ that pins charges at $r$ and $r+{e_x}$ to obtain an eigenstate $\ket{\Psi_{\{r,r+e_x\}}^{2c}}$
    \item[(iv)] Increase the charge separation by applying another pair creation operator $\ket{\psi^{2c}_{\{r-e_x,r+e_x\}}}=\sigma^x_{(r-e_x,r)}\ket{\Psi^{2c}_{\{r,r+e_x\}}}$
    \item[(v)] Run another few DMRG sweeps with different initial state $\ket{\psi^{2c}_{\{r-e_x,r+e_x\}}}$ and a new Hamiltonian $H_{\{r-e_x,r+e_x\}}=H-\mu_0 (V_{r-e_x}+V_{r+e_x})$ that pins charges at $r-e_x$ and $r+{e_x}$ to obtain an eigenstate $\ket{\Psi_{\{r-e_x,r+e_x\}}^{2c}}$
    \item[(vi)] Repeat points (iv) and (v) as many times as needed to separate charges by a distance $R$ and get $\ket{\Psi^{2c}_{\{r_0,r_1\}}}$ eigenstate of $H_{\{r_0,r_1\}}=H-\mu_0 (V_{r_0}+V_{r_1})$
\end{enumerate}
We remark that all of the above can be performed with both CAMPS and MPS states. In particular for the case of CAMPS the application of a local operator $O_r$ is achieved by commutating the operator through the Clifford and then applying it to the MPS part of the ansatz:
\begin{align}
    O C\ket{\psi}= C \tilde{O}\ket{\psi}\qquad\textrm{with} \qquad \tilde{O}=C^\dagger OC
\end{align}
Then the DMRG optimization can be done as explained in the previous subsection. However, we notice that optimizing the Clifford part of the ansatz at evert step (v) does not improve convergence and as such we just use the Clifford circuit optimized during the ground state search (i). A possible explanation for the string state to not be suitable for Clifford disentangling is the fact that $1+1$D free boson critical states have been shown difficult to disentangle under general dynamics~\cite{Frau_2025}. Nevertheless, since the main entanglement contribution comes from the ground state area law we find it crucial to keep at least the Clifford circuit optimized for the ground state.
\subsection{Performance results}
\label{subsec:Performanceresults}
\begin{figure*}
    \centering

    \begin{overpic}[width=\linewidth]{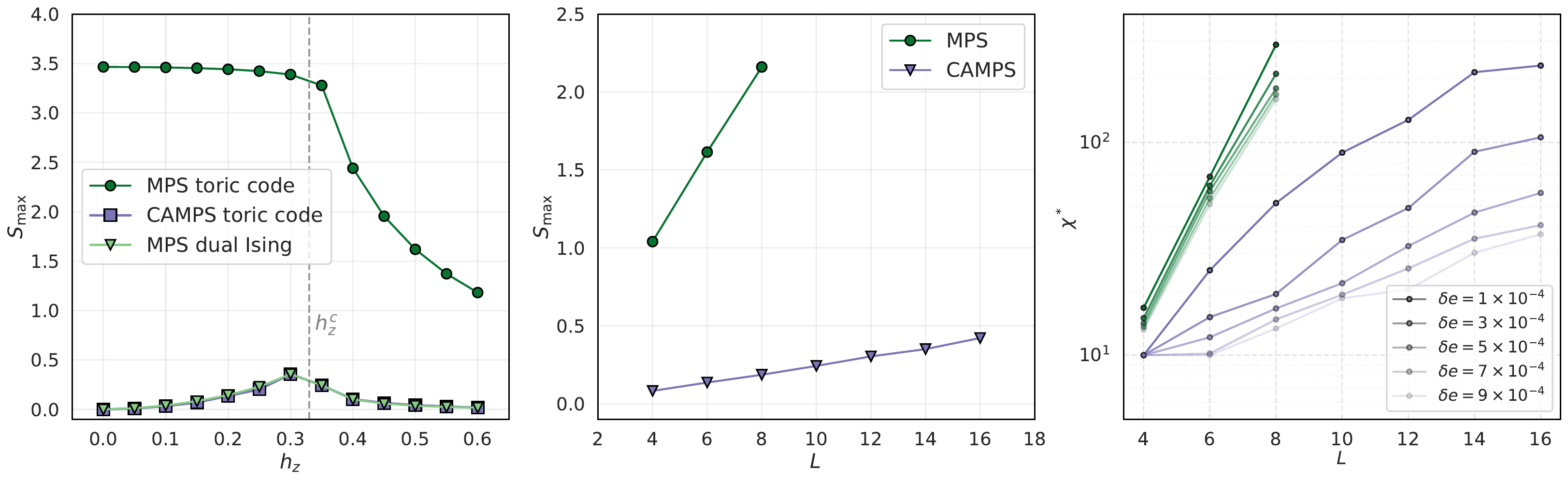}
        \put(5,28){\textbf{(a)}}
        \put(39,28){\textbf{(b)}}
        \put(73,28){\textbf{(c)}}
    \end{overpic}

    \caption{\textbf{Comparison of entanglement scaling and computational costs.} \textbf{(a)} Maximum bipartite entanglement entropy, $S_{\mathrm{max}}$, for the toric code and its dual Ising model as a function of the magnetic field $\vec{h} = (0,0,h_z)$. Ground states are obtained using both standard MPS and CAMPS.
    \textbf{(b)} Maximum bipartite entanglement entropy, $S_{\mathrm{max}}$, of the toric code [Eq.~\ref{eq:TC_hamiltonian}] in a magnetic field $\vec{h} = (0.2, 0.0, 0.5)$, comparing groundstates computed via standard MPS and CAMPS. 
    \textbf{(c)} Bond dimension needed to reach cylinders of size $L\times L$ for the two methods. Different lines correspond to different target accuracies in the groundstate energies. \\
    }
    \label{fig:ent_max}
\end{figure*}

We now wish to provide a systematic study of the performance of CAMPS against MPS as a function of system size in the model under study. We fix in this section $J_v=J_p=1.$, and a cylinder geometry with system size $L\times L$. 

Figure~\ref{fig:ent_max} compares the maximal entanglement entropy of states obtained using standard DMRG and Clifford-augmented DMRG. In Fig.~\ref{fig:ent_max}(b), we show results for various sizes at magnetic field $\vec{h} = (0.2, 0.0, 0.5)$, corresponding to confined phase with gauge-matter coupling. Other fields in the deconfined phase have been studied, showing very similar results. Both MPS and CAMPS ground states exhibit the expected area-law scaling of the entanglement entropy with system size. However, the disentangling action of the Clifford circuit in CAMPS systematically reduces the effective area-law coefficient, resulting in a parametric improvement. In other words, in addition to lowering the entanglement entropy by approximately an order of magnitude, CAMPS also yields a weaker growth with $L$. This reduction translates directly into improved numerical efficiency. With a bond dimension of $\chi=600$, CAMPS allows us to reach lattice sizes up to $16\times16$, which are challenging to access using standard MPS methods - in fact, for the present model, we are not aware of any MPS simulation beyond width of 8, and even those represent remarkable computational tasks. Fig.~\ref{fig:ent_max}(c) quantifies this advantage by showing the bond dimension $\chi^\ast$ required to achieve a given accuracy in the estimation of the energy. The accuracy is measured through the energy density error
\begin{align}
\delta e = \frac{E-E_0}{N_s},
\end{align}
where $N_s=2L(L-1)$ is the number of links, $E$ is the variational energy obtained with a given method and bond dimension, and $E_0$ is a reference energy computed with CAMPS at $\chi=600$.
For both methods, $\chi^\ast$ grows exponentially with system size, but the growth rate is significantly slower for CAMPS. Moreover, the relative advantage of CAMPS becomes more pronounced as the target accuracy is relaxed. 

We remark that the cost of CAMPS is very similar to that of MPS (see Sec. \ref{sec:method}) and that in practice the overhead from extra SVDs needed for the Clifford optimization makes the computational time $\sim 1.3$ larger for CAMPS at the largest fixed bond dimensions. While the results shown here correspond to the sign-free case, we demonstrate in Appendix~\ref{app:sign_problem_gain} that the exponential advantage of CAMPS persists also in the presence of a sign problem (finite $h_y$).

In Fig.~\ref{fig:ent_max}(a), we present results for a fixed system size of $18\times 6$ while varying the magnetic field $\vec{h} = (0, 0, h_z)$. The figure shows four data sets: two corresponding to simulations of the toric code using standard MPS and CAMPS, and two corresponding to simulations of its dual transverse-field Ising model (with spins located at the centers of the plaquettes), also computed using the same two methods. As expected, the entanglement entropy for the standard MPS simulation of the toric code is large in the deconfined (topological) phase for $h_z \lesssim 0.33$, and decreases upon entering the confined (paramagnetic) phase for $h_z \gtrsim 0.33$. In contrast, the CAMPS representation of the toric code exhibits a substantially reduced entanglement entropy across all values of $h_z$. In particular, it vanishes at $h_z = 0$, corresponding to the stabilizer point, and reaches a maximum in the vicinity of the critical point $h_z \approx 0.33$ (with finite-size shifts from the thermodynamic transition). 

Notably, the CAMPS entanglement entropy for the toric code in presence of a single $h_z$ field coincides with that obtained for its dual Ising model. This observation is consistent with the well-established role of duality transformations in entanglement renormalization schemes that incorporate gauge structure, as discussed in Ref.~\cite{entrenandgaugesymmetry}. In the present context, it is remarkable that this duality emerges naturally from the Clifford disentangling procedure, without any explicit prior knowledge of the underlying $\mathbb{Z}_2$ gauge symmetry. By contrast, for the CAMPS simulation of the dual Ising model, no significant improvement over standard MPS is observed. This behavior is consistent with earlier findings \cite{Frau_2025}, which show that certain regions of Hilbert space—including ground states of the one-dimensional Ising model—are not efficiently disentanglable via Clifford transformations. This suggests that the presence of an underlying $\mathbb{Z}_2$ gauge structure is a key ingredient for the effectiveness of Clifford-based disentangling approaches. We importantly stress that the exact mapping to the Ising model is possible only in absence of matter dynamics $h_x=0$, while the performance improvement of CAMPS is present also at $h_x\neq0$.

\begin{figure}
    \centering
    \includegraphics[width=\linewidth]{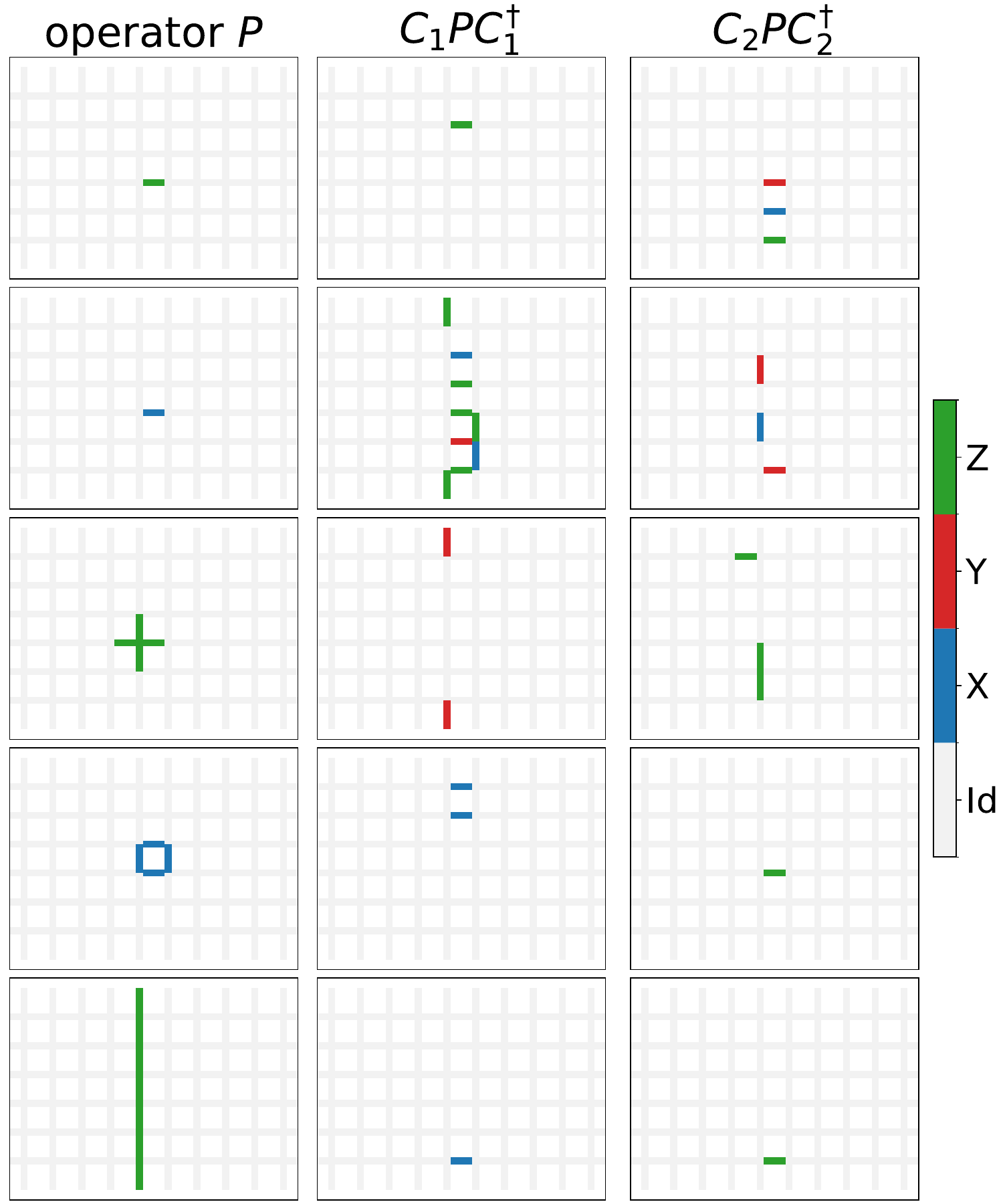}
    \caption{\textbf{Operator mappings under the optimized Clifford circuit in different phases.} Left column:  Few important initial operators $P$. Center column: Mapping of $P$ under the Clifford part of the CAMPS ansatz in the deconfined phase $h_x=0.2$, $h_z = 0.2$ ($C_1$)  Right column: Mapping of $P$ under the Clifford part of the CAMPS ansatz in the confined phase $h_x=0.2$, $h_z = 0.5$ ($C_2$) . The Clifford is non-local up to a range $L_y$, which is needed to reduce the area law prefactor in the entanglement entropy that the MPS part has to represent. }
    
    \label{fig:operatormapping}
\end{figure}

In order to gain further insight on the converged Clifford transformation, we illustrate in Fig.~\ref{fig:operatormapping} how several relevant observables are mapped to different Pauli strings. The emergent Clifford transformation $C_i$ depends on the phase, and we consider two representative cases in respectively deconfined and confined phases: $C_1$, corresponding to $\vec{h}_1 = (0.2, 0.0, 0.2)$, and $C_2$, corresponding to $\vec{h}_2 = (0.2, 0.0, 0.5)$. In both cases, we observe that non-local Wilson loop operators are mapped onto single Pauli operators, consistent with the duality mapping to the Ising model \cite{entrenandgaugesymmetry}. Plaquette and vertex operators are mapped into rather simple operators, while local $\sigma^x$ operators (which do not commute with the Wilson loops) can get scrambled over $\mathcal{O}(L)$ sites. Overall, the obtained Clifford transformations are not completely global; they connect sites at distances of order $L$ in the MPS and not of order $N_s$. 

It is useful to place the performance of CAMPS in the context of recent tensor-network studies of confinement and interface roughening in two spatial dimensions. Recent works have addressed complementary aspects of these problems using different MPS formulations. Simulations based on the dual transverse-field Ising representation have characterized string roughening and short- and intermediate-time dynamics in the pure-gauge $\mathbb{Z}_2$ theory \cite{DiMarcantonioRico_2026commphys_roughening}. MPS calculations for a $U(1)$ lattice gauge theory realized in triangular Rydberg arrays have demonstrated string roughening, the Lüscher correction with around 30\% accuracy, and string-breaking dynamics in quasi-one-dimensional geometries \cite{bombieri2026u1latticegaugetheory}, utilizing impressive values of the bond dimension. Infinite-DMRG calculations have characterized the roughening transition and its critical properties on infinite cylinders, including at finite gauge-matter coupling \cite{Pollman2025roughening}, for system as wide as 8 gauge links. Beyond MPS, a gauge-invariant PEPS formulation has recently been developed for pure Abelian lattice gauge theories, enabling large-scale ground-state calculations on lattices up to $32\times32$ and real-time dynamics of localized excitations directly on $10 \times 10$ lattices\cite{wu2025_pepsdynamics_purelgt}. Tree tensor-network methods have instead been applied to the dual Ising model to study interface roughening and domain-wall dynamics \cite{KrinitsinSchmitt_prl2025_rougheningising}, while tensor-network approaches to the three-dimensional classical Ising model have characterized equilibrium interface roughening in the corresponding dual statistical-mechanics setting \cite{uedatagliacozzo_2026interfaceroughening3dising}. Finally, neural-network quantum states have enabled large-scale ground-state simulations of the toric code  \cite{KufelYao_prl2025_nqs_toriccode} on sizes of up to $16\times 16$. However, these approaches focus primarily on ground states, localized excitations, or interface physics, rather than on the preparation, spectroscopy, and long-time dynamics of extended flux tubes.

CAMPS provides a more comprehensive framework for studying flux tubes. Working directly in the gauge-field representation, it naturally extends to finite gauge--matter coupling while the reduced entanglement enables substantially wider finite cylinders and long real-time dynamical simulations than standard MPS methods. This combination allows us to characterize long flux tubes, extract their low-energy spectrum, sound velocity, and improve the determination of the Lüscher correction through reduced transverse finite-size effects. Unlike conventional MPS calculations, however, the Clifford circuit obscures the bipartite entanglement structure of the physical state, making entanglement-based probes of string roughening less directly accessible.

\section{Static string results}\label{sec:statics}
In this section, we investigate the static properties of the electric-field string and compare our results with the predictions of Effective String Theory. Specifically, in Sec.~\ref{subsec:Lusher} we analyze the confining potential, while in Sec.~\ref{subsec:stringprofiles} we study the transverse profile of the string. We herby fix $J_v=2$ and $J_p=1$.
\subsection{Confining potential and Lüscher term}\label{subsec:Lusher}
\begin{figure}
    \centering
    \begin{overpic}[width=\linewidth]{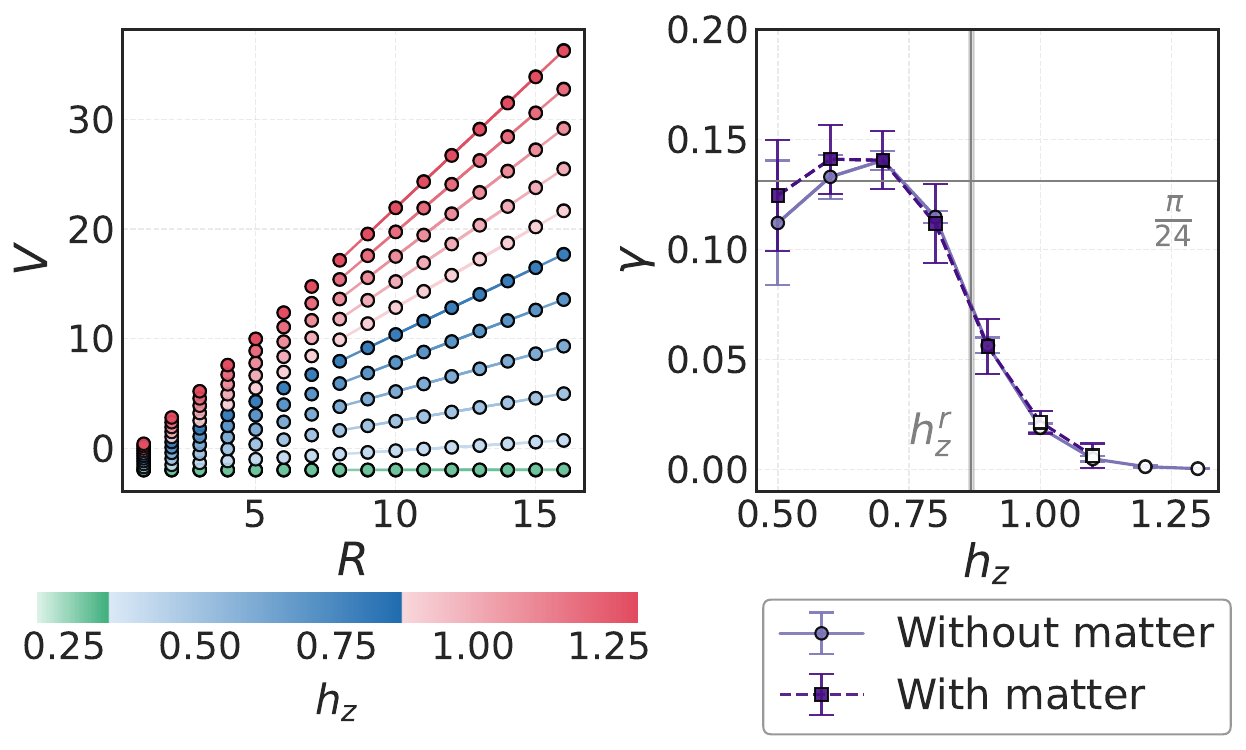}
    \put(12,53){\textbf{(a)}}
    \put(63,53){\textbf{(b)}}
    \end{overpic}
    \vspace{0.15cm}

{\small
\renewcommand{\arraystretch}{1.15}
\setlength{\tabcolsep}{0pt}
\begin{tabular*}{\linewidth}{@{\extracolsep{\fill}}c|ccccccccc@{}}
\toprule
$h_z$ & $0.50$ & $0.60$ & $0.70$ & $0.80$ & $0.90$ & $1.00$ & $1.10$ & $1.20$ & $1.30$ \\
$\sigma$ & $0.42$ & $0.69$ & $0.96$ & $1.22$ & $1.47$ & $1.71$ & $1.95$ & $2.17$ & $2.39$ \\
$v$ & $1.51(3)$ & $1.75(8)$ & $2.07(4)$ & $2.42(3)$ & $2.80(17)$ & $-$ & $-$ & $-$ & $-$ \\
\bottomrule
\end{tabular*}
}
    \caption{\textbf{Confining potential and L\"uscher term.} \textbf{(a)} Confining potential $V$ (Eq.\ref{eq:confining_potential}). Different curves correspond to values of the magnetic field $\vec{h}=(0,0,h_z)$, with $h_z$ uniformly spaced in the interval $[0.2,1.3]$ in steps of $0.1$.Values with $h_z < h_z^c \simeq 0.33$ lie in the deconfined phase (green lines), $h_z^c < h_z < h_z^r \simeq 0.87$ correspond to the weakly confined (rough) phase (blue lines), and $h_z > h_z^r$ to the strongly confined (stiff) phase (red lines). Data are obtained with a bond dimension of $\chi = 100$. The values of the string tension $\sigma$ obtained from the fits are summarized in the table.   \textbf{(b)} Lüscher term $\gamma = \gamma_0 / v$. The coefficient $\gamma_0$ is extracted by fitting Eq.~\ref{eq:confining_potential}. The two curves correspond to $h_x=0.0$ and $h_x=0.2$. The velocity $v$ is obtained from real-time dynamics simulations; its derivation is discussed in Sec.~\ref{sec:dynamics_normalmodes}, while the corresponding numerical values are summarized in the table. Since $v$ can only be reliably extracted in the rough phase and its immediate vicinity, the value $v=2.80$ obtained at $h_z=0.90$ is used for all values of $h_z>0.90$. These data points are indicated by open (white) markers in the plot. The horizontal gray line indicates the expected theoretical values of $\frac{\pi}{24}$, and the vertical gray line marks our estimate of the roughening transition $h_z^r = 0.868 \pm 0.002$ (see Fig.~\ref{fig:string_widths}). }
    \label{fig:stringproperties}
\end{figure}

In Fig.\ref{fig:stringproperties} we present our results for the confining potential $V$ for a cylinder of dimensions $24 \times 8$. The potential is defined as 
\begin{equation}
    V = E_R-E_{R=0 }, 
\end{equation} 
where $E_R$ is the energy of the state with two static charges pinned at separation $R$ (prepared as described in Sec.\ref{subsec:stringstateprep}) and $E_{R= 0}$ is the vacuum contribution. In particular, Fig.~\ref{fig:stringproperties}a shows the dependence of the confining potential on the charge separation $R$ for magnetic fields of the form $\vec{h}=(0,0,h_z)$. As expected, for $h_z<h_z^c\simeq0.33$ (green curves), corresponding to the deconfined phase, the potential remains constant with $R$. In this regime, separating the two charges translates in no additional energy cost, because of the absence of a confining electric-field string. After crossing the confinement transition at $h_z^c$, the potential develops a clear dependence on $R$, indicating the emergence of confinement. In both the weakly confined phase, $h_z^c<h_z<h_z^r\simeq0.868$ (blue curves), and the strongly confined phase, $h_z>h_z^r$ (red curves), the energy required to separate the charges increases with their distance, consistent with the formation of an electric-field string connecting them. We repeated this analysis in the presence of dynamical matter, considering the magnetic field $\vec{h}=(0.2,0.0,h_z)$. We found results consistent with those obtained for $h_x=0.0$. 
We also fit the confining potential $V$ using the functional form in Eq.~\ref{eq:confining_potential}, considering all fitting windows $[R_{\min}, R_{\max}]$ with $R_{\min}\in{7,\dots,12}$ and $R_{\max}\in{11,\dots,16}$. As expected, the extracted string tension $\sigma$ vanishes in the deconfined phase and increases linearly with the magnetic field $h_z$ in the confined regime. 

The Lüscher coefficient $\gamma$ is shown in Fig.~\ref{fig:stringproperties}b. The data points correspond to averages over all fitting windows. For comparison with the theoretical prediction in the rough phase, $\gamma = \frac{\pi}{24}$, we extract the corresponding velocities $v$ for each value of $h_z$ from dynamical simulations, as detailed in Sec.~\ref{sec:dynamics_normalmodes}. The error bars therefore incorporate both the standard deviation of $\gamma_0$ across fitting windows and the uncertainty in the determination of $v$.

We find that the rescaled quantity $\gamma = \gamma_0 / v$ exhibits a clear plateau in the weakly confined phase ($h_z < h_z^r$) at $h_x=0.0$, consistent with the predicted value $\pi/24$, while, as expected, it vanishes in the strongly confined phase. This improved resolution of the plateau, compared to previous results, crucially relies on the extraction of $v$ from dynamical simulations, which reduces systematic uncertainties in the rescaling procedure. It is further enabled by the lower-entangled representation of the string state provided by CAMPS. In addition, we observe that the L\"uscher term displays the expected behavior also in the presence of a finite gauge-matter coupling $h_x = 0.2$.

This result is consistent with the findings of Refs.~\cite{Bonati:2020orj,Bonati:2021vbc}, where the authors studied the behaviour of the flux tube in the $SU(2)$ lattice gauge theory in the presence of bosonic matter and found that the predictions of Effective String Theory, in particular the Lüscher term, remain unaffected by the coupling to bosonic matter.

\begin{figure}
    \centering

    \begin{overpic}[width=\linewidth]{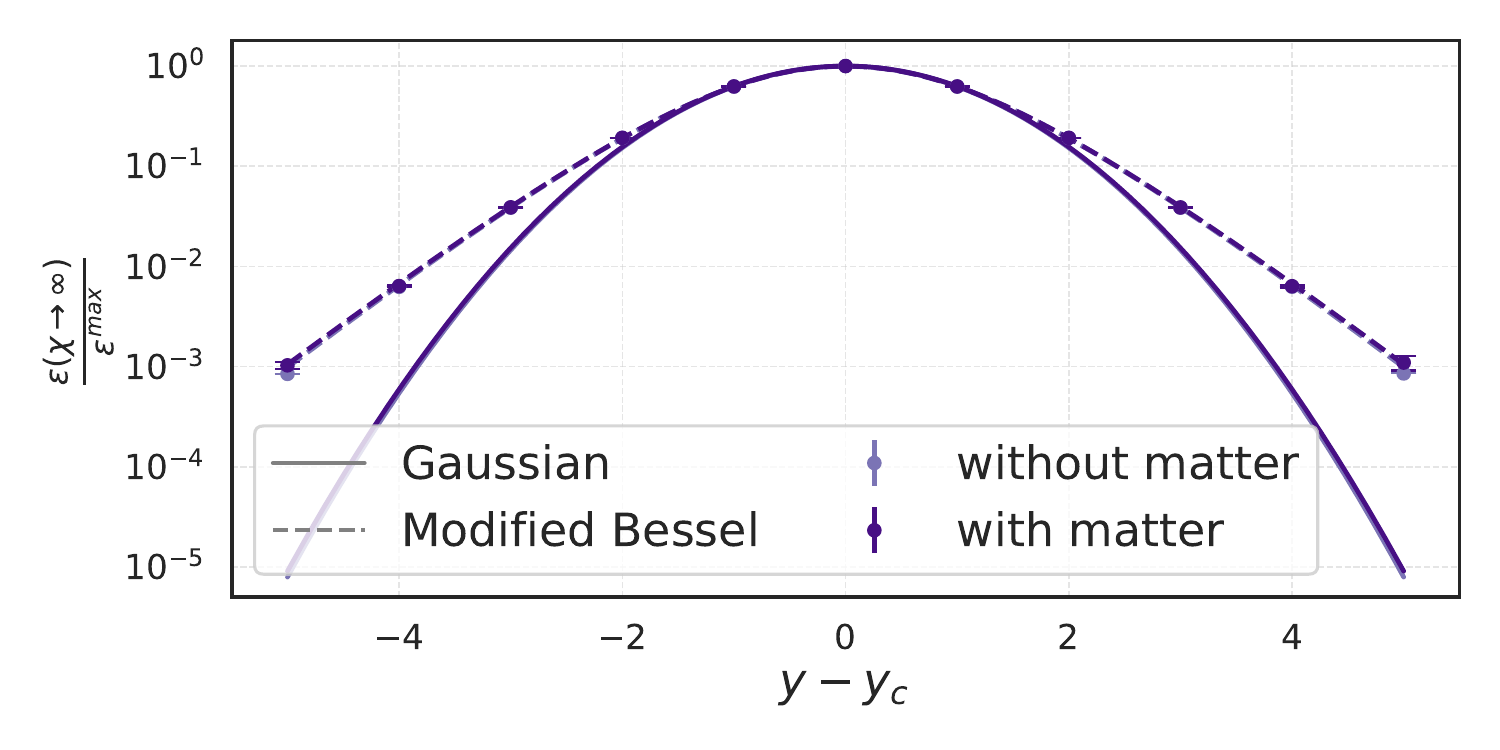}
        \put(17,43){\textbf{(a)}}
    \end{overpic}

    \vspace{0.15cm}

{\small
\renewcommand{\arraystretch}{1.15}
\setlength{\tabcolsep}{0pt}
\begin{tabular*}{0.9\linewidth}{@{\extracolsep{\fill}}ccc@{}}
\toprule
$\chi^2_{\mathrm{red}, K_0}/\chi^2_{\mathrm{red}, G}$ & $w^2$ & $\lambda$ \\
\midrule
$0.002$
& $1.129$ & $0.495$ \\
$0.002$
& $1.127 \pm 0.001$ & $0.505 \pm 0.001$ \\
\bottomrule
\end{tabular*}
}

    \vspace{0.2cm}

    \begin{overpic}[width=\linewidth]{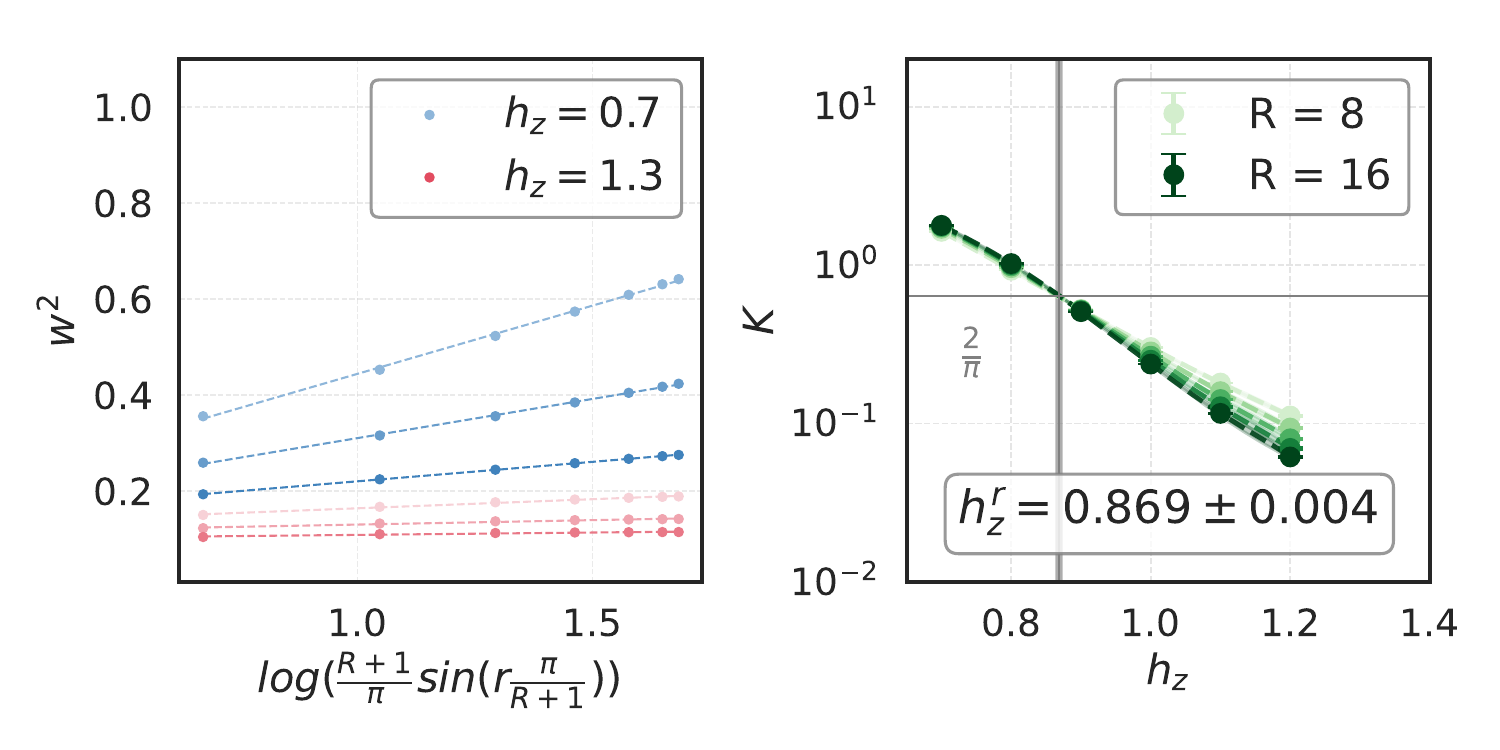}
        \put(13,42){\textbf{(b)}}
        \put(61,42){\textbf{(c)}}
    \end{overpic}

    \caption{\textbf{String profile and widths.}
    \textbf{(a)} Transverse string profiles as a function of position, for size $30\times12$, and for two values of magnetic field: $\vec{h}_1 = (0.0, 0.0, 0.5)$ and $\vec{h}_2 = (0.2, 0.0, 0.5)$. The profile is fitted with two functional forms: a Gaussian, shown by solid lines, and a modified Bessel function, shown by dashed lines. The table below panel (a) reports, for $\vec{h}_1$ (first row) and $\vec{h}_2$ (second row), the ratio $\chi^2_{\mathrm{red}, K_0}/\chi^2_{\mathrm{red}, G}$, together with the fitted $K_0$-ansatz parameters $w^2$ and $\lambda$. Since the modified-Bessel ansatz yields a reduced $\chi^2$ approximately three orders of magnitude smaller than that of the Gaussian ansatz, only the parameters of the former are reported.Errors smaller than $10^{-3}$ are omitted.
    \textbf{(b)} Widths of the string profile extracted from the modified Bessel fits, as a function of distance between the charges, on a $24\times10$ cylinder. Different curves correspond to values of the magnetic field $\vec{h}=(0,0,h_z)$ across the confined phase. $h_z$ is uniformly spaced in the interval $[0.5,1.3]$ in steps of $0.1$. Dashed lines correspond to fits with Eq.~\ref{eq:widths}.
    \textbf{(c)} Luttinger parameter $K$, obtained from the fits in panel (b), as a function of $h_z$ for different charge separations $R = 8, 10, 12, 14, 16$. For clarity, only $R = 8$ and $R = 16$ are labeled in the legend. The horizontal line indicates the theoretical prediction $K_c = 2/\pi$. The crossing of each curve with this critical value is determined via cubic spline interpolation, yielding an estimate of the critical point $h_z^r = 0.868 \pm 0.004$.}
    \label{fig:string_widths}
\end{figure}

\subsection{String profile}\label{subsec:stringprofiles}

In this section, we characterize the confining electric-field string. We first investigate its transverse profile and compare it with phenomenological models of confinement. We then study how the string width evolves across the phase diagram and use this behavior to locate the roughening transition.

Throughout this section, we characterize the string using the electric field on the link
$l=(r,r+\mathbf e_y)$, defined relative to the vacuum configuration in the absence of
external charges,
\begin{align}
\epsilon_l
=
\mathcal E_{\rm str}
-
\mathcal E_{\rm vac},
\label{eq:electric_field}
\end{align}
where the bare electric field is
\begin{align}
\mathcal E_l=-\langle\sigma_l^z\rangle.
\end{align}
\subsubsection{Intrinsic width of the flux tube}


Figure\ref{fig:string_widths}(a) displays the transverse profiles of two strings on a $30\times12$ cylinder with charge separation $R=20$. The profiles are obtained by taking a vertical cut through the center of the string, at $x=R/2$, for two values of the magnetic field in the weakly confined regime: $\vec{h}_1=(0.0,0.0,0.5)$ and $\vec{h}_2=(0.2,0.0,0.5)$. 

The markers correspond to the extrapolated values obtained from fits in the bond dimension $\chi$, while the error bars represent the uncertainty on the asymptotic value extracted from the corresponding exponential fits. The profiles are rescaled by their maximum value $\epsilon_{\max}$, so that the peak of each curve is set to unity. Details of the bond-dimension extrapolation procedure are provided in App.~\ref{app:conv_static}.

Looking at Fig.~\ref{fig:string_widths}(a), we see that the profile of the flux tube differs significantly from the Gaussian shape predicted by the Nambu--Goto string. In fact, fitting the data with a Gaussian function results in very poor $\chi^2$ values. This is, however, a well-known phenomenon. As discussed in Sec. \ref{sec:effective_stringtheory}, the EST describes the flux tube as an infinitely thin string, whereas the physical flux tube has a well defined finite transverse thickness, commonly referred to as its "intrinsic width". Such an intrinsic width is expected to be present in any lattice gauge theory and has recently been investigated in detail for the $SU(2)$ gauge theory in Ref.~\cite{caselle2026intrinsicwidthfluxtube}.

Finding a theoretical model  able to combine this finite intrinsic width into the Nambu--Goto EST remains an important open problem, and its solution would certainly provide valuable insight into the microscopic mechanism responsible for confinement. In this respect a  particularly interesting approach is provided by the so-called Clem ansatz~\cite{Clem:1975ohd}, which is rooted in Polyakov's original picture of confinement as a consequence of monopole condensation. Although this idea was first developed for compact $U(1)$ gauge theory in $(2+1)$ dimensions, it was later generalized to non-Abelian gauge theories, leading to the well-known dual superconductor picture of confinement~\cite{Mandelstam:1974pi,tHooft:1981bkw}.

Within this framework, Refs.~\cite{Cea:2012qw,Cea:2017ocq} proposed a phenomenological expression for the transverse profile of the flux tube based on the model introduced by Clem~\cite{Clem:1975ohd} to describe the magnetic vortex profile in an ordinary superconductor. In our notation, the Clem profile reads
\begin{equation}
\label{eq:clem}
    \rho(y) = A \, K_0 \! \left( \frac{\sqrt{y^2 + \xi^2}}{\lambda} \right),
\end{equation}

where $K_0$ denotes the modified Bessel function of the second kind of order zero.

Besides the overall normalization constant $A$, the profile depends on two independent length scales, $\xi$ and $\lambda$. Within the dual superconductor picture, these correspond respectively to the vortex core radius and to the London penetration depth. It is straightforward to verify that Eq.~\eqref{eq:clem} interpolates smoothly between a Gaussian-like behaviour near the core ($\epsilon \propto \exp(-\frac{y}{2w^2})$)  for $y\ll\xi$
and an exponentially decaying tail,($\epsilon \propto \exp(-\frac{|y|}{\lambda})$) for $y\gg\xi$

The fit results are summarized in the table of Fig.~\ref{fig:string_widths}. The reduced chi-squared ratio $\chi^2_{\rm red,K_0}/\chi^2_{\rm red,G}=0.002$ demonstrates that the Clem ansatz provides a dramatically better description of the transverse profile than a Gaussian fit. Accordingly, only the parameters extracted from the Clem fits are reported. Introducing a finite gauge--matter coupling, $h_x=0.2$, has no appreciable effect on either the transverse profile or the fitted parameters.

As discussed in Ref.~\cite{caselle2026intrinsicwidthfluxtube}, the exponential decay of the flux-tube tails is not determined solely by the effective string dynamics, but also reflects the interplay with the degrees of freedom of the underlying gauge theory. In particular, the slowest screening mode of the vacuum is expected to control the large-distance behavior of the transverse profile.

As a consequence, the parameter $\lambda$ governing the exponential decay of the transverse profile is expected to be related to the inverse of the lightest glueball mass ($0^+$ in this case \cite{Agostini:1996xy}). In our setup, this mass can be extracted from the decay of the connected correlator of plaquette operators $B_{r_1}$ and $B_{r_2}$, which, as discussed in App. \ref{app:glueball}. For a relativistic massive dispersion in 2+1D, this is expected to behave as
\begin{equation*}
\langle B_{r_1} B_{r_2} \rangle_c \propto \frac{1}{|r_1-r_2|}\exp\left(-\frac{|r_1-r_2|}{\lambda_{\mathrm{GB}}}\right).
\end{equation*}
We have also tested pure exponential behaviour (see App. \ref{app:glueball}) finding that it is preferred only away from the continuum limit.
At the specific point $h_z=0.5$, we obtain $\lambda_{\mathrm{GB}} = 0.512 \pm 0.005$, roughly consistent with the fitted values of the intrinsic width $\lambda$ for both $\vec{h}_1$ and $\vec{h}_2$. 

It is also interesting to study the dependence of this length scale $\lambda_{GB}$ as a function of $h_z$ which diverges as the continuum limit is approached (see App. \ref{app:glueball}). Defining a scale invariant quantity $\lambda_{GB}/\sqrt{\sigma}$, we obtain in the continuum limit:
\begin{align}
    h_z\to h_z^c : \; \; \lambda_{GB}^{-1}/\sqrt{\sigma}=3.19 \pm 0.07
\end{align}

This value is very close to that reported in Ref.~\cite{Agostini:1996xy}, namely 3.08(3). The small positive difference could be due to residual contamination of the plaquette operator by heavier glueball states. This is suggested by the fact that including shorter range data in the correlation function fit slightly increases the extrapolated continuum value to $3.36\pm0.06$ (see App.\ref{app:glueball}). 


\subsubsection{Roughening transition}
Having established the transverse profile of the confining string, we now investigate how its width evolves across the phase diagram.

In Fig.~\ref{fig:string_widths}(b), we show the spatial dependence of the string width $w^2$ along the flux tube for two charges pinned at a distance $R=16$ on a $24\times10$ cylinder. The width is extracted at positions $r\in{1,\ldots,R/2}$ by fitting the transverse profile $\epsilon(y)$ with the Clem ansatz. The markers represent the resulting values of $w^2$, obtained using a bond dimension $\chi=100$. Different colors correspond to magnetic fields of the form $\vec{h}=(0.0,0.0,h_z)$, with $h_z$ ranging from $0.5$ to $1.3$ in steps of $0.1$. As discussed above, introducing a small finite gauge-matter coupling $h_x$ does not affect the estimation of $w^2$ and is therefore omitted from the present analysis. Dashed lines represent the fits with Eq.\ref{eq:widths} for each magnetic field. We observe that, for $0.5\leq h_z<h_z^r$, the width of the strings grows logarithmically with the charge-separation $R$, in agreement with the Effective String Theory prediction for the fluctuations of the height-field. Beyond the roughening transition, located at $h_z^r \simeq 0.868$ (a more detailed explanation of the determination of the transition point is presented in the next panel), the flux-tube width no longer exhibits logarithmic growth with $R$ and instead saturates to an approximately constant value. This behavior signals the onset of a stiff flux-tube regime, in which transverse fluctuations are strongly suppressed.

In Fig.~\ref{fig:string_widths}(c), we show the Luttinger parameter $K$ for five different charge separations $R\in[8, 10, 12,14, 16]$. Here, the markers denote the values of the Luttinger parameter obtained from the fits, while the error bars represent the corresponding uncertainties on these fitted parameters. The roughening transition belongs to the Berezinskii-Kosterlitz-Thouless (BKT) universality class, for which the Luttinger parameter takes the critical value $K_c=2/\pi$ at the transition. Consequently, the critical field $h_z^r$ can be identified as the point where the curves $K(h_z)$ cross this threshold. The curves corresponding to different charge separations intersect in the vicinity of the same point, providing evidence for an underlying scale-invariant critical field. Small deviations from an exact crossing are expected due to finite-size effects and the characteristic logarithmic corrections of BKT transitions, which induce a slow drift of the apparent critical point with $R$.

To estimate the critical field $h_z^r$, we perform a bootstrap-based interpolation of the data. Specifically, for each charge separation $R$, we generate $N_s=100$ resampled datasets of the form
\begin{equation*}
K_{\mathrm{sample}} = K + \mathcal{N}(0, K_{\mathrm{err}}),
\end{equation*}
where $K$ denotes the fitted values of the Luttinger parameter and $K_{\mathrm{err}}$ their corresponding uncertainties. For each sample, we compute a cubic spline interpolation of $K^*_{\mathrm{(i)}}(h_z)$ and we identify the crossing value ${h_r^{(i)}}_R$. The estimate of ${h_z^r}_R$ at fixed $R$ is then obtained as the mean over the 100 bootstrap realizations, with the corresponding uncertainty given by their standard deviation. The results for $L_y=8$ and $L_y=10$ are
\[
\begin{array}{c|c|c|c}
R & {h_z^r}_R& {h_z^r}_R& {h_z^r}_R\\
  & L_y=8;h_x = 0.0 &L_y=10;h_x = 0.0& L_y=8; h_x =0.2\\
\hline
8 & 0.8658(2) & 0.8659(2) & 0.8668(2)\\
10 & 0.8682(2) & 0.8681(2) & 0.8691(2)\\
12 & 0.8694(2)& 0.8692(2) & 0.8703(1)\\
14 & 0.8700(2) & 0.8697(1) & 0.8709(1)\\
16 & 0.8702(2) & 0.8699(1) & 0.8711(1)
\end{array}
\]

The results for $L_y=8$ and $L_y=10$ are in very good agreement for all charge separations, with differences remaining below $3\times10^{-4}$ and in any case within uncertainties. This indicates that the estimate of the critical field is effectively converged with respect to the transverse system size $L_y$, and we therefore use the larger-$L_y$ results without applying a further finite-$L_y$ correction. Above we also report finite $R$ crossings values at $L_y=8$ for finite gauge-matter coupling $h_x=0.2$ converged at $\chi=300$. 

Finally, we average the results over $R$ to obtain the final estimate 
\[
h_z^{r} = 0.868 \pm 0.004
\]
where the error is taken as the difference between the maximum and minimum estimates obtained for the three charge separations. Other recent MPS studies reported similar values. Ref. \cite{DiMarcantonioRico_2026commphys_roughening} reported $h_z^r\simeq(0.91)^2\simeq0.83$ at $h_x=0$ by studying the kink mass of the string. Ref. \cite{Pollman2025roughening} instead studied the transition with an infinite string length and finite transverse direction $L_y=6-8$ and including dynamical matter, reporting  $h_z^r\simeq0.855-0.865$ at $h_x=0.$ and $h_z^r\simeq0.875-0.88$ at $h_x=0.2$ and $J_v=2$. 

\section{String spectroscopy}\label{sec:dynamics}
In this section we analyze dynamical properties of the electric field string connecting charges in the confined phase. In particular we start by describing the excitation protocol (Sec. \ref{sec:dynamics_protocol}). Then we focus respectively on early time dynamics (Sec. \ref{sec:dynamics_earlytime}) and longer times (Sec. \ref{sec:dynamics_normalmodes}) from which the normal modes spectrum of the string can be inferred.
\subsection{String-pull protocol}\label{sec:dynamics_protocol}

\begin{figure*}
    \centering
    \begin{overpic}
[width=0.24\linewidth]{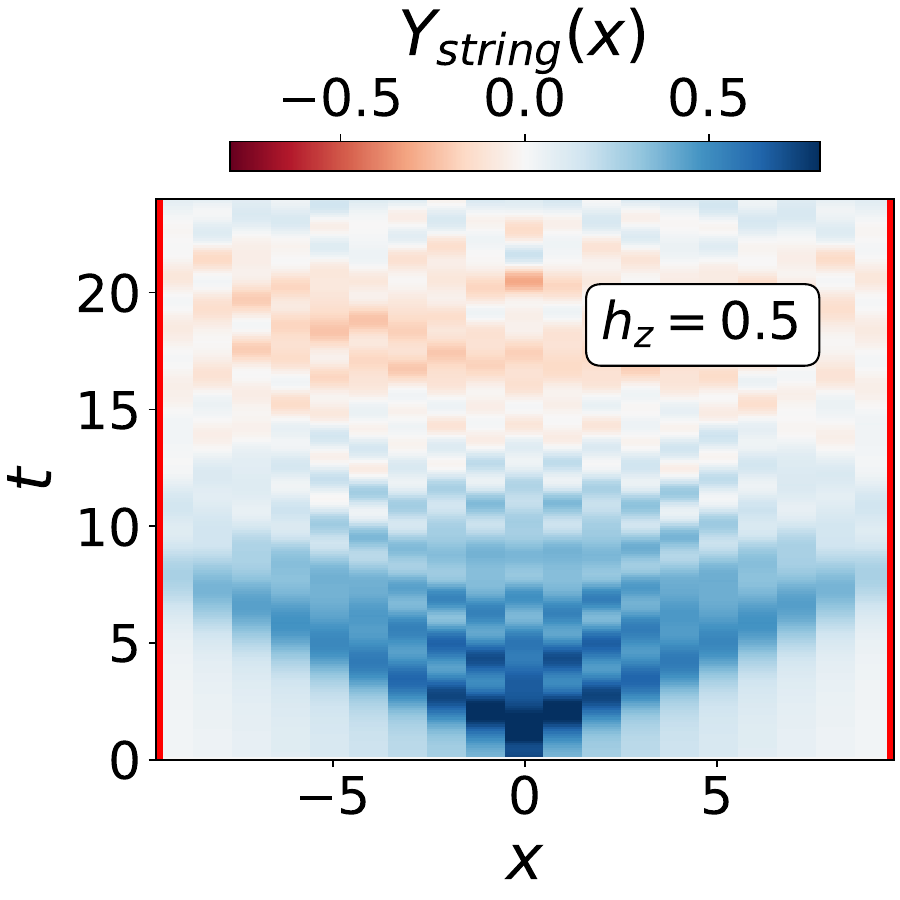}
        \put(20,68){\textbf{(a)}}
    \end{overpic}
        \begin{overpic}
[width=0.24\linewidth]{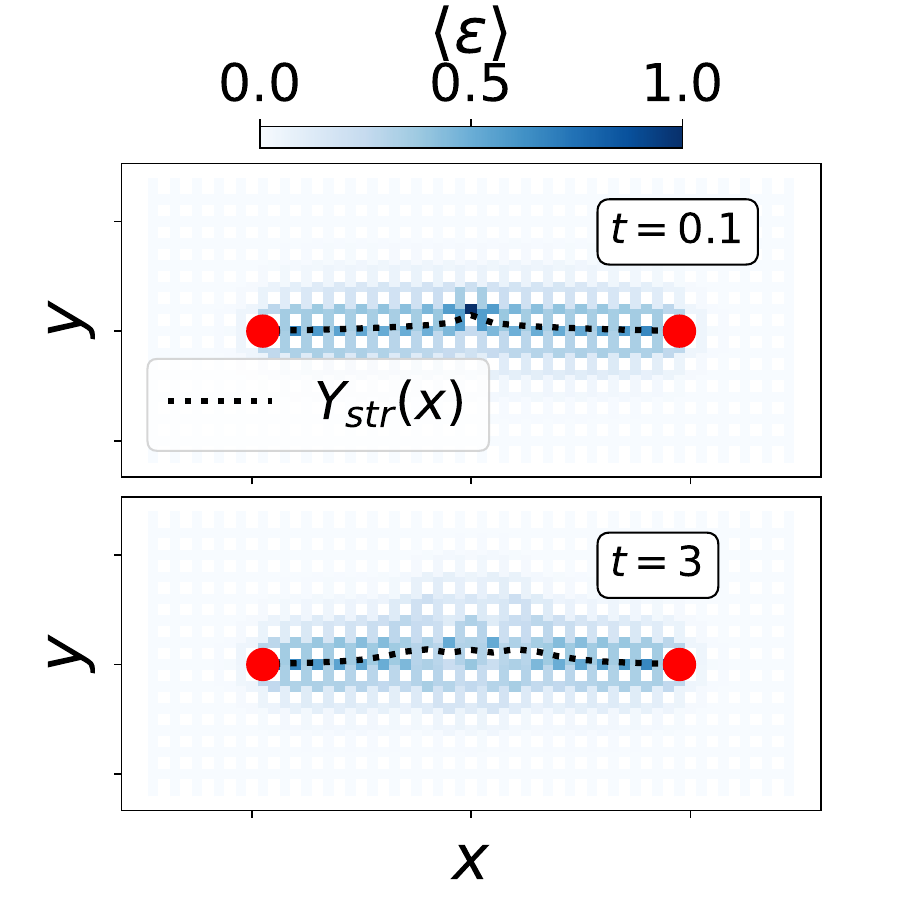}
            \put(15,72){\textbf{(c)}}
            \put(15,35){\textbf{(b)}}

    \end{overpic}
    \begin{overpic}
[width=0.24\linewidth]{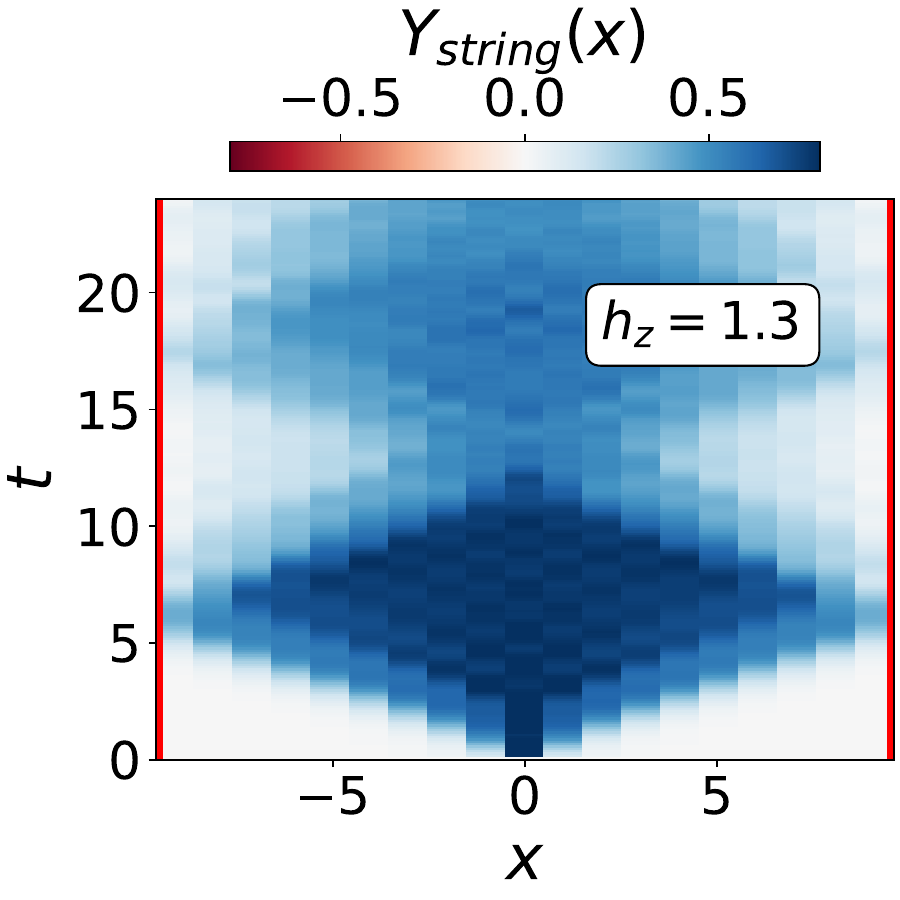}   
\put(20,68){\textbf{(d)}}
    \end{overpic}
    \begin{overpic}
[width=0.24\linewidth]{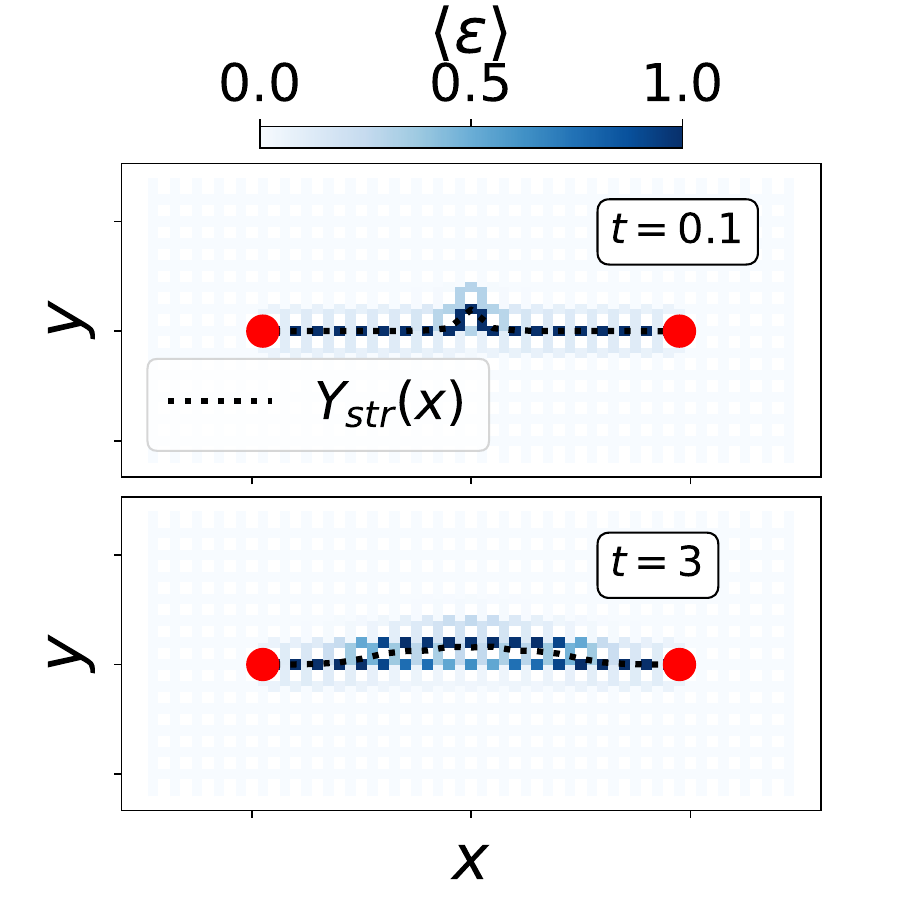}
            \put(15,35){\textbf{(f)}}
            \put(15,72){\textbf{(e)}}
            \end{overpic}
    \caption{\textbf{String-pull dynamics} Real time dynamics after the string-pull protocol in the weakly confined phase $h_z=0.5$ \textbf{(a,b,c)} and in the strongly confined phase $h_z=1.3$ \textbf{(d,e,f)}. In panel \textbf{(a)} and \textbf{(d)} we show the the mean transverse position $Y_{string}(x)$ as function of longitudinal coordinate and time. In panels \textbf{(b,c,e,f)} we instead show the full snapshots of the electric field at two different times, right after the string has been pulled $t=0.1$ \textbf{(b,e)} and after a small time $t=3$ \textbf{(c,f)}. System size $30\times12$ and string length $R=19$; $\chi=100$.}
    \label{fig:propagation}
\end{figure*}
The string-pull protocol used in this section aims at exciting the string connecting a pair of charges in the transverse direction. As schematically shown in Fig. \ref{fig:fig1}(e), the protocol consists in the following steps:
\begin{enumerate}
    \item Prepare a string state $\ket{\psi}$ of length $R$ with charges at $r_1=(-R/2,0)$ and $r_2=(R/2,0)$, eigenstate of $H$;
    \item Project onto an excited string state $\ket{\tilde{\psi}}=P_{l_m}\ket{\psi}$ with a projector $P_{l_m}=\left(1-\sigma^z_{l_m}\right)/2$ such that the string must pass through the link $l_m=\left([-\frac{1}{2},1],[\frac{1}{2},1]\right)$ (see Fig. \ref{fig:propagation}(b) and (e));
    \item Perform a unitary evolution $\ket{\tilde{\psi}(t)}=e^{-iHt}\ket{\tilde{\psi}}$ and measure observables (see Fig. \ref{fig:propagation}(c) and (f)).
\end{enumerate}
The physical interpretation of the above protocol is that of an instantaneous \textit{pull} of the string by one lattice site with respect to its rest position. This sets the transverse position in an out-of-equilibrium configuration, hence populating all transverse modes with the correct symmetry (odd modes of Eq. \eqref{eq:normal_modes} are not excited). The observable we will use to keep track of the dynamics is indeed the transverse string position:
\begin{align}\label{eq:def_Yav}
    Y_{str}(x) = \frac{\sum_{y} y\,\epsilon_{(r,r+e_x)} }{\sum_{y}\epsilon_{(r,r+e_x)}}
\end{align}
where $r=[x-0.5 ,y]$ such that $x$ is link centered. We center the string around $x=0$.

We now explain more in detail how all of the above are implemented numerically. The first step corresponds to the string state preparation used in section \ref{sec:statics}. The CAMPS state is obtained via DMRG with the Clifford unitary being optimized only on the ground state in absence of charges. The second step projects the electric field expectation value to be $\langle \mathcal{E}_{l}\rangle =+1$ at the specific link $l_m$. Note that when the state $\ket{\psi}$ is represented with a CAMPS $\ket{\psi}=C\ket{\psi_{0}}$ ($C$ the Clifford unitary and $\ket{\psi_0}$ the MPS) then the projector $\Pi_l=(1-\sigma^z_{l_m})$ has to be commuted through the Clifford to be applied to the MPS part:
\begin{align}
    \ket{\tilde{\psi}} =\Pi_l \,C\ket{\psi_0}= C(C^\dagger \Pi_lC)\ket{\psi_0} = C\ket{\tilde{\psi}_0}
\end{align}
where $\ket{\tilde{\psi}_0}=(C^\dagger \Pi_lC)\ket{\psi_0} $. The Clifford unitary transform the local projection into a non-local one which can be represented by an MPO of bond dimension 2. This can be then simply applied to the MPS part which can be later compressed back. The third step is then performed via a standard TDVP algorithm on the MPS part with a Clifford transformed Hamiltonian:
\begin{align}
   \ket{\tilde{\psi}(t)}= e^{-iHt}C\ket{\tilde{\psi}_0}= Ce^{-i\left(C^\dagger H C\right)t}\ket{\tilde{\psi}_0}
\end{align}
Differently from what has been tried in other works \cite{qian2024clifford, mello2024clifford}, we keep the Clifford unitary the same at every time step, as we expect the string dynamics to not generate too much entanglement, keeping the Clifford unitary optimized for the original ground state is enough. Finally, we remark that similar protocols have been used to probe very high energy dynamics in experiments, in a context that is not directly related to effective string theory \cite{Cochran2025_visualizingstring_google}.

\subsection{Real-time dynamics}\label{sec:dynamics_earlytime}
We now present the real time dynamics following the string-pull protocol introduced in the previous section. We focus mostly on the static matter scenario ($h_x=0.$), but similar results can be obtained for the case of  finite gauge-matter coupling (see App. \ref{app:string_dyn_withmatter}) and study systems on cylinders of up to $30\times12$.\\

The dynamical response of the string to a pull strongly differs in the rough and stiff phases. This is explicitly shown by the snapshots in Fig. \ref{fig:propagation} representing the electric field string connecting the pinned charges. In the case of weak confinement ($h_z=0.5$) the string is free to fluctuate after the pull and at a later time its average transverse position is passing through 0 having started from 1 at $t=0^+$. On the other hand for strong confinement ($h_z=1.3$) the string is tightly bound to the lattice and its average position $Y_{str}$ in the pulling point remains positive. These can also be seen in Fig.\ref{fig:propagation} where the transverse string position is tracked for all times as a function of longitudinal coordinate $x$. Here it is also clear that transverse string excitations propagate with a light-cone, however we find that the effective velocity of such spreading strongly depends on the string length $R$ (not shown). This can be explained by the fact that the string-pull is highly localized in space and hence has high overlap with high-momentum modes where the dispersion is not linear and lattice corrections become important (see also Sec. \ref{sec:dynamics_normalmodes}).

\begin{figure}
    \centering
    \begin{overpic}[width=\linewidth]{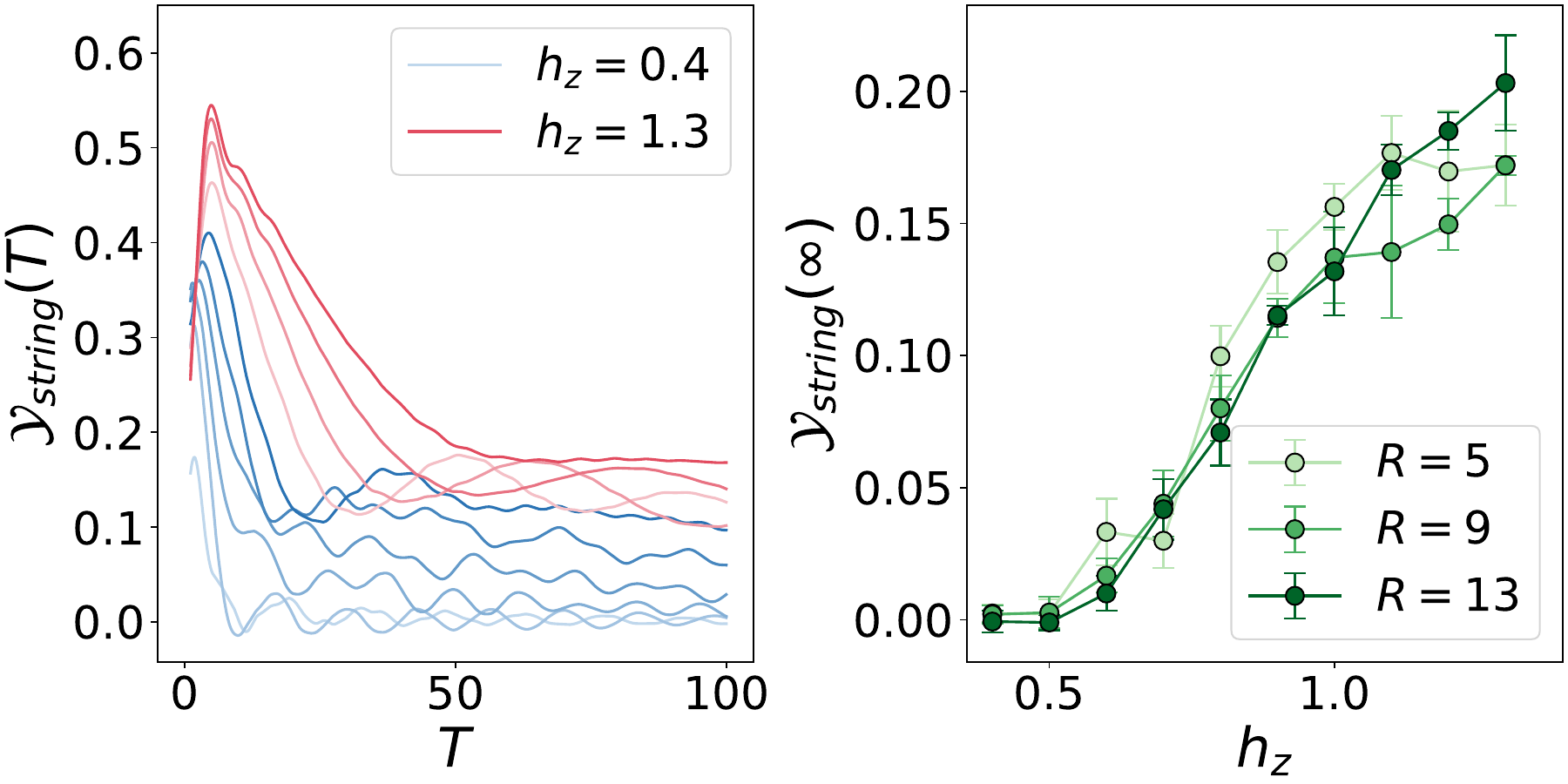}
    \put(11,45){\textbf{(a)}}
    \put(63,45){\textbf{(b)}}
    \end{overpic}
    \caption{\textbf{String lattice-locking:} \textbf{(a)} Time-space averaged string position $\mathcal{Y}_{string}(T)$ (see Eq. \eqref{eq:Ytav_def}) as a function of the integrating time $T$ and for different values of $h_z$ across the roughening transition. \textbf{(b)} Long-time value of $\mathcal{Y}_{string}$ for different string lengths $R$ across the transition; errorbars are estimated as variations at late times $T\in[80,100]$. System size $18\times 6$; $\chi=100$. }
    \label{fig:string_locking}
\end{figure}

Nevertheless, the different long-time dynamical behaviors of the string after a pull can be used to track the roughening transition. This is investigated in Fig. 
\ref{fig:string_locking} where we show the time-space-averaged string transverse position:
\begin{align}\label{eq:Ytav_def}
    \mathcal{Y}_{string}(T)=\frac{1}{RT}\int_0^Tdt\,\sum_x Y_{string}(x,t)
\end{align}
We choose to use an all-time average instead of a running-time average in order to be less sensitive to slow modes which are indeed present (see later discussion on Fig. \ref{fig:normal_modes}). At large integrating times $T$ this quantity will detect whether after the pull the string oscillates around its rest position or not. The latter scenario is realized in the strongly confined phase, where the lattice effectively locks the string. While we do see a lack of thermalization in the available times in the stiff phase, we expect a complete thermalization to eventually occur at larger times. Complementary dynamical signatures have been recently discussed in the roughening dynamics of interfaces in the Ising model \cite{KrinitsinSchmitt_prl2025_rougheningising}. 

\subsection{String normal modes}\label{sec:dynamics_normalmodes}
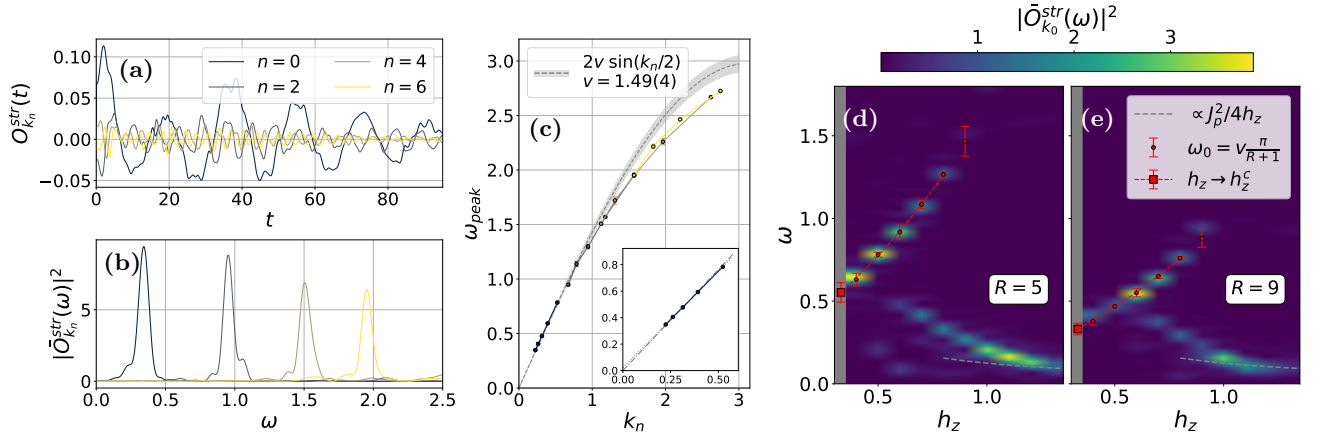
\begin{figure*}
    \centering
    \begin{overpic}[width=0.55\linewidth]{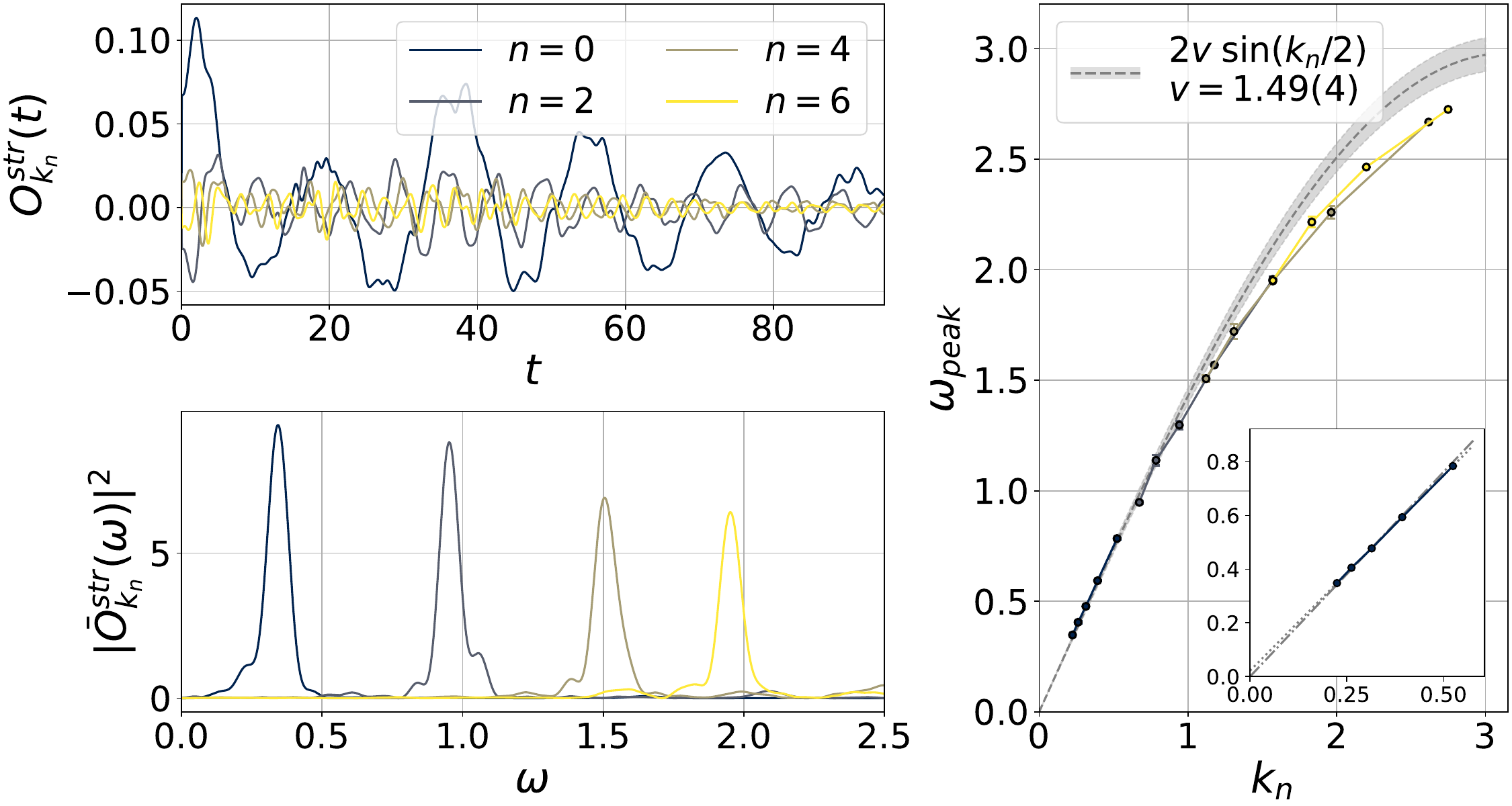}
    \put(15,48){\textbf{(a)}}
    \put(13,22){\textbf{(b)}}
    \put(70,40){\textbf{(c)}}
    \end{overpic}
    \begin{overpic}
        [width=0.4\linewidth]{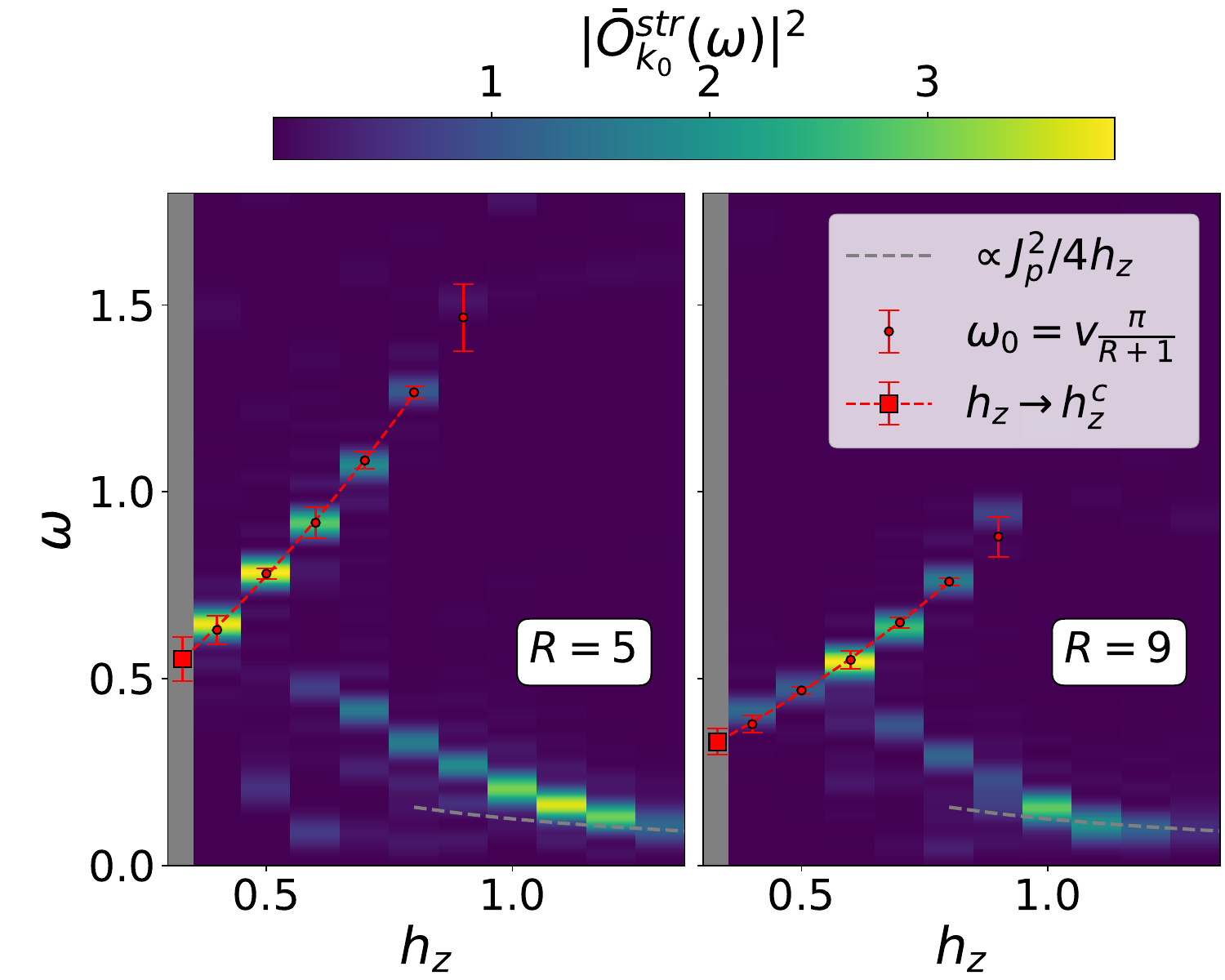}
        \put(15,57){\textcolor{white}{\textbf{(d)}}}
        \put(59,57){\textcolor{white}{\textbf{(e)}}}

        \end{overpic}
    \caption{\textbf{String normal modes:} Real time expectation value \textbf{(a)} and its normalized fourier transform \textbf{(b)} for the even modes $n=0,2,4,6$ with profile $\phi_{k_n}(x)$ at $R=13$ at $h_z=0.5$. \textbf{(c)} Energy extracted for $n=0,2,4,6$ at different $R=5,7,9,11,13$ as a function of $k_n=\pi(n+1)/(R+1)$ for $h_z=0.5$. The free-boson dispersion is shown as a gray dashed line with a velocity extracted from the fundamental mode $n=0$. The inset shows the linear extrapolation used to extract the velocity (dotted $\omega=vk+c$ and linear dotted-dashed $\omega=vk$). Panel \textbf{(d)} and \textbf{(e)} show the evolution of the fundamental mode spectra across the roughening transition for $R=5$ and $R=9$. The extracted velocity is shown in red, while the gray dashed line is a guide to the eye for the transverse fluctuations activated by the string-pull protocol. The gray region signals the continuum limit at $h_z^c\simeq0.33$. System size in \textbf{(a,b,c)} is $18\times8$ while in \textbf{(d,e)} is $18\times 6$; $\chi=100$.}
    \label{fig:normal_modes}
\end{figure*}
The string dynamics after the local string-pull can be used to extract important information regarding its transverse modes, expected to be described by the effective model described in Sec. \ref{sec:effective_stringtheory}. 

To do so, we track the projection of the string transverse position $Y_{str}(x)$ against the expected open boundary modes for a free boson $\phi_n(x)$ (see Eq. \eqref{eq:normal_modes}) as:
\begin{align}
    O_n^{str}(t)= \frac{1}{R}\sum_x\phi_n(x)Y_{str}(x,t)
\end{align}
The result is shown in Fig. \ref{fig:normal_modes}(a) for a string of length $R=13$. As all odd $n$ modes have a zero in the middle point where the string is pulled, their population remains low at all times ($<10^{-3}$) and is thus not shown. Even modes $n=0,2,4,6$ instead have a large population which oscillates with a well resolved periodicity. To extract their frequency, we take a fourier transform of the signal $ O_n^{str}(t)$ in a window $t\in [t_1,t_2]$ and normalize it such that its modulus square integrates to 1:
\begin{align}
    \bar{O}^{str}_n(\omega)=\frac{1}{\mathcal{N}_n}\int _{t_1}^{t_2}dt \,e^{i\omega t}O_n^{str}(t)
\end{align}
This is shown in Fig. \ref{fig:normal_modes}(b) where $t_1=15$ to avoid early time physics and $t_2=100$ the longest available time. All signals show a strongly peaked fourier decomposition, consistently with the fact that each mode is an eigenstate of the evolution. The width of the peaks is limited by the finite time evolution. Finally, by studying different modes $n=0,2,4,5$ at different string lengths $R=5,7,9,11,13$ we can faithfully reproduce the expected linear dispersion $\omega_n\simeq vk_n$ in Fig. \ref{fig:normal_modes}(c). In order to account for the lattice discretization in the $x$ direction, we compare the data with a free boson lattice dispersion $\omega_n=2v \sin(k_n/2)$ which in any case recovers the expected continuum result as small $k_n$. At large momenta $k_n\sim 1$ non-universal correction kick in, likely coming from the presence of a lattice in the $y$ direction (the cosine term in the effective model  \eqref{eq:EST_hamiltonian}). The velocity $v=1.49(4)$ is extracted from a linear fit of the lowest mode frequency $n=0$ as shown in the inset of Fig. \ref{fig:normal_modes}(c). The errorbar in the velocity is taken as the difference between a constrained fit $\omega=vk_n$ and an unconstrained one $\omega=vk_n+c$. 

This method can be used to extract the effective velocity of the string throughout the rough phase. We do so in Fig. \ref{fig:normal_modes}(d,e) where we show the lowest mode fourier decomposition across the rough and stiff phase for two string length $R=5$ and $R=9$. In the rough phase there is a clear mode corresponing to the fundamental mode whose energy increases towards the roughening transition $h_z\simeq0.9$ and then disappears. Comparing the two sizes clarifies that such mode is of collective nature and its energy goes down with $R$. From these two sizes we extract a velocity, whose value is used in Fig. \ref{fig:normal_modes}(d,e) to reproduce the values of the fundamental frequency (red circle markers). 

We further extrapolate the fitted velocity to the continuum limit, that is the confinment transition $h_z^c\simeq0.328$. By taking only points within the rough phase $h_z=0.4,..,0.8$ and extrapolating with $v(h_z)=v_{cont} +a(h_z-h_z^c)+b(h_z-h_z^c)^2$ we get:
\begin{align}
    v_{cont}=1.05\pm0.11
\end{align}
which agrees with a continuum expectation of $v=1$, based on arguments of emergent Lorentz invariance at the critical point. Note that this differs from a suggested \cite{DiMarcantonioRico_2026commphys_roughening}constant value of $v\sim2$ in the rough phase extracted from purely static observables. The errorbar here includes both statistical error of the fit and the error on each velocity as previously discussed. The resulting extrapolated fundamental mode frequency is shown in Fig. \ref{fig:normal_modes}(d,e) as a square red marker on a gray shaded region. A similar continuum limit can also be taken with matter dynamics (see App. \ref{app:string_dyn_withmatter}), giving compatible results:
\begin{align}
    v_{cont}^{h_x=0.2}=1.26\pm0.17
\end{align}
although with a larger errorbar. Note that string-breaking phenomena are here highly suppressed as the energy injected by the string pull protocol is not enough to break the string (see App. \ref{app:string_dyn_withmatter}).

In the stiff (strongly confined) phase there is another very low frequency mode, which is roughly independent of $R$. We identify this with a local transverse fluctuations of the string which is activated by the string-pull protocol. The reasoning is the following. In the large $h_z\gg J_p$ regime the string is described by a classical configuration on top of which plaquette terms generate localized transverse fluctuations with an energy cost $\sim 4h_z$, that is the cost of increasing the string length by 2. When the string is pulled on the positive $y$ side, the local fluctuations induced by the plaquette generate a second order process which tunnels the excitation to negative $y$ passing through an off-resonant intermediate state which correspond to the string at rest. The expected strength of such process is of order $J_p^2/4h_z$ which we show as dashed gray lines in Fig. \ref{fig:normal_modes}(d,e). Indeed, as the fundamental mode expectation value has a finite overlap with the average transverse position of the string, it will show up in its fourier decomposition.

It is interesting to use these large scale results to interpret the trotterized string dynamics observed in Ref. \cite{Cochran2025_visualizingstring_google}. By accessing long enough strings and times, we are able to clearly distinguish collective string modes from the aforementioned localized oscillations which dominate the strongly confined regime. Furthermore the string-lattice-locking is revealed to be a very robust feature also of long strings and at long times $tJ_p\simeq 100$.

\section{Conclusions and outlook}

In this work, we used Clifford-augmented matrix product states to investigate the static and real-time properties of confining strings in a $(2+1)$-dimensional $\mathbb{Z}_2$-Higgs lattice gauge theory. The reduced entanglement carried by the MPS component of the ansatz allowed us to access extended strings and evolution times that are challenging for conventional MPS simulations. Our main results can be summarized as follows:

\begin{itemize}

\item[-] We showed that the optimized Clifford circuit substantially reduces the area-law coefficient of the entanglement entropy. Such a reduction is enabled by the fact that Clifford dressing is able to take care of entanglement due to gauge invariance and duality, a feature that guides future application to other gauge theories. Our approach leads to a {\it parametrically} increasing computational advantage with the transverse system size, and allows simulations of cylinders up to widths that would require computational resources (both memory and time) larger by at least 2 orders of magnitude for the volumes considered here, and more for larger ones. 

\item[-] We obtained a quantitative characterization of the static confining string. From the intercharge potential, and using an independently determined string velocity, we recovered the universal L\"uscher coefficient $\pi/24$ throughout the weakly confined rough phase. The agreement persists at finite gauge-matter coupling, within the regime where string breaking remains suppressed.

\item[-] We characterized the transverse structure of the flux tube across the roughening transition. Its width displays the logarithmic broadening predicted by Effective String Theory in the rough phase, while its microscopic profile is strongly non-Gaussian and is accurately described by a modified-Bessel form. The associated intrinsic scale is connected to the bulk glueball correlation length.

\item[-] We introduced a string-pull protocol that directly excites and resolves the transverse normal modes of the string. In the rough phase, the resulting dynamics is collective and is described at long wavelengths by an approximately linear gapless dispersion. This allowed us to extract the renormalized propagation velocity and to obtain a continuum extrapolation compatible with the emergent
relativistic value $v=1$. Our predictions put on firm ground Effective String theory as a tool to not only predict vacuum properties, but also dynamical ones, that are inaccessible to Monte Carlo simulations, and that we verify here for the first time.

\item[-] In the strongly confined regime, the dynamics changes qualitatively. The string remains predominantly locked to the lattice and the response is governed by long-lived local transverse processes rather than propagating collective modes. The characteristic frequency is consistent with a perturbative
scale of order $J_p^2/(4h_z)$.

\end{itemize}

Taken together, these results establish a quantitative
connection between Hamiltonian tensor-network simulations and the effective low-energy description of confining strings. The static calculations provide stringent benchmarks against Effective String Theory and Euclidean lattice results, while the string-pull protocol accesses dynamical information that is difficult to obtain through imaginary-time methods. In particular, the crossover from propagating collective modes to lattice-dominated local motion provides a dynamical
characterization of the roughening phenomenon complementary to conventional equilibrium observables.

More broadly, our findings demonstrate that Clifford-augmented tensor networks offer a promising route toward controlled real-time simulations of extended objects in two-dimensional gauge theories. The fact that the optimized Clifford circuit naturally captures part of the underlying gauge and duality structure suggests that similar approaches may be effective in other constrained quantum systems - a key aspect, since for traditional statistical mechanics models, parametric gain over MPS has only been observed in one-dimensional gapless phases. The protocol introduced here is also formulated in terms of local state preparation and measurements, making it directly relevant for quantum simulation experiments.

Several extensions can be pursued. An important next step is to enter regimes where matter-pair production and string breaking occur on accessible time scales, and to study the resulting fragmentation and scattering dynamics. It would also be interesting to consider different local quenches, collisions between strings and charges, and dynamics close to the deconfinement and roughening transitions. Generalizations to other gauge groups may be pursued: in particular, our findings provide key insights on which lattice regularization might be better suited. For instance, $SU(2)$ shall use a qubit realization, while for $SU(3)$, whose center is not compatible with $\mathbb{Z}_2$, a qudit regularization shall be employed. Having access to $\mathbb{Z}_3$ would already open the possibility to study barionic matter. These can be tackled by generalizing the CAMPS ansatz to qudits by changing the Clifford group to that of qudits. Also, it would be interesting to understand whether the systematic advantage of CAMPS over MPS carries over to 3D systems, thanks to the fact that the Hilbert space reduction due to Gauss law is more drastic than in 2D, and if Clifford augmentation on other tensor network structure could help~\cite{felser2021efficient,yosprakob2026cliffordcircuitsaugmentedgrassmann}, possibly even in digitized string dynamics~\cite{banerjee2022nematic}.

Finally, we hope that our work can stimulate a set of new questions in effective string theory, based on the fact that one can now probe the entire string dynamics at the wave function level. In particular, it would be interesting to attack this problem from the point of view of quantum correlations, such as entanglement, and test its relation to confinement at the dynamical level ~\cite{Klebanov_2008,tu2020einstein,Amorosso:2024leg, Amorosso:2026mdo}.

\section{Acknowledgments}
The tensor network part of the numerical simulations has been carried out with ITensor \cite{ITensor}. We thank Gerald Fux for providing codes that integrate the stabilizer formalism with ITensor. We thank Markus Heyl, Julien Vidal for insightful discussions.
M.C. would like to thank T. Canneti,  A. Mariani, A.Nada, M. Panero, D. Panfalone and L. Verzichelli for many insightful discussions.
The work of M. C. was supported by
the Simons Foundation grant 994300 (Simons Collaboration on Confinement and QCD
Strings) and by the SFT Scientific Initiative of INFN.
LT acknowledges the support from the Proyecto Sinérgico CAM Programa TEC-2024/COM-84 QUITEMAD-CM; from the CSIC Research Platform on Quantum Technologies PTI-001 and from the grant PID2024-160172NB-I00 funded by MICIU/AEI/10.13039/501100011033 and by FEDER,UE.
M.~D. was partly supported by the EU-Flagship programme Pasquans2, by INFN Iniziativa Specifica Quantum, by ONR Global (grant number: 14608147), and by the ERC Consolidator grant WaveNets (Grant agreement ID: 101087692).

\appendix 

\section{Performance of CAMPS with $h_y$}
\label{app:sign_problem_gain}
In this appendix we demonstrate that the advantages of CAMPS persist in the presence of a sign problem, namely for finite $h_y$.

\begin{figure}
    \centering

    \begin{overpic}[width=\linewidth]{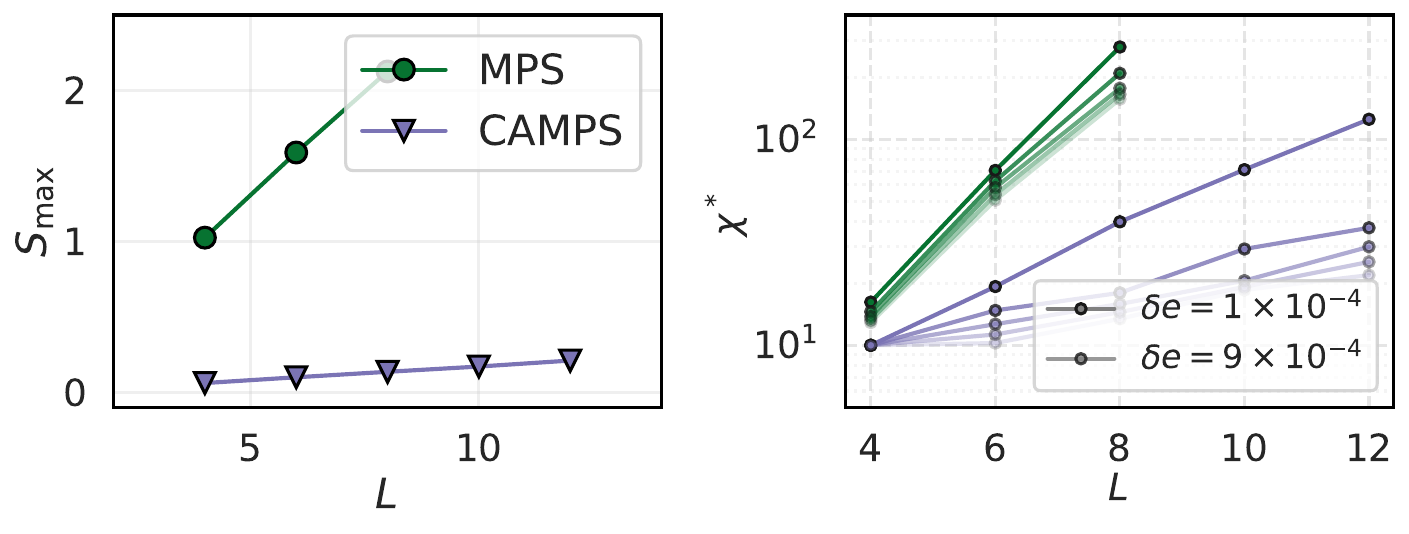}
        \put(9,32){\textbf{(a)}}
        \put(62,32){\textbf{(b)}}
    \end{overpic}
    \caption{\textbf{Comparison of entanglement scaling and computational costs in presence of $h_y$.} \textbf{(a)} Maximum bipartite entanglement entropy, $S_{\mathrm{max}}$, of the toric code [Eq.~\ref{eq:TC_hamiltonian}] in a magnetic field $\vec{h} = (0.0, 0.2, 0.5)$, comparing ground states computed via standard MPS and CAMPS. 
    \textbf{(b)} Bond dimension needed to reach cylinders of size $L\times L$ in a magnetic field $\vec{h} = (0.0, 0.2, 0.5)$ for the two methods. Different lines correspond to different target accuracies in the ground-state energies. \\
    }
    \label{fig:ent_max_with_hy}
\end{figure}

Figure~\ref{fig:ent_max_with_hy} reports the same analysis presented in the main text, now performed at finite $h_y$, specifically for $\vec{h} = (0.0, 0.2, 0.5)$. Fig.\ref{fig:ent_max_with_hy}(a) shows that, as in the sign-free case, the ground states obtained with both standard DMRG and Clifford-augmented DMRG obey the expected area-law scaling of the entanglement entropy with system size. Nevertheless, the Clifford circuit continues to act as an efficient disentangler, systematically reducing the area-law coefficient. Consequently, CAMPS produces states with substantially lower entanglement entropy and a weaker dependence on the system size than conventional MPS, despite the presence of a sign problem.

This reduction in entanglement translates directly into improved numerical efficiency. Figure~\ref{fig:ent_max_with_hy}(b) shows the bond dimension $\chi^\ast$ required to achieve a given target accuracy. As in the sign-free case, $\chi^\ast$ grows exponentially with the system size for both methods. However, the growth rate is significantly smaller for CAMPS, demonstrating that its exponential advantage is robust against the onset of the sign problem. These results indicate that the disentangling mechanism underlying CAMPS is largely insensitive to the sign structure of the wavefunction, making the method broadly applicable beyond sign-problem-free models.
\section{Glueball mass}\label{app:glueball}
In this appendix we provide additional discussion on how the first glueball mass is extracted. Glueballs are pure gauge excitations and can be in general classified in terms of their symmetries, i.e. angular momentum, parity etc. In the context of the $Z_2$ LGT in 2+1D the lowest mass is the $0^+$ glueball ($0$ angular momentum and $+$ for parity)  \cite{Agostini:1996xy}. While the exact wavefunction of this excitation is not fixed, the simplest gauge configuration that fits these quantum number is a simple flip of electric field links on a plaquette, that is directly generated by $J_p$. 

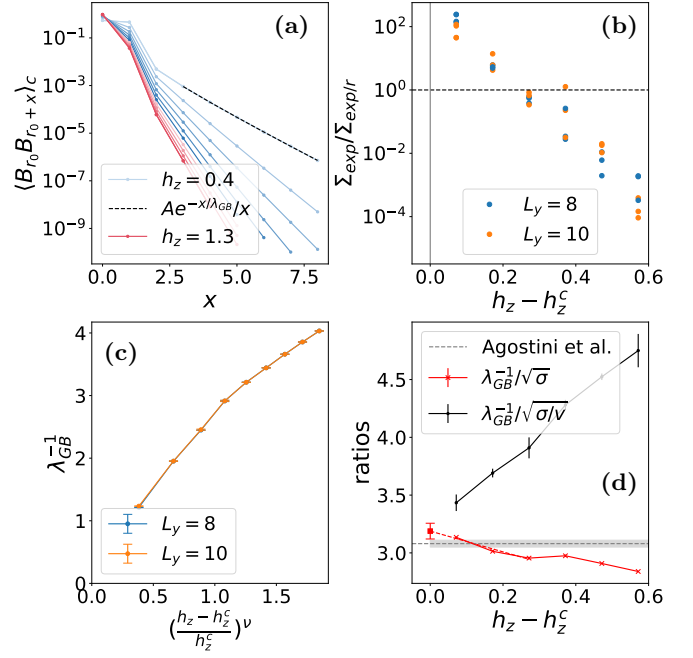
\begin{figure}
    \centering
        \begin{overpic}[width=\linewidth]{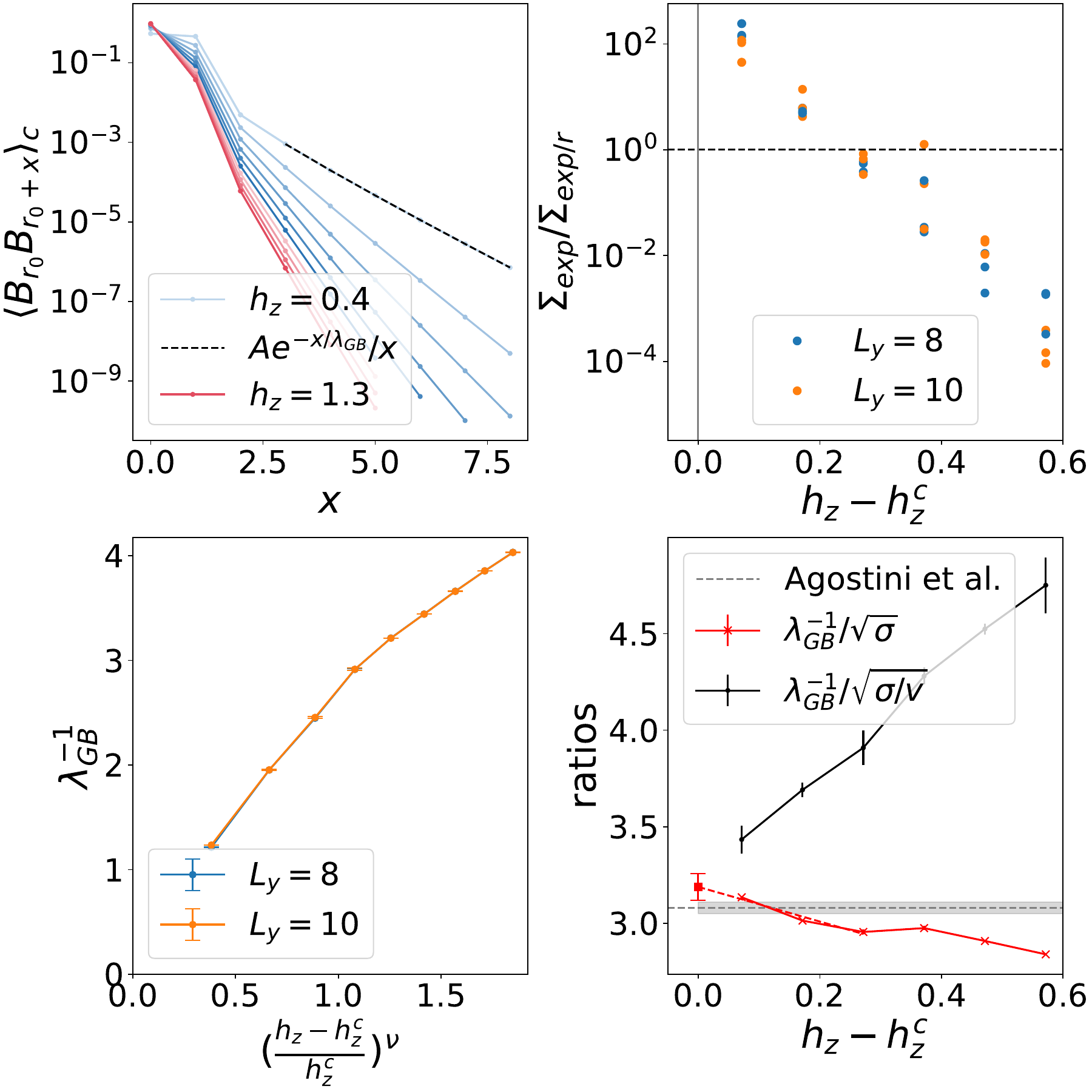}
    \put(40,95){\textbf{(a)}}
    \put(14,45){\textbf{(c)}}
    \put(90,25){\textbf{(d)}}
    \put(90,95){\textbf{(b)}}

    \end{overpic}

    \caption{\textbf{Glueball mass and continuum limit} \textbf{(a)} Plaquette connected correlator for different fields $h_z$, corrected exponential fit shown as a dashed line. \textbf{(b)} Comparison of residues between pure exponential $\Sigma_{exp}$ and corrected exponential $\Sigma_{exp}/r$  fits for different $L_y$ and $h_z$. \textbf{(c)} Approach of the fitted glueball mass to the critical point ($\nu=0.63$ for 3D Ising) at different sizes $L_y$.  \textbf{(d)} Scale invariant ratios as a function of distance from the critical point $h_z^c$ and extrapolation (square marker, dashed red line). The gray dashed line reports the value obtained by Agostini et al. \cite{Agostini:1996xy} for the lowest glueball mass ratio. System of size $30\times10$ in \textbf{(a,d)} and $30\times L_y$ in \textbf{(b,c)}  }
    \label{fig:glueball_mass}
\end{figure}
In order to extract its mass $\Delta_{GB}$ (in energy units), one would need in principle to perform real-time spectroscopy of the plaquette operator to reconstruct a related dynamical structure factor. Instead, we compute its connected equal time two point correlator $\langle B_{r}B_{r'}\rangle_c$, as shown in Fig. \ref{fig:glueball_mass}(a) for different $h_z$ and fixed system size $30\times10$. For a free massive particle in the continuum, the equal-time two point function should behave as:
\begin{align}
    \langle B_{0}B_{\boldsymbol{r}}\rangle \propto \int d^2\boldsymbol{q} \,e^{i\boldsymbol{q}\cdot\boldsymbol{r}}\frac{1}{E(|\boldsymbol{q}|)}\propto \frac{e^{-|\boldsymbol{r}|/\lambda_{GB}}}{|\boldsymbol{r}|}
\end{align}
where we have used $E(q)=\sqrt{\Delta_{GB}^2+(v_{GB}q)^2}$ with $\lambda_{GB}^{-1}=\Delta_{GB}/v_{GB}$, i.e. the relativistic energy dispersion expected to be valid close to the continuum limit or at low momenta. On the lattice corrections at high momenta can give rise to corrections which are in principle non-universal and that we neglect. Note that we have introduced two scales, a mass gap $\Delta_{GB}$ and a velocity $v_{GB}$ which does not in principle need to coincide with the string velocity $v$ discussed in the main text (at least away from the continuum limit).

To confirm this functional form, we compare the following log-spaced residues:
\begin{align}
    \Sigma_f = \sum_x [\log(y)-\log(f(x))]^2
\end{align}
for functions $f(x)=Ae^{-x/\lambda_{GB}}$ and $f(x)=Ae^{-x/\lambda_{GB}}/x$; here $y$ are the data points. Fig. \ref{fig:glueball_mass}(b) shows the ratio $\Sigma_{exp}/\Sigma_{exp/r}$ for two cylinder circumferences $L_y=8,10$ and for different data sets that only depend on the starting point of the correlation function $r_0$. It is clear that approaching the continuum limit the power law corrected exponential is better than the bare exponential fit (ratio $>1$) while away from it the pure exponential gives a better fit (ratio $<1$). This guarantees that our data are precise enough to distinguish the actual 2+1D behaviour close to the continuum limit.

We then extrapolate this length scale, as well as the string tension $\sigma$ (not shown), to the continuum limit $h_z^c$. This is done in Fig.\ref{fig:glueball_mass}(c,d). First in panel (c) we show, for different cylinder circumferences, the glueball mass $\lambda^{-1}_{GB}$ approaching the continuum limit as function of the rescaled distance $((h_z-h_z^c)/h_z^c)^\nu$ with $\nu=0.63$ of the 3D Ising universality class. Then in panel (d), we focus on two scale invariant ratios $\lambda^{-1}_{GB}/\sqrt{\sigma/v}$ (black) and $\lambda^{-1}_{GB}/\sqrt{\sigma}$ (red) \cite{Agostini:1996xy} where $v$ is the effective string velocity obtained in Sec. \ref{sec:dynamics}. While the former has no dimensions but requires the extrapolation of the velocity, the latter can depend in principle on a space-time rescaling but does not require the extraction of a velocity. We use the ratio without the velocity as its errorbars are smaller given that no dynamical calculation has to be performed, furthermore the value of the velocity in the continuum is compatible with 1. We further choose to use only the three closest points to the transition to perform the extrapolations as these are compatible with the power law corrected behavior (panel (b)). The linear extrapolation gives a continuum value:
\begin{align}
    \lambda_{GB}^{-1}/\sqrt{\sigma} \to 3.19\pm0.07
\end{align}
The obtained value for this ratio is compatible with the value $3.08\pm0.03$ obtained with Monte Carlo \cite{Agostini:1996xy} and reported as a dashed line in panel (d). We note that choosing a fitting window far enough from $x=0$ is needed to avoid corrections from more massive glueball excitations. Choosing as a fitting window $x\in[2,6]$ for example gives a continuum value of $3.36\pm0.06$.

\section{String breaking}\label{app:string_breaking}
In this appendix we expand on string breaking phenomena in presence of matter dynamics $|h_x|>0$. 
\begin{figure}
    \centering
    \begin{overpic}
[width=\linewidth]{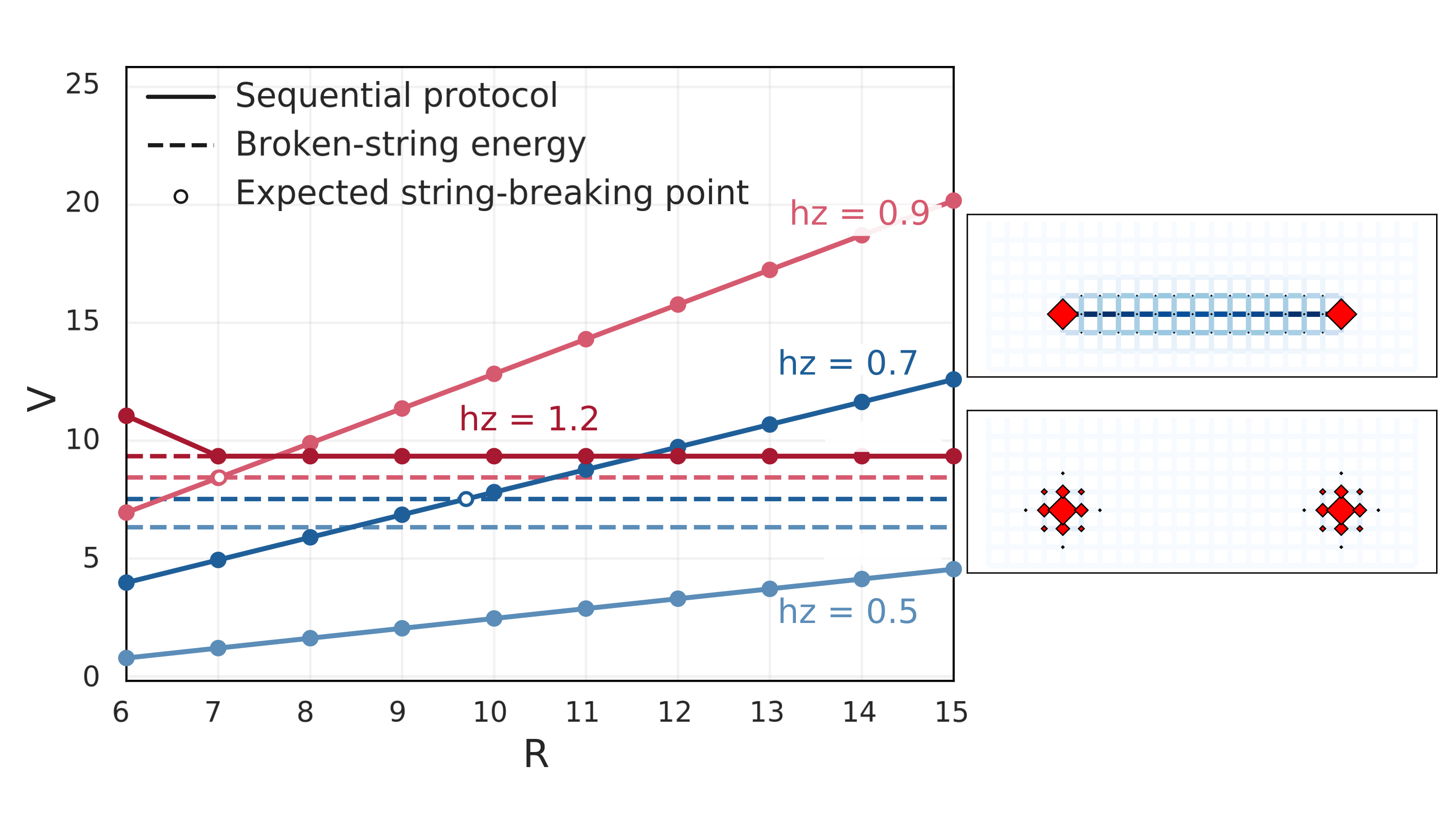}
         \put(10,54){(\textbf{a)}}
            \put(68,44){\textbf{(b)}}
            \put(68,12){\textbf{(c)}}

    \end{overpic}
    \caption{\textbf{String breaking} \textbf{(a)} Energy of string states obtained with the sequential protocol vs string length $R$ (full lines) compared to energies of string broken states obtained by a random initial state (dahsed lines). Example of string state \textbf{(b)} and string broken state \textbf{(c)} for $h_z = 0 .7$, electric field shown with a blue colorbar and charge density with red diamonds. System size $24\times 8$, $h_x=0.2$ and $J_v=2$.}
    \label{fig:string_breaking}
\end{figure}
In this case, having very long strings $E\sim\sigma R$ can become energetically unfavorable compared to the creation of two charges at cost $4J_v$. The sequential procedure explained in the main text (Sec. \ref{subsec:stringstateprep}) biases the DMRG with good initial states such that string states are actually converged. Avoiding this procedure and starting with a random state instead converges a broken string state. This is shown in Fig. \ref{fig:string_breaking}. In particular we show the energy of different string states as a function of $R$ obtained with a sequential protocol (full lines) and by a random initial state (dashed lines). The electric field and charge density of these two scenarios are shown in panel (b) and (c). In the latter a charge of $\sim 2$ is accumulated around both of the two end-points
determined by a further chemical potential ( Sec. \ref{subsec:stringstateprep}). The expected string breaking point (open circles in Fig. \ref{fig:string_breaking}(a) ) strongly depends on the string tension (controlled by $h_z$). For intermediate $h_z$ we see how the sequential protocol allows for a faithful study of the string state, despite it being metastable. At larger $h_z\sim1.2$ instead we note that even the sequential protocol converges a string broken state (full dark red line) whose energy is independent of $R$. 

We understand this in terms of an expected exponentially small matrix element between the broken string and the full string $\sim (h_x/J_v)^R$; as one needs to go through many intermediate states that create a pair of charge and move it. Thus when the crossing point happens at large enough $R$, the string broken state cannot be reached from the metastable string by local updates present in DMRG.

\section{String dynamics with matter}\label{app:string_dyn_withmatter}
In this appendix we report on results of the string-pull protocol described in Sec. \ref{sec:dynamics} in presence of matter dynamics, i.e. $h_x=0.2$ and $J_v=2.$. The large value $J_v$ suppresses string breaking phenomena in the regimes of $R$ and $h_z$ explored. 

\begin{figure}
    \centering
    \includegraphics[width=0.66\linewidth]{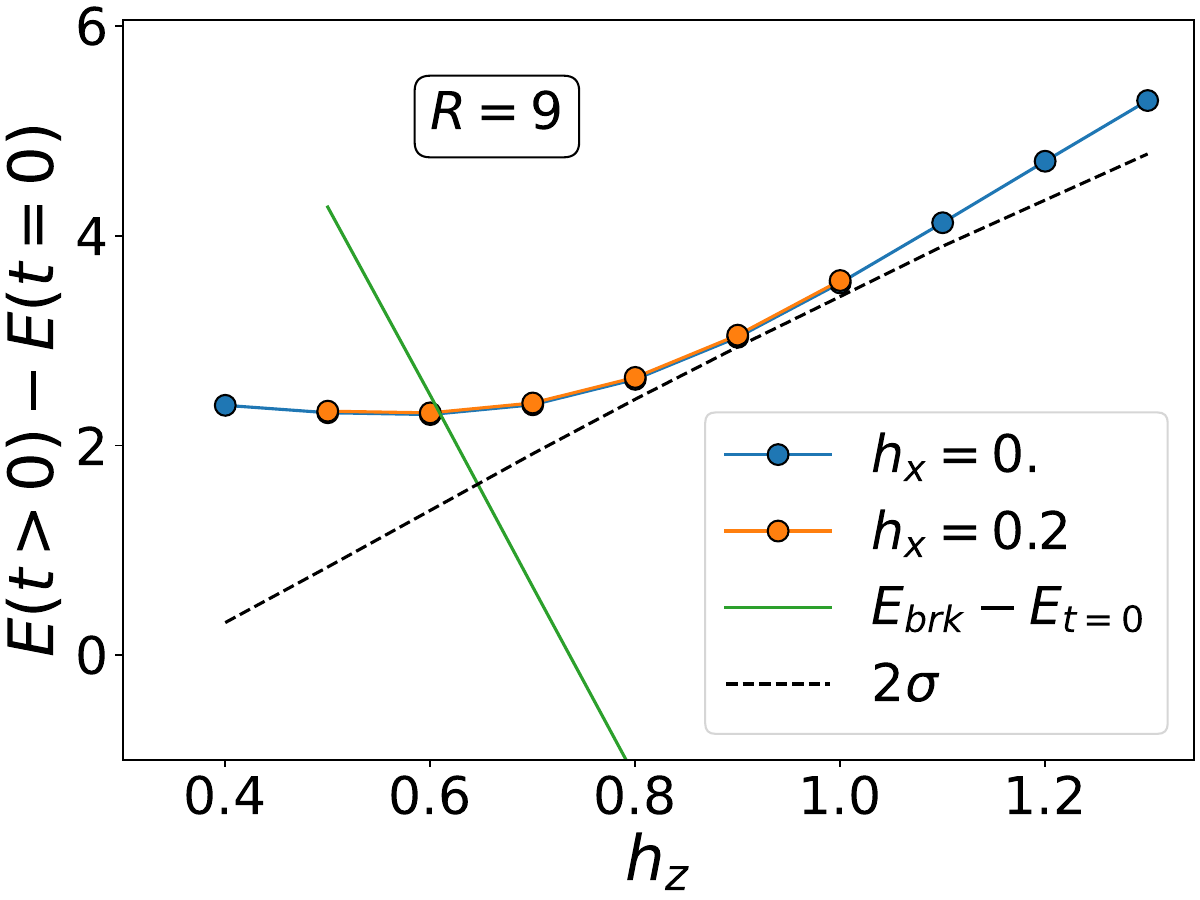}
    \caption{\textbf{Energy injected by the pull} Energy difference between the state befor the pull $t=0$ and after the pull $t>0$ as a function of $h_z$ for both $h_x=0.$ and $h_x=0.2$. The dashed black line represent twice the string tension $\sigma$, the energy of broken string states is show as a green line. System size $18\times 6$  }
    \label{fig:energy_injection}
\end{figure}

To verify this we compare the energy injected by the string pull protocol to the energy of the broken string states discussed in App. \ref{app:string_breaking}. In Fig. \ref{fig:energy_injection} in particular we show the injected energy as a function of $h_z$ for $R=9$ for both $h_x=0.$ and $h_x=0.2$. First of all we see that the injected energy is very similar between the two cases, signaling that most of the energy is injected into excitations of the string and not in creation of pairs of charges. Then we see how in the stiff phase the injected energy roughly follows twice the string tension $\sigma$; meaning that a large part of the energy is carried by an effective increase of the string length. On the other hand, the energy in the rough phase is mainly distributed in the transverse normal modes, giving an overall energy roughly independent of $h_z$. We further show with a green line (Fig.\ref{fig:energy_injection}) the energy of the broken string state (see App. \ref{app:string_breaking}). For the string length considered here $R=9$ this becomes negative at $h_z\sim0.8$, meaning that the initial string state is metastable. There is then another important crossing at around $h_z\sim0.6$ where the energy injected by the string pull protocol becomes large enough to in principle break the string. However this does not happen dynamically in the time-scales observed here. We only observe a small increase of the total charge with time which remains small ($\lesssim0.04$ at $T=100$).   

\begin{figure}
    \centering
    \begin{overpic}[width=\linewidth]{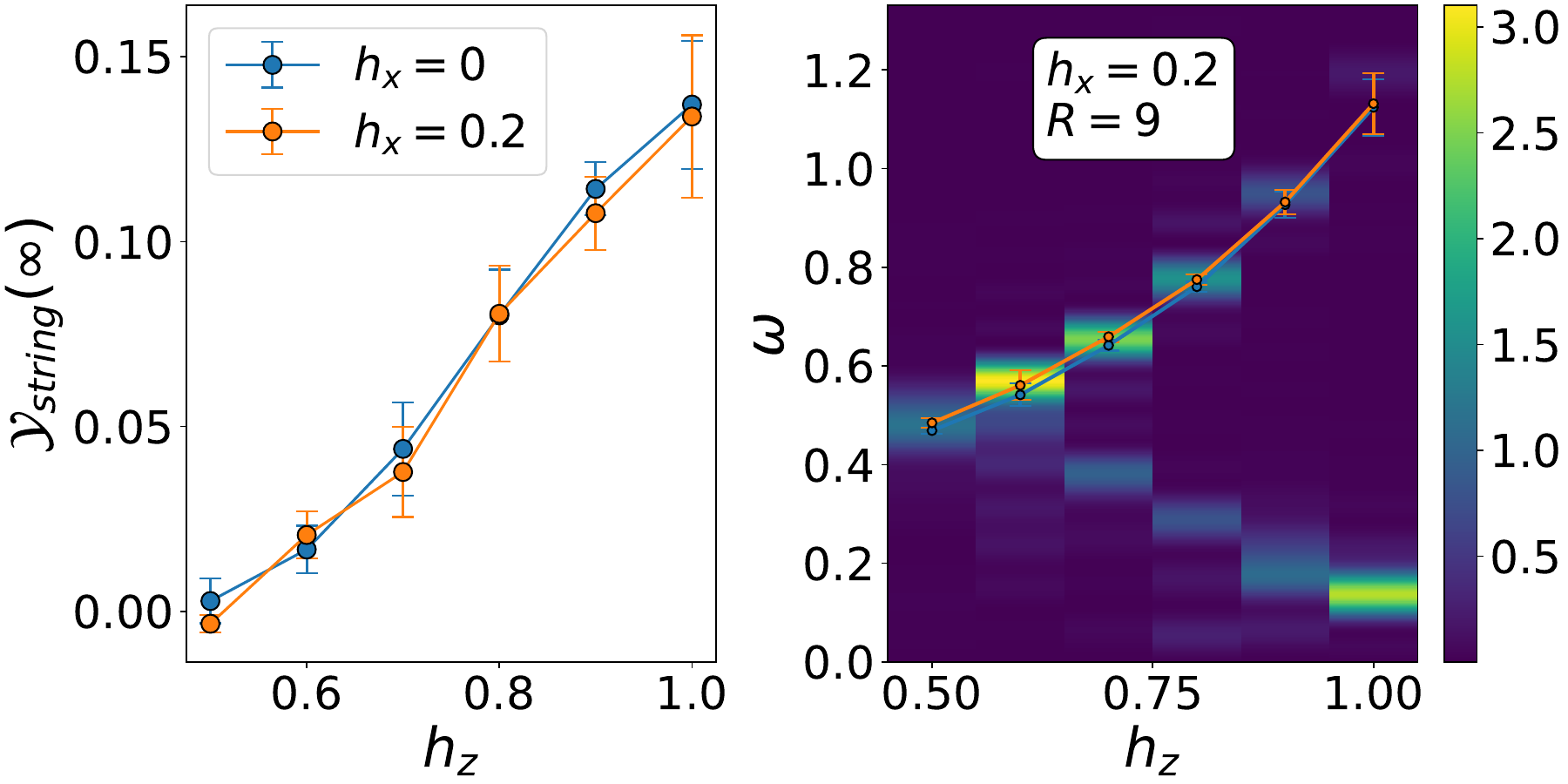}
    \put(13,35){\textbf{(a)}}
    \put(58,45){\textbf{\textcolor{white}{(b)}}}

    \end{overpic}
    \caption{\textbf{String-pull protocol with matter} (a) Long-time average of the string transverse position with (orange) and without (blue) matter as a function of $h_z$. Data without matter dynamics are also shown in Fig. \ref{fig:string_locking}. (b) Fundamental mode occupation for $R=9$ and $h_x=0.2$, transformed in frequency domain. The extracted velocity at $h_x=0.2$ is shown in orange, while the values at $h_x=0.$ are reported in blue. No statistically significant difference is seen.}
    \label{fig:dynamics_withmatter}
\end{figure}

Once string breaking phenomena are excluded, we focus on the string dynamical properties. In Fig. \ref{fig:dynamics_withmatter} we show how including matter dynamics $h_x=0.2$ does not produce appreciable differences in the properties of the transverse modes. In panel (a) we show the long-time averaged string position for a string of length $R=9$ (data for $h_x=0.$ are also shown in Fig. \ref{fig:string_locking}). In panel (b) instead we show the fundamental mode spectra, from which a velocity can be extracted (see data for $h_x=0.$ and discussion on how to extract it around Fig. \ref{fig:normal_modes}). Neither quantity shows a statistically significant difference between $h_x=0$ and $h_x=0.2$.

\section{Velocity from early-times}

\begin{figure}
    \centering
    \begin{overpic}
        [width=\linewidth]{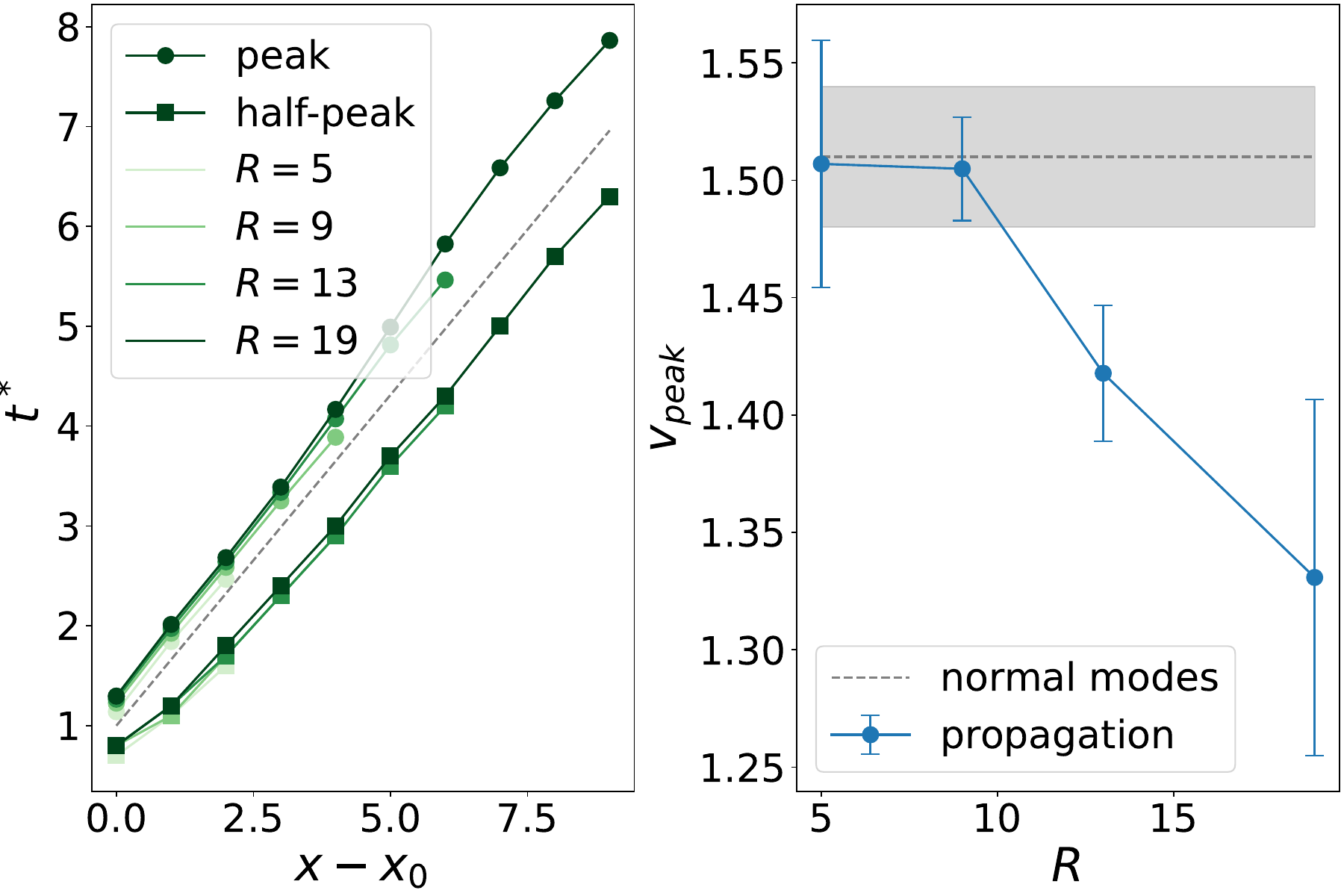}
        \put(40,10){\textbf{(a)}}
        \put(93,56){\textbf{(b)}}
    \end{overpic}
    \caption{\textbf{Velocity from early-time dynamics} \textbf{(a)} Arrival of the fist peak (dots) and half peak (square) in the transverse position of the string $Y_{string}(x)$ as a function of the distance from the point where the pull occurs $x_0$. Different string length are shown, all with system size $30\times12$. \textbf{(b)} Drift of the extracted velocity from the peak propagation, against the normal mode velocity extracted in Sec. \ref{sec:dynamics_normalmodes}  }
    \label{fig:velocity_earlytime}
\end{figure}
We here discuss more in details the extraction of the velocity from early time propagation of the excitation on the string. In Fig. \ref{fig:velocity_earlytime}(a) we track the arrival time of the excitation $t^*(x)$ as the peak of the transverse string position $Y_{string}(x,t)$ (circles). Here we also show the arrival time of the half-peak (squares). The extracted velocity $v_{peak}$ is shown in Fig. \ref{fig:velocity_earlytime}(b) as a function of the string length from which the fit is carried out. We also show with a dashed gray line the velocity extracted from normal modes ($v=1.49\pm0.04$ for $h_z=0.5$). Although qualitatively agreeing, the extracted propagation velocity does not match the velocity extracted from direct spectroscopy and actually seem to drift to lower values for large $R$. We tentatively attribute this shift to a possible mixing with high momentum modes which do not follow the free boson dispersion at velocity $v$. Indeed the string-pull protocol creates an highly localized excitation which has overlap on all momenta. The negative correction to the peak velocity is coherent with the fact that the observed dispersion $\omega(k)$ (Fig. \ref{fig:normal_modes}(c)) seem to show a smaller group velocity at high momenta. Overall, while tracking the early time propagation can give good results for the string velocity, it does not provide the level accuracy and stability that is provided by direct spectroscopy from long-time oscillations.

\section{Convergence of static results}\label{app:conv_static}
In this appendix, we assess the convergence of the static results with the MPS bond dimension $\chi$ and the cylinder circumference $L_y$. 

\begin{figure}
    \centering

    \begin{overpic}[width=\linewidth]{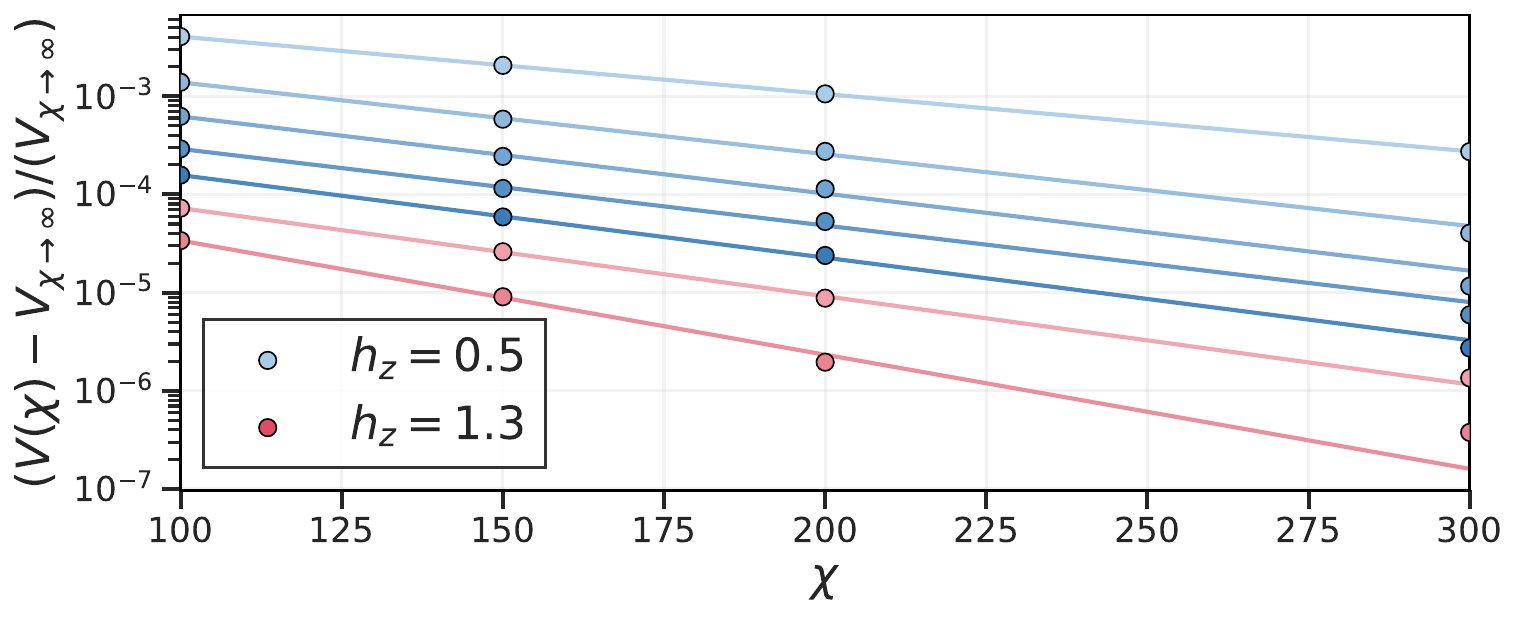}
        \put(14,35){\textbf{(a)}}
    \end{overpic}

    \vspace{0.15cm}

    \begin{overpic}[width=\linewidth]{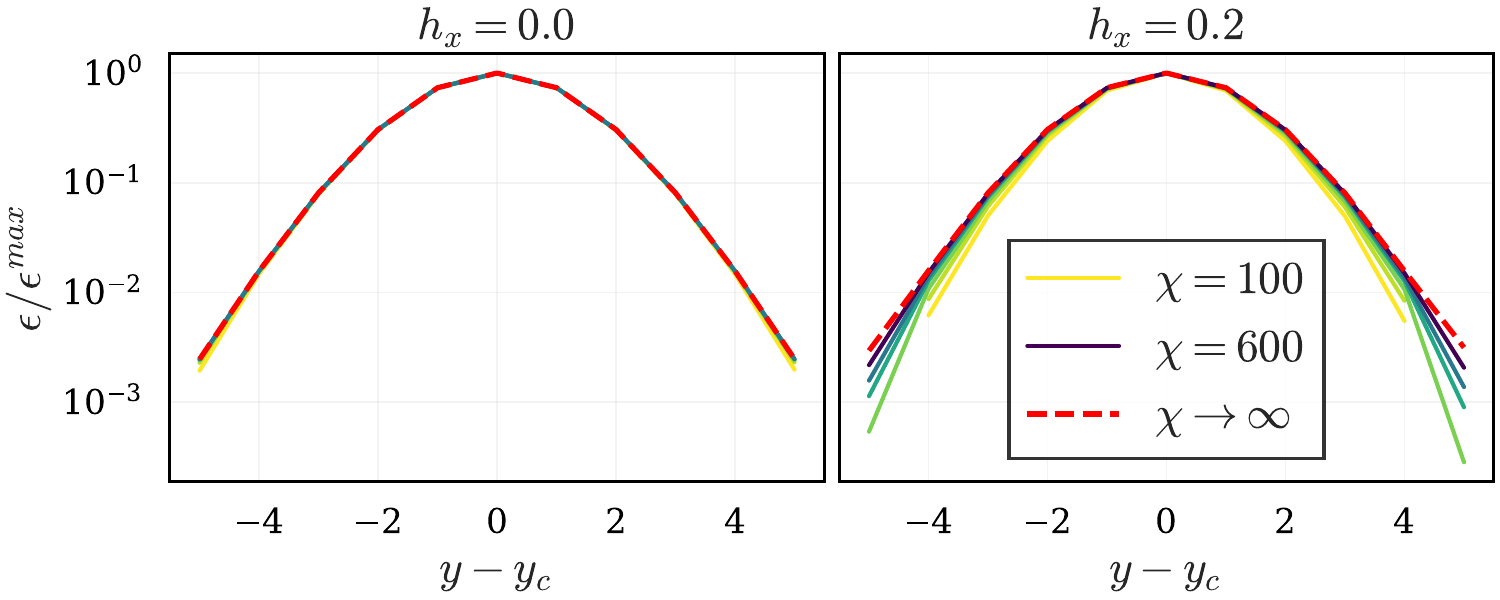}
        \put(13,33){\textbf{(b)}}
        \put(57,33){\textbf{(c)}}
    \end{overpic}

     \begin{overpic}[width=\linewidth]{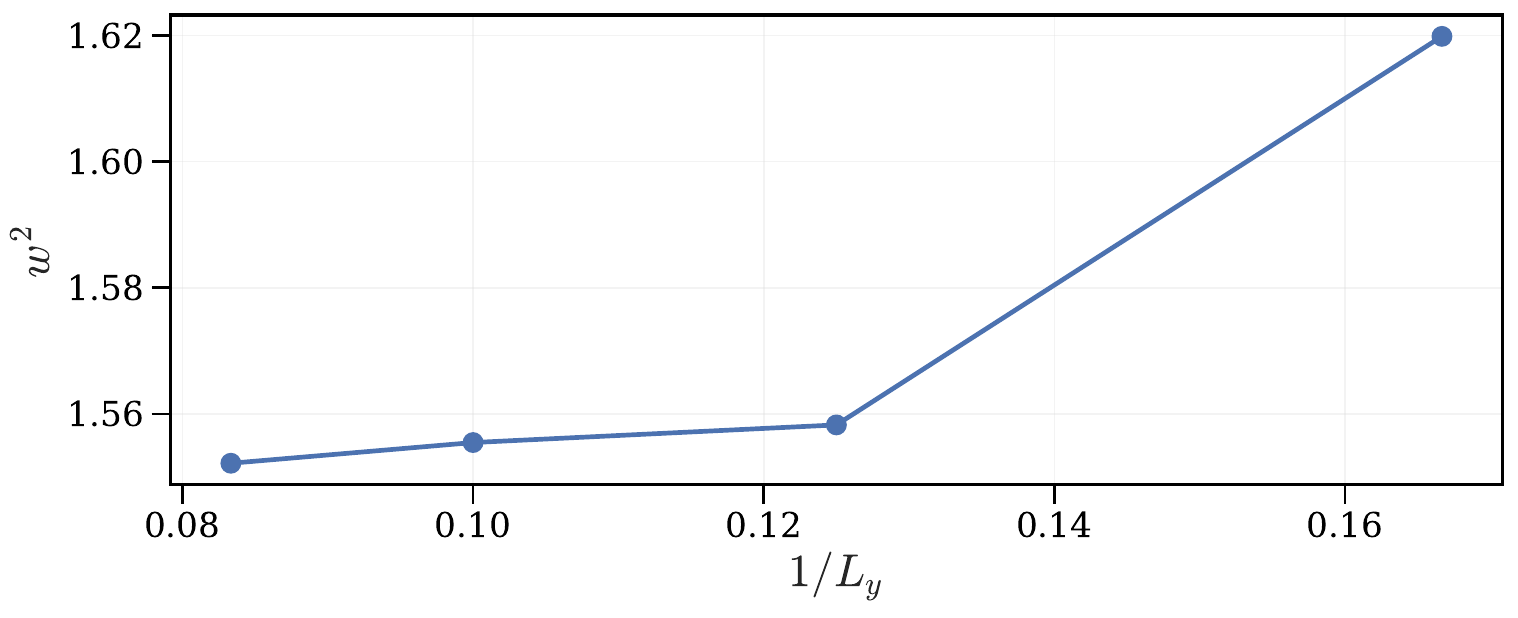}
        \put(14,35){\textbf{(d)}}
    \end{overpic}
    \caption{\textbf{Convergence of static results in bond dimension and $L_y$.}
    \textbf{(a)}: Relative deviation of the confining potential from its extrapolated infinite-bond-dimension value, as a function of bond dimension, for a  $24 \times 8$ cylinder charge separation $R = 16$ and gauge-matter coupling $h_x = 0.2$. Different lines correspond to different values of confining field $h_z\in[0.5, 1.1]$ with steps of $0.1$.
    \textbf{(b)} and \textbf{(c)}:Transverse profile of the string for cylinder of size $30\times12$, $h_z = 0.5$, and charge separation $R = 20$, for $h_x = 0.0$, $h_x = 0.2$, respectively. Different lines correspond to different bond dimensions $\chi \in [100, 600]$. Dashed red curves are the profiles extrapolated with an exponential fit in $\chi$ at each $y$. \textbf{(d)} Scaling of the string width $w^2$ with the inverse cylinder circumference $1/L_y$ for $R=16$. The width is obtained by fitting the string profiles with a modified Bessel function of the second kind, $K_0$, as in Eq.\ref{eq:clem}, and then extrapolated to infinite bond dimension .}
    \label{fig:string_conv}
\end{figure}

Figure~\ref{fig:string_conv}(a) shows the convergence of the confining potential $V(R)$ on a $24\times8$ cylinder at $h_x=0.2$ and charge separation $R=16$ corresponding to the largest separation used to extract the Lüscher term in Sec.~\ref{subsec:Lusher}. We consider values of the longitudinal field in the range $h_z\in[0.4,1.2]$, spanning from the weakly to the strongly confined regimes. As expected, convergence with $\chi$ is slower for smaller $h_z$, corresponding to broader strings. For each value of $h_z$, we estimate the $\chi\rightarrow\infty$ limit by fitting the bond-dimension dependence to an exponential function. The fit are indicated by the dashed lines. This extrapolation is performed for every charge separation $R$ used in Fig.~\ref{fig:stringproperties}, although for clarity we only show here the largest separation, where the finite-$\chi$ effects are most pronounced.

In Fig.~\ref{fig:string_conv}(b), (c) we show the transverse profile of the electric string for a representative simulation on a $30\times12$ cylinder at $h_z=0.5$ and charge separation $R=20$. Different lines represent different bond dimensions in the range $\chi \in [100, 600]$. While the profile at $h_x=0$ is already well converged for the bond dimensions considered, the case $h_x=0.2$ exhibits a more pronounced dependence on $\chi$, particularly in the tails of the profile. This slower convergence motivates an extrapolation to the $\chi\to\infty$ limit. At each transverse coordinate $y$, we fit the $\chi$ dependence to the exponential form where $\epsilon_\infty(y)$ is the extrapolated infinite-bond-dimension value. The resulting extrapolated profiles are shown as dashed red curves in the figure. 

Finally, Fig.~\ref{fig:string_conv}(d) shows the scaling of the extracted string width $w^2$ with the inverse cylinder circumference $1/L_y$. The widths are obtained by fitting the bond-dimension-extrapolated string profiles with a modified Bessel function of the second kind, $K_0$. While the narrowest cylinder ($L_y = 6$) exhibits a noticeable finite-size correction, the change in the extracted width between $L_y = 8$ and $L_y = 10$ is an order of magnitude smaller. This indicates that the cylinder circumference $L_y = 10$, employed throughout the main text, is already sufficiently large to obtain a reliable estimate of the roughening transition point.

\section{Convergence of time-dependent results}\label{app:conv_timedep}

\begin{figure}[ht!]
    \centering
    \vspace{0.3cm}
    \begin{overpic}
[width=\linewidth]{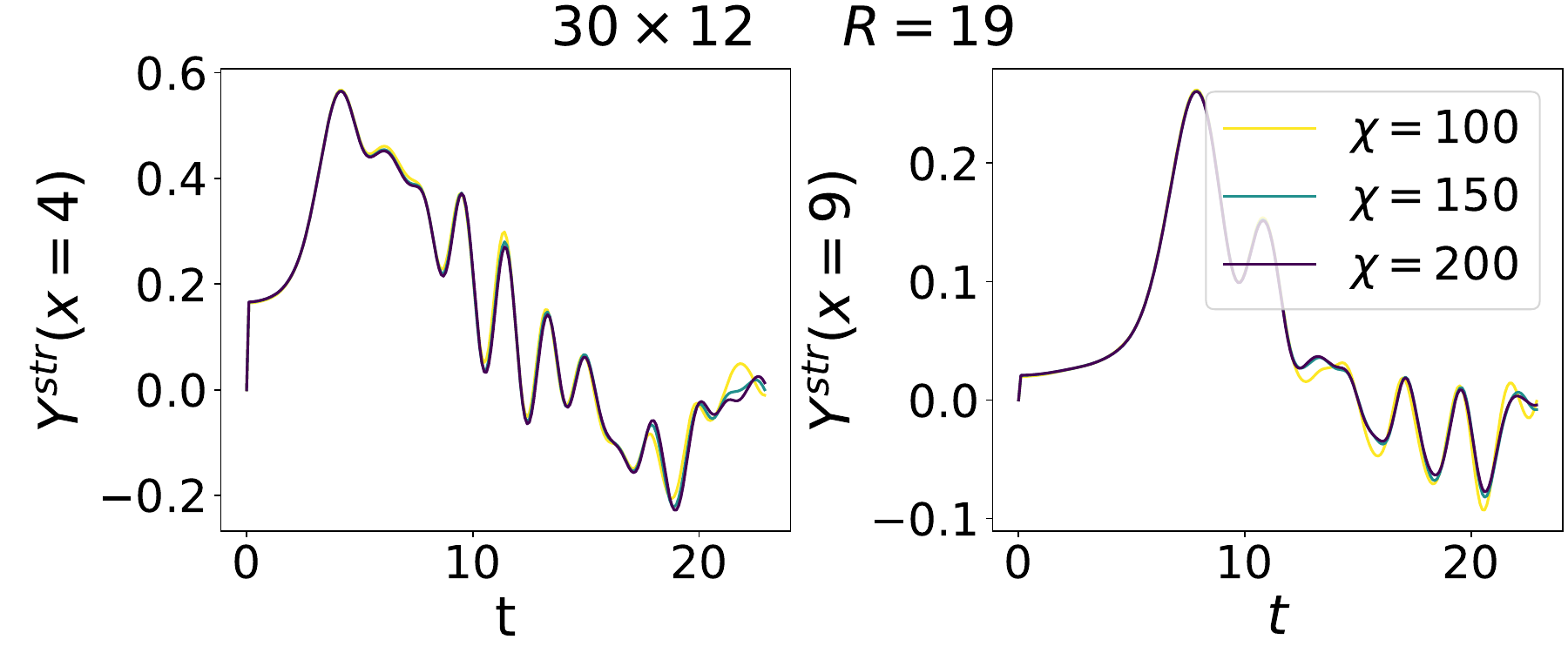}  
\put(15,10){\textbf{(a)}}
\put(62,10){\textbf{(b)}}

    \end{overpic}
        
    \begin{overpic}[width=\linewidth]{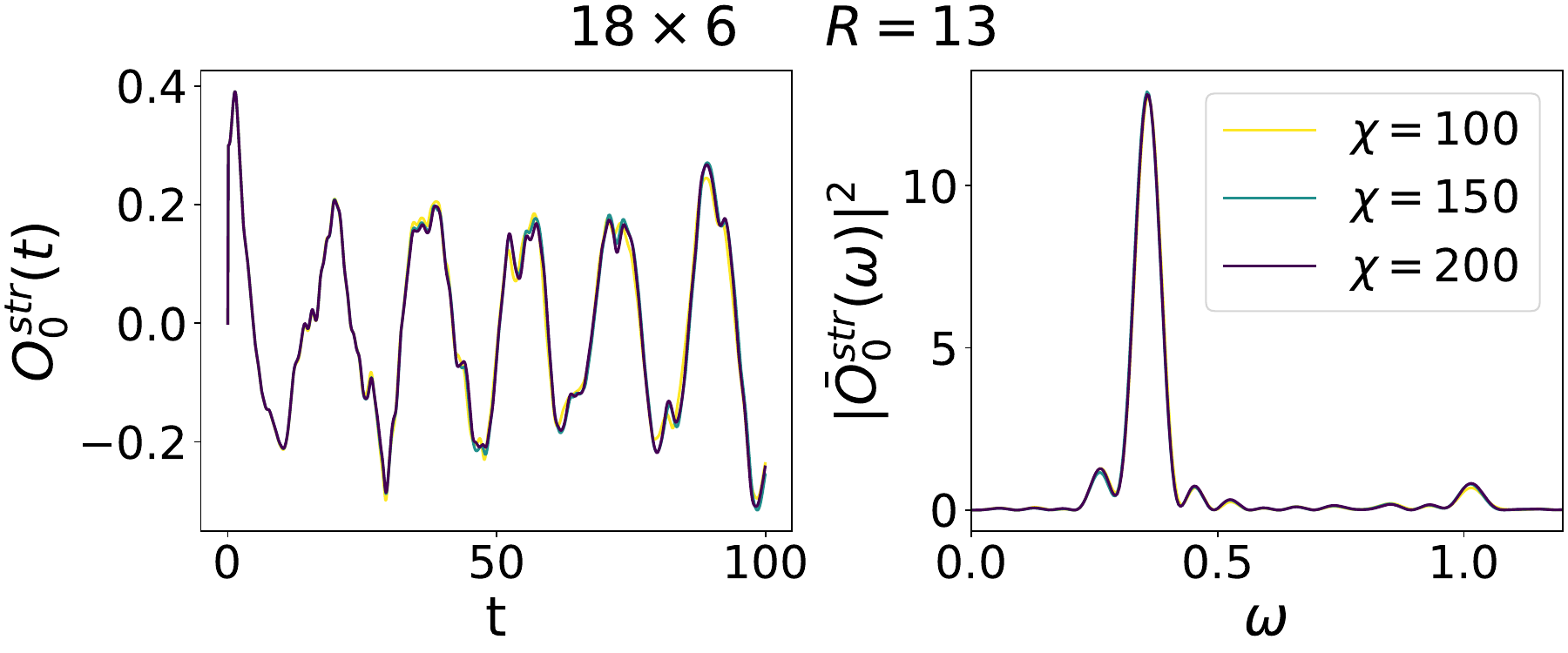}
    \put(15,10){\textbf{(c)}}
\put(62,10){\textbf{(d)}}

    \end{overpic}
    \caption{\textbf{Convergence of time evolution with bond dimension} String-pull protocol for early \textbf{(a,b)} and late times \textbf{(c,d)} at $h_z=0.5$ $h_x=0$. At early times we focus on the large system $30\times 12$ with a string of length $R=19$, while at late times on a smaller $18\times 6$ system with string length $R=13$. The observables used to extract the string properties are already converged at $\chi=100$.}
    \label{fig:convergence_chi}
\end{figure}

In this section we provide a numerical analysis on the convergence of the dynamics simulations. First of all, we investigate the convergence with the maximum bond dimension used for the string dynamics. In figure \ref{fig:convergence_chi} we test the convergence of the early time dynamics for the large system size $30\times12$, and the later time normal mode oscillations for a smaller size $18\times 6$ for $R=13$.

\begin{figure}[h!]
    \centering
    \begin{overpic}[width=\linewidth]{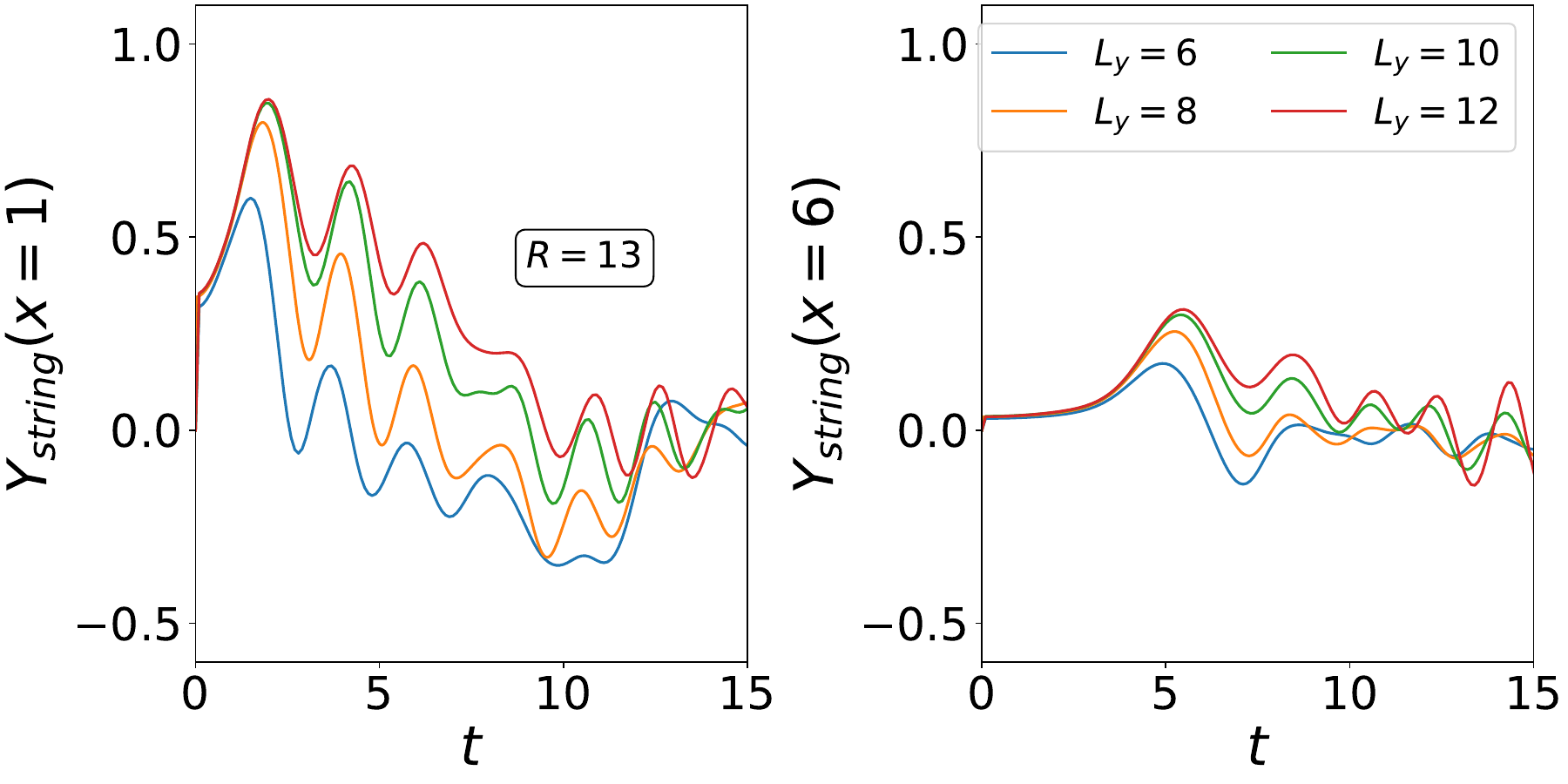}
    \put(20,45){\textbf{(a)}}
    \put(70,35){\textbf{(b)}}
    \end{overpic}
    \begin{overpic}[width=\linewidth]{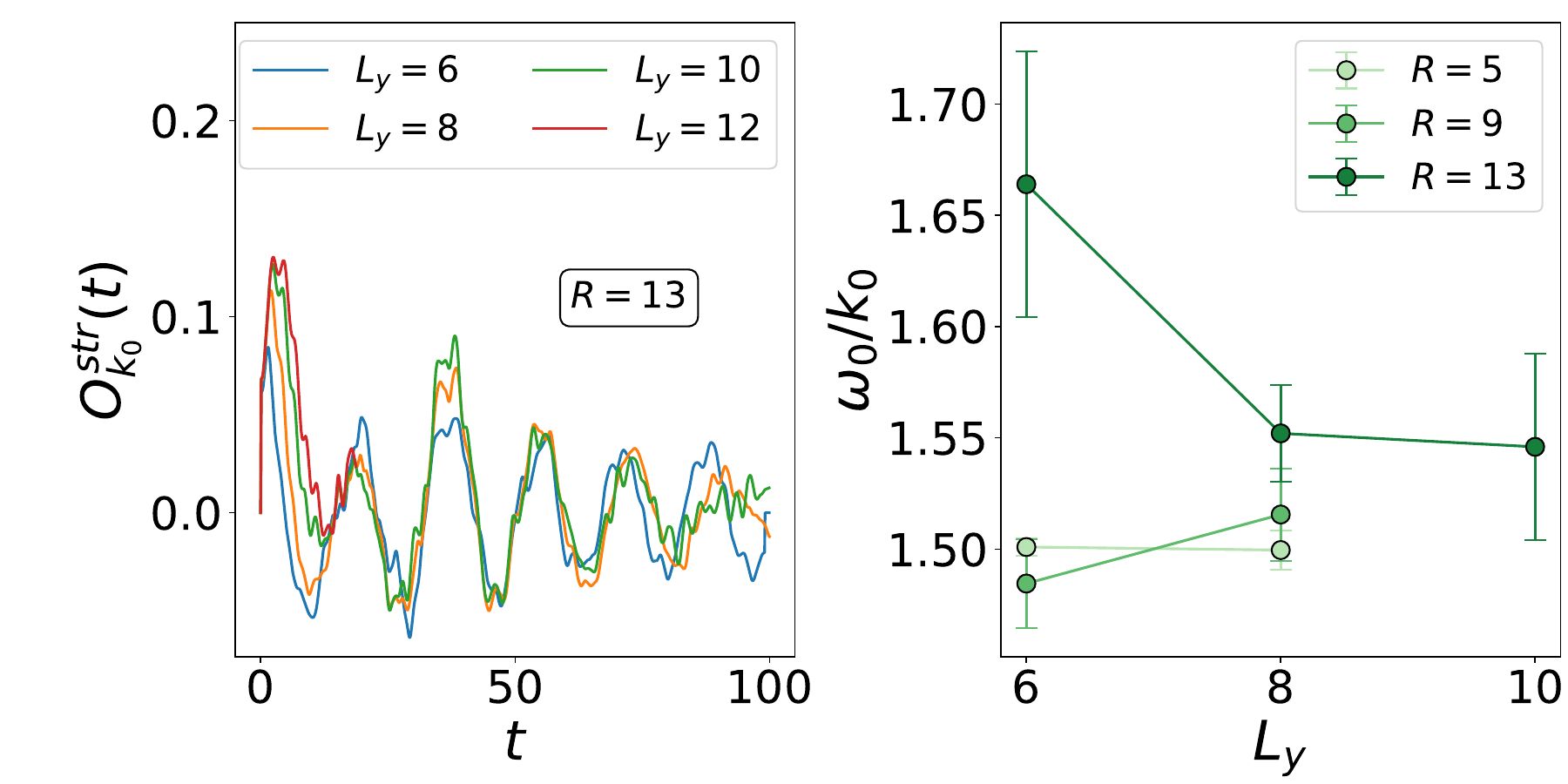}
    \put(20,35){\textbf{(c)}}
    \put(70,40){\textbf{(d)}}
    \end{overpic}
    \caption{\textbf{Convergence of time evolution with circumference $L_y$} \textbf{(a,b)} Short time dynamics of transverse position at different distances from the excitation point $x=1$ and $x=6$ for a string of length $R=13$. Fundamental mode occupation as a function of time \textbf{(c)} and the extracted normal mode frequency \textbf{(d)}. In all panels we fix $h_z=0.5$ and $h_x=0$. Panel \textbf{(c)} also fixes the string length to $R=13$ while in panel \textbf{(d)} we show how only for the largest strings effects on the normal mode from finite $L
    _y$ can be resolved.   }
    \label{fig:conv_Ly_dyn}
\end{figure}
We also investigate the circumference $L_y$ dependence on the transverse string dynamics. This is done in Fig. \ref{fig:conv_Ly_dyn} for a reference value $h_z=0.5$ where transverse fluctuations are large. First, we see in panel (a,b) that the short time dynamics of the transverse string position depends on $L_y$. In particular the arrival time $t^*$, that is the position of the first peak, shifts to later times for both positions $x$ shown. This suggests that very large systems are needed to faithfully reconstruct the short-time dynamics. Intuitively, at short times the transverse oscillation is larger in magnitude as it is focused in $x$-space. Hence part of its tails start to feel the finite transverse direction $L_y$ and leak into the opposite part of the system, reducing $Y_{string}$. At $L_y=12$ the first arrival time is however almost converged and hence this circumference is used to track the early time dynamics.

Shifting the focus on longer times, we look at the occupation of the fundamental mode in Fig. \ref{fig:conv_Ly_dyn}(c,d). The early time dynamics again do show a dependence on $L_y$ which however is not critical for the long-time behavior and the extraction of its frequency. A small frequency shift can be resolved only for $L_y=6$ and the large string length $R=13$, while shorter strings show converged results already at $L_y=6$. This can be attributed to the fact that transverse fluctuations are large and the finite cylinder circumference becomes important. For this reasons we do not use $L_y=6$ but $L_y=8$ in the case $R=13$. We also use $L_y=8$ for $h_z=0.4$ where transverse fluctuations are larger.

\newpage
\bibliography{main.bbl}

\begin{thebibliography}{110}%
\makeatletter
\providecommand \@ifxundefined [1]{%
 \@ifx{#1\undefined}
}%
\providecommand \@ifnum [1]{%
 \ifnum #1\expandafter \@firstoftwo
 \else \expandafter \@secondoftwo
 \fi
}%
\providecommand \@ifx [1]{%
 \ifx #1\expandafter \@firstoftwo
 \else \expandafter \@secondoftwo
 \fi
}%
\providecommand \natexlab [1]{#1}%
\providecommand \enquote  [1]{``#1''}%
\providecommand \bibnamefont  [1]{#1}%
\providecommand \bibfnamefont [1]{#1}%
\providecommand \citenamefont [1]{#1}%
\providecommand \href@noop [0]{\@secondoftwo}%
\providecommand \href [0]{\begingroup \@sanitize@url \@href}%
\providecommand \@href[1]{\@@startlink{#1}\@@href}%
\providecommand \@@href[1]{\endgroup#1\@@endlink}%
\providecommand \@sanitize@url [0]{\catcode `\\12\catcode `\$12\catcode `\&12\catcode `\#12\catcode `\^12\catcode `\_12\catcode `\%12\relax}%
\providecommand \@@startlink[1]{}%
\providecommand \@@endlink[0]{}%
\providecommand \url  [0]{\begingroup\@sanitize@url \@url }%
\providecommand \@url [1]{\endgroup\@href {#1}{\urlprefix }}%
\providecommand \urlprefix  [0]{URL }%
\providecommand \Eprint [0]{\href }%
\providecommand \doibase [0]{https://doi.org/}%
\providecommand \selectlanguage [0]{\@gobble}%
\providecommand \bibinfo  [0]{\@secondoftwo}%
\providecommand \bibfield  [0]{\@secondoftwo}%
\providecommand \translation [1]{[#1]}%
\providecommand \BibitemOpen [0]{}%
\providecommand \bibitemStop [0]{}%
\providecommand \bibitemNoStop [0]{.\EOS\space}%
\providecommand \EOS [0]{\spacefactor3000\relax}%
\providecommand \BibitemShut  [1]{\csname bibitem#1\endcsname}%
\let\auto@bib@innerbib\@empty
\bibitem [{\citenamefont {Wilson}(1974)}]{Wilson1974}%
  \BibitemOpen
  \bibfield  {author} {\bibinfo {author} {\bibfnamefont {K.~G.}\ \bibnamefont {Wilson}},\ }\bibfield  {title} {\bibinfo {title} {Confinement of quarks},\ }\href {https://doi.org/10.1103/PhysRevD.10.2445} {\bibfield  {journal} {\bibinfo  {journal} {Physical Review D}\ }\textbf {\bibinfo {volume} {10}},\ \bibinfo {pages} {2445} (\bibinfo {year} {1974})}\BibitemShut {NoStop}%
\bibitem [{\citenamefont {Kogut}(1979)}]{Kogut1979}%
  \BibitemOpen
  \bibfield  {author} {\bibinfo {author} {\bibfnamefont {J.~B.}\ \bibnamefont {Kogut}},\ }\bibfield  {title} {\bibinfo {title} {An introduction to lattice gauge theory and spin systems},\ }\href {https://doi.org/10.1103/RevModPhys.51.659} {\bibfield  {journal} {\bibinfo  {journal} {Reviews of Modern Physics}\ }\textbf {\bibinfo {volume} {51}},\ \bibinfo {pages} {659} (\bibinfo {year} {1979})}\BibitemShut {NoStop}%
\bibitem [{\citenamefont {Montvay}\ and\ \citenamefont {M\"{u}nster}(1994)}]{montvay1994quantum}%
  \BibitemOpen
  \bibfield  {author} {\bibinfo {author} {\bibfnamefont {I.}~\bibnamefont {Montvay}}\ and\ \bibinfo {author} {\bibfnamefont {G.}~\bibnamefont {M\"{u}nster}},\ }\href {https://doi.org/10.1017/cbo9780511470783} {\emph {\bibinfo {title} {Quantum Fields on a Lattice}}}\ (\bibinfo  {publisher} {Cambridge University Press},\ \bibinfo {year} {1994})\BibitemShut {NoStop}%
\bibitem [{\citenamefont {Wen}(2017)}]{Wen_rmp2017}%
  \BibitemOpen
  \bibfield  {author} {\bibinfo {author} {\bibfnamefont {X.-G.}\ \bibnamefont {Wen}},\ }\bibfield  {title} {\bibinfo {title} {Colloquium: Zoo of quantum-topological phases of matter},\ }\href {https://doi.org/10.1103/RevModPhys.89.041004} {\bibfield  {journal} {\bibinfo  {journal} {Rev. Mod. Phys.}\ }\textbf {\bibinfo {volume} {89}},\ \bibinfo {pages} {041004} (\bibinfo {year} {2017})}\BibitemShut {NoStop}%
\bibitem [{\citenamefont {Kitaev}(2003)}]{Kitaev2003}%
  \BibitemOpen
  \bibfield  {author} {\bibinfo {author} {\bibfnamefont {A.~Y.}\ \bibnamefont {Kitaev}},\ }\bibfield  {title} {\bibinfo {title} {Fault-tolerant quantum computation by anyons},\ }\href {https://doi.org/10.1016/S0003-4916(02)00018-0} {\bibfield  {journal} {\bibinfo  {journal} {Annals of Physics}\ }\textbf {\bibinfo {volume} {303}},\ \bibinfo {pages} {2} (\bibinfo {year} {2003})}\BibitemShut {NoStop}%
\bibitem [{\citenamefont {Moessner}\ and\ \citenamefont {Moore}(2021)}]{moessner1945topological}%
  \BibitemOpen
  \bibfield  {author} {\bibinfo {author} {\bibfnamefont {R.}~\bibnamefont {Moessner}}\ and\ \bibinfo {author} {\bibfnamefont {J.~E.}\ \bibnamefont {Moore}},\ }\href {https://doi.org/10.1017/9781316226308} {\emph {\bibinfo {title} {Topological Phases of Matter}}}\ (\bibinfo  {publisher} {Cambridge University Press},\ \bibinfo {year} {2021})\BibitemShut {NoStop}%
\bibitem [{\citenamefont {Savary}\ and\ \citenamefont {Balents}(2016)}]{savary2017quantum}%
  \BibitemOpen
  \bibfield  {author} {\bibinfo {author} {\bibfnamefont {L.}~\bibnamefont {Savary}}\ and\ \bibinfo {author} {\bibfnamefont {L.}~\bibnamefont {Balents}},\ }\bibfield  {title} {\bibinfo {title} {Quantum spin liquids: a review},\ }\href {https://doi.org/10.1088/0034-4885/80/1/016502} {\bibfield  {journal} {\bibinfo  {journal} {Reports on Progress in Physics}\ }\textbf {\bibinfo {volume} {80}},\ \bibinfo {pages} {016502} (\bibinfo {year} {2016})}\BibitemShut {NoStop}%
\bibitem [{\citenamefont {Creutz}(1980)}]{Creutz_prd1980_montecarlo}%
  \BibitemOpen
  \bibfield  {author} {\bibinfo {author} {\bibfnamefont {M.}~\bibnamefont {Creutz}},\ }\bibfield  {title} {\bibinfo {title} {Monte carlo study of quantized su(2) gauge theory},\ }\href {https://doi.org/10.1103/PhysRevD.21.2308} {\bibfield  {journal} {\bibinfo  {journal} {Phys. Rev. D}\ }\textbf {\bibinfo {volume} {21}},\ \bibinfo {pages} {2308} (\bibinfo {year} {1980})}\BibitemShut {NoStop}%
\bibitem [{\citenamefont {Kronfeld}(2012)}]{Kronfeld2012_montecarlorev}%
  \BibitemOpen
  \bibfield  {author} {\bibinfo {author} {\bibfnamefont {A.~S.}\ \bibnamefont {Kronfeld}},\ }\bibfield  {title} {\bibinfo {title} {Twenty-first century lattice gauge theory: Results from the quantum chromodynamics lagrangian},\ }\href {https://doi.org/10.1146/annurev-nucl-102711-094942} {\bibfield  {journal} {\bibinfo  {journal} {Annual Review of Nuclear and Particle Science}\ }\textbf {\bibinfo {volume} {62}},\ \bibinfo {pages} {265–284} (\bibinfo {year} {2012})}\BibitemShut {NoStop}%
\bibitem [{\citenamefont {Fodor}\ and\ \citenamefont {Hoelbling}(2012)}]{fodor_rmp2012_qcd}%
  \BibitemOpen
  \bibfield  {author} {\bibinfo {author} {\bibfnamefont {Z.}~\bibnamefont {Fodor}}\ and\ \bibinfo {author} {\bibfnamefont {C.}~\bibnamefont {Hoelbling}},\ }\bibfield  {title} {\bibinfo {title} {Light hadron masses from lattice qcd},\ }\href {https://doi.org/10.1103/RevModPhys.84.449} {\bibfield  {journal} {\bibinfo  {journal} {Rev. Mod. Phys.}\ }\textbf {\bibinfo {volume} {84}},\ \bibinfo {pages} {449} (\bibinfo {year} {2012})}\BibitemShut {NoStop}%
\bibitem [{\citenamefont {Wiese}(2013)}]{wiese2013ultracold}%
  \BibitemOpen
  \bibfield  {author} {\bibinfo {author} {\bibfnamefont {U.-J.}\ \bibnamefont {Wiese}},\ }\bibfield  {title} {\bibinfo {title} {Ultracold quantum gases and lattice systems: quantum simulation of lattice gauge theories},\ }\href {https://doi.org/https://doi.org/10.1002/andp.201300104} {\bibfield  {journal} {\bibinfo  {journal} {Annalen der Physik}\ }\textbf {\bibinfo {volume} {525}},\ \bibinfo {pages} {777} (\bibinfo {year} {2013})}\BibitemShut {NoStop}%
\bibitem [{\citenamefont {Gattringer}\ and\ \citenamefont {Langfeld}(2016)}]{Gattringer2016_signproblem}%
  \BibitemOpen
  \bibfield  {author} {\bibinfo {author} {\bibfnamefont {C.}~\bibnamefont {Gattringer}}\ and\ \bibinfo {author} {\bibfnamefont {K.}~\bibnamefont {Langfeld}},\ }\bibfield  {title} {\bibinfo {title} {Approaches to the sign problem in lattice field theory},\ }\href {https://doi.org/10.1142/s0217751x16430077} {\bibfield  {journal} {\bibinfo  {journal} {International Journal of Modern Physics A}\ }\textbf {\bibinfo {volume} {31}},\ \bibinfo {pages} {1643007} (\bibinfo {year} {2016})}\BibitemShut {NoStop}%
\bibitem [{\citenamefont {Dalmonte}\ and\ \citenamefont {Montangero}(2016)}]{DalmonteMontangero2016_rev}%
  \BibitemOpen
  \bibfield  {author} {\bibinfo {author} {\bibfnamefont {M.}~\bibnamefont {Dalmonte}}\ and\ \bibinfo {author} {\bibfnamefont {S.}~\bibnamefont {Montangero}},\ }\bibfield  {title} {\bibinfo {title} {Lattice gauge theory simulations in the quantum information era},\ }\href {https://doi.org/10.1080/00107514.2016.1151199} {\bibfield  {journal} {\bibinfo  {journal} {Contemporary Physics}\ }\textbf {\bibinfo {volume} {57}},\ \bibinfo {pages} {388–412} (\bibinfo {year} {2016})}\BibitemShut {NoStop}%
\bibitem [{\citenamefont {Bañuls}\ \emph {et~al.}(2020)\citenamefont {Bañuls}, \citenamefont {Blatt}, \citenamefont {Catani}, \citenamefont {Celi}, \citenamefont {Cirac}, \citenamefont {Dalmonte}, \citenamefont {Fallani}, \citenamefont {Jansen}, \citenamefont {Lewenstein}, \citenamefont {Montangero}, \citenamefont {Muschik}, \citenamefont {Reznik}, \citenamefont {Rico}, \citenamefont {Tagliacozzo}, \citenamefont {Van~Acoleyen}, \citenamefont {Verstraete}, \citenamefont {Wiese}, \citenamefont {Wingate}, \citenamefont {Zakrzewski},\ and\ \citenamefont {Zoller}}]{banuls2020simulating}%
  \BibitemOpen
  \bibfield  {author} {\bibinfo {author} {\bibfnamefont {M.~C.}\ \bibnamefont {Bañuls}}, \bibinfo {author} {\bibfnamefont {R.}~\bibnamefont {Blatt}}, \bibinfo {author} {\bibfnamefont {J.}~\bibnamefont {Catani}}, \bibinfo {author} {\bibfnamefont {A.}~\bibnamefont {Celi}}, \bibinfo {author} {\bibfnamefont {J.~I.}\ \bibnamefont {Cirac}}, \bibinfo {author} {\bibfnamefont {M.}~\bibnamefont {Dalmonte}}, \bibinfo {author} {\bibfnamefont {L.}~\bibnamefont {Fallani}}, \bibinfo {author} {\bibfnamefont {K.}~\bibnamefont {Jansen}}, \bibinfo {author} {\bibfnamefont {M.}~\bibnamefont {Lewenstein}}, \bibinfo {author} {\bibfnamefont {S.}~\bibnamefont {Montangero}}, \bibinfo {author} {\bibfnamefont {C.~A.}\ \bibnamefont {Muschik}}, \bibinfo {author} {\bibfnamefont {B.}~\bibnamefont {Reznik}}, \bibinfo {author} {\bibfnamefont {E.}~\bibnamefont {Rico}}, \bibinfo {author} {\bibfnamefont {L.}~\bibnamefont {Tagliacozzo}}, \bibinfo {author} {\bibfnamefont {K.}~\bibnamefont {Van~Acoleyen}}, \bibinfo {author} {\bibfnamefont
  {F.}~\bibnamefont {Verstraete}}, \bibinfo {author} {\bibfnamefont {U.-J.}\ \bibnamefont {Wiese}}, \bibinfo {author} {\bibfnamefont {M.}~\bibnamefont {Wingate}}, \bibinfo {author} {\bibfnamefont {J.}~\bibnamefont {Zakrzewski}},\ and\ \bibinfo {author} {\bibfnamefont {P.}~\bibnamefont {Zoller}},\ }\bibfield  {title} {\bibinfo {title} {Simulating lattice gauge theories within quantum technologies},\ }\href {http://dx.doi.org/10.1140/epjd/e2020-100571-8} {\bibfield  {journal} {\bibinfo  {journal} {The European Physical Journal D}\ }\textbf {\bibinfo {volume} {74}} (\bibinfo {year} {2020})}\BibitemShut {NoStop}%
\bibitem [{\citenamefont {Bauer}\ \emph {et~al.}(2023)\citenamefont {Bauer}, \citenamefont {Davoudi}, \citenamefont {Balantekin}, \citenamefont {Bhattacharya}, \citenamefont {Carena}, \citenamefont {de~Jong}, \citenamefont {Draper}, \citenamefont {El-Khadra}, \citenamefont {Gemelke}, \citenamefont {Hanada}, \citenamefont {Kharzeev}, \citenamefont {Lamm}, \citenamefont {Li}, \citenamefont {Liu}, \citenamefont {Lukin}, \citenamefont {Meurice}, \citenamefont {Monroe}, \citenamefont {Nachman}, \citenamefont {Pagano}, \citenamefont {Preskill}, \citenamefont {Rinaldi}, \citenamefont {Roggero}, \citenamefont {Santiago}, \citenamefont {Savage}, \citenamefont {Siddiqi}, \citenamefont {Siopsis}, \citenamefont {Van~Zanten}, \citenamefont {Wiebe}, \citenamefont {Yamauchi}, \citenamefont {Yeter-Aydeniz},\ and\ \citenamefont {Zorzetti}}]{bauer2023quantum}%
  \BibitemOpen
  \bibfield  {author} {\bibinfo {author} {\bibfnamefont {C.~W.}\ \bibnamefont {Bauer}}, \bibinfo {author} {\bibfnamefont {Z.}~\bibnamefont {Davoudi}}, \bibinfo {author} {\bibfnamefont {A.~B.}\ \bibnamefont {Balantekin}}, \bibinfo {author} {\bibfnamefont {T.}~\bibnamefont {Bhattacharya}}, \bibinfo {author} {\bibfnamefont {M.}~\bibnamefont {Carena}}, \bibinfo {author} {\bibfnamefont {W.~A.}\ \bibnamefont {de~Jong}}, \bibinfo {author} {\bibfnamefont {P.}~\bibnamefont {Draper}}, \bibinfo {author} {\bibfnamefont {A.}~\bibnamefont {El-Khadra}}, \bibinfo {author} {\bibfnamefont {N.}~\bibnamefont {Gemelke}}, \bibinfo {author} {\bibfnamefont {M.}~\bibnamefont {Hanada}}, \bibinfo {author} {\bibfnamefont {D.}~\bibnamefont {Kharzeev}}, \bibinfo {author} {\bibfnamefont {H.}~\bibnamefont {Lamm}}, \bibinfo {author} {\bibfnamefont {Y.-Y.}\ \bibnamefont {Li}}, \bibinfo {author} {\bibfnamefont {J.}~\bibnamefont {Liu}}, \bibinfo {author} {\bibfnamefont {M.}~\bibnamefont {Lukin}}, \bibinfo {author} {\bibfnamefont
  {Y.}~\bibnamefont {Meurice}}, \bibinfo {author} {\bibfnamefont {C.}~\bibnamefont {Monroe}}, \bibinfo {author} {\bibfnamefont {B.}~\bibnamefont {Nachman}}, \bibinfo {author} {\bibfnamefont {G.}~\bibnamefont {Pagano}}, \bibinfo {author} {\bibfnamefont {J.}~\bibnamefont {Preskill}}, \bibinfo {author} {\bibfnamefont {E.}~\bibnamefont {Rinaldi}}, \bibinfo {author} {\bibfnamefont {A.}~\bibnamefont {Roggero}}, \bibinfo {author} {\bibfnamefont {D.~I.}\ \bibnamefont {Santiago}}, \bibinfo {author} {\bibfnamefont {M.~J.}\ \bibnamefont {Savage}}, \bibinfo {author} {\bibfnamefont {I.}~\bibnamefont {Siddiqi}}, \bibinfo {author} {\bibfnamefont {G.}~\bibnamefont {Siopsis}}, \bibinfo {author} {\bibfnamefont {D.}~\bibnamefont {Van~Zanten}}, \bibinfo {author} {\bibfnamefont {N.}~\bibnamefont {Wiebe}}, \bibinfo {author} {\bibfnamefont {Y.}~\bibnamefont {Yamauchi}}, \bibinfo {author} {\bibfnamefont {K.}~\bibnamefont {Yeter-Aydeniz}},\ and\ \bibinfo {author} {\bibfnamefont {S.}~\bibnamefont {Zorzetti}},\ }\bibfield  {title}
  {\bibinfo {title} {Quantum simulation for high-energy physics},\ }\href {http://dx.doi.org/10.1103/PRXQuantum.4.027001} {\bibfield  {journal} {\bibinfo  {journal} {PRX Quantum}\ }\textbf {\bibinfo {volume} {4}} (\bibinfo {year} {2023})}\BibitemShut {NoStop}%
\bibitem [{\citenamefont {Di~Meglio}\ \emph {et~al.}(2024)\citenamefont {Di~Meglio}, \citenamefont {Jansen}, \citenamefont {Tavernelli}, \citenamefont {Alexandrou}, \citenamefont {Arunachalam}, \citenamefont {Bauer}, \citenamefont {Borras}, \citenamefont {Carrazza}, \citenamefont {Crippa}, \citenamefont {Croft}, \citenamefont {de~Putter}, \citenamefont {Delgado}, \citenamefont {Dunjko},\ and\ \citenamefont {et~al.}}]{di2024quantum}%
  \BibitemOpen
  \bibfield  {author} {\bibinfo {author} {\bibfnamefont {A.}~\bibnamefont {Di~Meglio}}, \bibinfo {author} {\bibfnamefont {K.}~\bibnamefont {Jansen}}, \bibinfo {author} {\bibfnamefont {I.}~\bibnamefont {Tavernelli}}, \bibinfo {author} {\bibfnamefont {C.}~\bibnamefont {Alexandrou}}, \bibinfo {author} {\bibfnamefont {S.}~\bibnamefont {Arunachalam}}, \bibinfo {author} {\bibfnamefont {C.~W.}\ \bibnamefont {Bauer}}, \bibinfo {author} {\bibfnamefont {K.}~\bibnamefont {Borras}}, \bibinfo {author} {\bibfnamefont {S.}~\bibnamefont {Carrazza}}, \bibinfo {author} {\bibfnamefont {A.}~\bibnamefont {Crippa}}, \bibinfo {author} {\bibfnamefont {V.}~\bibnamefont {Croft}}, \bibinfo {author} {\bibfnamefont {R.}~\bibnamefont {de~Putter}}, \bibinfo {author} {\bibfnamefont {A.}~\bibnamefont {Delgado}}, \bibinfo {author} {\bibfnamefont {V.}~\bibnamefont {Dunjko}},\ and\ \bibinfo {author} {\bibnamefont {et~al.}},\ }\bibfield  {title} {\bibinfo {title} {Quantum computing for high-energy physics: State of the art and challenges},\
  }\href {http://dx.doi.org/10.1103/PRXQuantum.5.037001} {\bibfield  {journal} {\bibinfo  {journal} {PRX Quantum}\ }\textbf {\bibinfo {volume} {5}} (\bibinfo {year} {2024})}\BibitemShut {NoStop}%
\bibitem [{\citenamefont {Davoudi}(2026)}]{davoudi2026quantum}%
  \BibitemOpen
  \bibfield  {author} {\bibinfo {author} {\bibfnamefont {Z.}~\bibnamefont {Davoudi}},\ }\href {https://arxiv.org/abs/2605.20417} {\bibinfo {title} {Quantum simulation of gauge theories for particle and nuclear physics}} (\bibinfo {year} {2026}),\ \Eprint {https://arxiv.org/abs/2605.20417} {arXiv:2605.20417 [hep-lat]} \BibitemShut {NoStop}%
\bibitem [{\citenamefont {Halimeh}\ \emph {et~al.}(2025)\citenamefont {Halimeh}, \citenamefont {Mueller}, \citenamefont {Knolle}, \citenamefont {Papić},\ and\ \citenamefont {Davoudi}}]{halimeh2025_quantumsimulationoutofequilibriumdynamics}%
  \BibitemOpen
  \bibfield  {author} {\bibinfo {author} {\bibfnamefont {J.~C.}\ \bibnamefont {Halimeh}}, \bibinfo {author} {\bibfnamefont {N.}~\bibnamefont {Mueller}}, \bibinfo {author} {\bibfnamefont {J.}~\bibnamefont {Knolle}}, \bibinfo {author} {\bibfnamefont {Z.}~\bibnamefont {Papić}},\ and\ \bibinfo {author} {\bibfnamefont {Z.}~\bibnamefont {Davoudi}},\ }\href {https://arxiv.org/abs/2509.03586} {\bibinfo {title} {Quantum simulation of out-of-equilibrium dynamics in gauge theories}} (\bibinfo {year} {2025}),\ \Eprint {https://arxiv.org/abs/2509.03586} {arXiv:2509.03586 [quant-ph]} \BibitemShut {NoStop}%
\bibitem [{\citenamefont {Martinez}\ \emph {et~al.}(2016)\citenamefont {Martinez}, \citenamefont {Muschik}, \citenamefont {Schindler}, \citenamefont {Nigg}, \citenamefont {Erhard}, \citenamefont {Heyl}, \citenamefont {Hauke}, \citenamefont {Dalmonte}, \citenamefont {Monz}, \citenamefont {Zoller},\ and\ \citenamefont {Blatt}}]{Martinez_nat2016_quantumsimulationgauge}%
  \BibitemOpen
  \bibfield  {author} {\bibinfo {author} {\bibfnamefont {E.~A.}\ \bibnamefont {Martinez}}, \bibinfo {author} {\bibfnamefont {C.~A.}\ \bibnamefont {Muschik}}, \bibinfo {author} {\bibfnamefont {P.}~\bibnamefont {Schindler}}, \bibinfo {author} {\bibfnamefont {D.}~\bibnamefont {Nigg}}, \bibinfo {author} {\bibfnamefont {A.}~\bibnamefont {Erhard}}, \bibinfo {author} {\bibfnamefont {M.}~\bibnamefont {Heyl}}, \bibinfo {author} {\bibfnamefont {P.}~\bibnamefont {Hauke}}, \bibinfo {author} {\bibfnamefont {M.}~\bibnamefont {Dalmonte}}, \bibinfo {author} {\bibfnamefont {T.}~\bibnamefont {Monz}}, \bibinfo {author} {\bibfnamefont {P.}~\bibnamefont {Zoller}},\ and\ \bibinfo {author} {\bibfnamefont {R.}~\bibnamefont {Blatt}},\ }\bibfield  {title} {\bibinfo {title} {Real-time dynamics of lattice gauge theories with a few-qubit quantum computer},\ }\href {https://doi.org/10.1038/nature18318} {\bibfield  {journal} {\bibinfo  {journal} {Nature}\ }\textbf {\bibinfo {volume} {534}},\ \bibinfo {pages} {516–519} (\bibinfo {year}
  {2016})}\BibitemShut {NoStop}%
\bibitem [{\citenamefont {De}\ \emph {et~al.}(2024)\citenamefont {De}, \citenamefont {Lerose}, \citenamefont {Luo}, \citenamefont {Surace}, \citenamefont {Schuckert}, \citenamefont {Bennewitz}, \citenamefont {Ware}, \citenamefont {Morong}, \citenamefont {Collins}, \citenamefont {Davoudi}, \citenamefont {Gorshkov}, \citenamefont {Katz},\ and\ \citenamefont {Monroe}}]{deMonroe_2024_1Dobservationstringbreakingdynamicsquantum}%
  \BibitemOpen
  \bibfield  {author} {\bibinfo {author} {\bibfnamefont {A.}~\bibnamefont {De}}, \bibinfo {author} {\bibfnamefont {A.}~\bibnamefont {Lerose}}, \bibinfo {author} {\bibfnamefont {D.}~\bibnamefont {Luo}}, \bibinfo {author} {\bibfnamefont {F.~M.}\ \bibnamefont {Surace}}, \bibinfo {author} {\bibfnamefont {A.}~\bibnamefont {Schuckert}}, \bibinfo {author} {\bibfnamefont {E.~R.}\ \bibnamefont {Bennewitz}}, \bibinfo {author} {\bibfnamefont {B.}~\bibnamefont {Ware}}, \bibinfo {author} {\bibfnamefont {W.}~\bibnamefont {Morong}}, \bibinfo {author} {\bibfnamefont {K.~S.}\ \bibnamefont {Collins}}, \bibinfo {author} {\bibfnamefont {Z.}~\bibnamefont {Davoudi}}, \bibinfo {author} {\bibfnamefont {A.~V.}\ \bibnamefont {Gorshkov}}, \bibinfo {author} {\bibfnamefont {O.}~\bibnamefont {Katz}},\ and\ \bibinfo {author} {\bibfnamefont {C.}~\bibnamefont {Monroe}},\ }\href {https://arxiv.org/abs/2410.13815} {\bibinfo {title} {Observation of string-breaking dynamics in a quantum simulator}} (\bibinfo {year} {2024}),\ \Eprint
  {https://arxiv.org/abs/2410.13815} {arXiv:2410.13815 [quant-ph]} \BibitemShut {NoStop}%
\bibitem [{\citenamefont {Meth}\ \emph {et~al.}(2025)\citenamefont {Meth}, \citenamefont {Zhang}, \citenamefont {Haase}, \citenamefont {Edmunds}, \citenamefont {Postler}, \citenamefont {Jena}, \citenamefont {Steiner}, \citenamefont {Dellantonio}, \citenamefont {Blatt}, \citenamefont {Zoller}, \citenamefont {Monz}, \citenamefont {Schindler}, \citenamefont {Muschik},\ and\ \citenamefont {Ringbauer}}]{meth2025simulating}%
  \BibitemOpen
  \bibfield  {author} {\bibinfo {author} {\bibfnamefont {M.}~\bibnamefont {Meth}}, \bibinfo {author} {\bibfnamefont {J.}~\bibnamefont {Zhang}}, \bibinfo {author} {\bibfnamefont {J.~F.}\ \bibnamefont {Haase}}, \bibinfo {author} {\bibfnamefont {C.}~\bibnamefont {Edmunds}}, \bibinfo {author} {\bibfnamefont {L.}~\bibnamefont {Postler}}, \bibinfo {author} {\bibfnamefont {A.~J.}\ \bibnamefont {Jena}}, \bibinfo {author} {\bibfnamefont {A.}~\bibnamefont {Steiner}}, \bibinfo {author} {\bibfnamefont {L.}~\bibnamefont {Dellantonio}}, \bibinfo {author} {\bibfnamefont {R.}~\bibnamefont {Blatt}}, \bibinfo {author} {\bibfnamefont {P.}~\bibnamefont {Zoller}}, \bibinfo {author} {\bibfnamefont {T.}~\bibnamefont {Monz}}, \bibinfo {author} {\bibfnamefont {P.}~\bibnamefont {Schindler}}, \bibinfo {author} {\bibfnamefont {C.}~\bibnamefont {Muschik}},\ and\ \bibinfo {author} {\bibfnamefont {M.}~\bibnamefont {Ringbauer}},\ }\bibfield  {title} {\bibinfo {title} {Simulating two-dimensional lattice gauge theories on a qudit quantum
  computer},\ }\href {https://doi.org/10.1038/s41567-025-02797-w} {\bibfield  {journal} {\bibinfo  {journal} {Nature Physics}\ }\textbf {\bibinfo {volume} {21}},\ \bibinfo {pages} {570–576} (\bibinfo {year} {2025})}\BibitemShut {NoStop}%
\bibitem [{\citenamefont {Joshi}\ \emph {et~al.}(2026)\citenamefont {Joshi}, \citenamefont {Tian}, \citenamefont {Hemery}, \citenamefont {Srivatsa}, \citenamefont {Osborne}, \citenamefont {Dreyer}, \citenamefont {Rinaldi},\ and\ \citenamefont {Halimeh}}]{joshiHalimeh_2026observationgenuine21dstring}%
  \BibitemOpen
  \bibfield  {author} {\bibinfo {author} {\bibfnamefont {R.}~\bibnamefont {Joshi}}, \bibinfo {author} {\bibfnamefont {Y.}~\bibnamefont {Tian}}, \bibinfo {author} {\bibfnamefont {K.}~\bibnamefont {Hemery}}, \bibinfo {author} {\bibfnamefont {N.~S.}\ \bibnamefont {Srivatsa}}, \bibinfo {author} {\bibfnamefont {J.~J.}\ \bibnamefont {Osborne}}, \bibinfo {author} {\bibfnamefont {H.}~\bibnamefont {Dreyer}}, \bibinfo {author} {\bibfnamefont {E.}~\bibnamefont {Rinaldi}},\ and\ \bibinfo {author} {\bibfnamefont {J.~C.}\ \bibnamefont {Halimeh}},\ }\href {https://arxiv.org/abs/2604.07436} {\bibinfo {title} {Observation of genuine $2+1$d string dynamics in a u$(1)$ lattice gauge theory with a tunable plaquette term on a trapped-ion quantum computer}} (\bibinfo {year} {2026}),\ \Eprint {https://arxiv.org/abs/2604.07436} {arXiv:2604.07436 [quant-ph]} \BibitemShut {NoStop}%
\bibitem [{\citenamefont {Bernien}\ \emph {et~al.}(2017)\citenamefont {Bernien}, \citenamefont {Schwartz}, \citenamefont {Keesling}, \citenamefont {Levine}, \citenamefont {Omran}, \citenamefont {Pichler}, \citenamefont {Choi}, \citenamefont {Zibrov}, \citenamefont {Endres}, \citenamefont {Greiner}, \citenamefont {Vuletić},\ and\ \citenamefont {Lukin}}]{bernien2017probing}%
  \BibitemOpen
  \bibfield  {author} {\bibinfo {author} {\bibfnamefont {H.}~\bibnamefont {Bernien}}, \bibinfo {author} {\bibfnamefont {S.}~\bibnamefont {Schwartz}}, \bibinfo {author} {\bibfnamefont {A.}~\bibnamefont {Keesling}}, \bibinfo {author} {\bibfnamefont {H.}~\bibnamefont {Levine}}, \bibinfo {author} {\bibfnamefont {A.}~\bibnamefont {Omran}}, \bibinfo {author} {\bibfnamefont {H.}~\bibnamefont {Pichler}}, \bibinfo {author} {\bibfnamefont {S.}~\bibnamefont {Choi}}, \bibinfo {author} {\bibfnamefont {A.~S.}\ \bibnamefont {Zibrov}}, \bibinfo {author} {\bibfnamefont {M.}~\bibnamefont {Endres}}, \bibinfo {author} {\bibfnamefont {M.}~\bibnamefont {Greiner}}, \bibinfo {author} {\bibfnamefont {V.}~\bibnamefont {Vuletić}},\ and\ \bibinfo {author} {\bibfnamefont {M.~D.}\ \bibnamefont {Lukin}},\ }\bibfield  {title} {\bibinfo {title} {Probing many-body dynamics on a 51-atom quantum simulator},\ }\href {https://doi.org/10.1038/nature24622} {\bibfield  {journal} {\bibinfo  {journal} {Nature}\ }\textbf {\bibinfo {volume} {551}},\
  \bibinfo {pages} {579–584} (\bibinfo {year} {2017})}\BibitemShut {NoStop}%
\bibitem [{\citenamefont {González-Cuadra}\ \emph {et~al.}(2025)\citenamefont {González-Cuadra}, \citenamefont {Hamdan}, \citenamefont {Zache}, \citenamefont {Braverman}, \citenamefont {Kornjača}, \citenamefont {Lukin}, \citenamefont {Cantú}, \citenamefont {Liu}, \citenamefont {Wang}, \citenamefont {Keesling}, \citenamefont {Lukin}, \citenamefont {Zoller},\ and\ \citenamefont {Bylinskii}}]{GonzlezCuadra2025_nat2025_observationstringbreaking}%
  \BibitemOpen
  \bibfield  {author} {\bibinfo {author} {\bibfnamefont {D.}~\bibnamefont {González-Cuadra}}, \bibinfo {author} {\bibfnamefont {M.}~\bibnamefont {Hamdan}}, \bibinfo {author} {\bibfnamefont {T.~V.}\ \bibnamefont {Zache}}, \bibinfo {author} {\bibfnamefont {B.}~\bibnamefont {Braverman}}, \bibinfo {author} {\bibfnamefont {M.}~\bibnamefont {Kornjača}}, \bibinfo {author} {\bibfnamefont {A.}~\bibnamefont {Lukin}}, \bibinfo {author} {\bibfnamefont {S.~H.}\ \bibnamefont {Cantú}}, \bibinfo {author} {\bibfnamefont {F.}~\bibnamefont {Liu}}, \bibinfo {author} {\bibfnamefont {S.-T.}\ \bibnamefont {Wang}}, \bibinfo {author} {\bibfnamefont {A.}~\bibnamefont {Keesling}}, \bibinfo {author} {\bibfnamefont {M.~D.}\ \bibnamefont {Lukin}}, \bibinfo {author} {\bibfnamefont {P.}~\bibnamefont {Zoller}},\ and\ \bibinfo {author} {\bibfnamefont {A.}~\bibnamefont {Bylinskii}},\ }\bibfield  {title} {\bibinfo {title} {Observation of string breaking on a (2 + 1)d rydberg quantum simulator},\ }\href
  {https://doi.org/10.1038/s41586-025-09051-6} {\bibfield  {journal} {\bibinfo  {journal} {Nature}\ }\textbf {\bibinfo {volume} {642}},\ \bibinfo {pages} {321–326} (\bibinfo {year} {2025})}\BibitemShut {NoStop}%
\bibitem [{\citenamefont {Xiang}\ \emph {et~al.}(2025)\citenamefont {Xiang}, \citenamefont {Zhou}, \citenamefont {Liu}, \citenamefont {Liu}, \citenamefont {Zhang}, \citenamefont {Yuan}, \citenamefont {Zhang}, \citenamefont {Xu}, \citenamefont {Dalmonte}, \citenamefont {Deng},\ and\ \citenamefont {Li}}]{xiang2025real}%
  \BibitemOpen
  \bibfield  {author} {\bibinfo {author} {\bibfnamefont {D.-S.}\ \bibnamefont {Xiang}}, \bibinfo {author} {\bibfnamefont {P.}~\bibnamefont {Zhou}}, \bibinfo {author} {\bibfnamefont {C.}~\bibnamefont {Liu}}, \bibinfo {author} {\bibfnamefont {H.-X.}\ \bibnamefont {Liu}}, \bibinfo {author} {\bibfnamefont {Y.-W.}\ \bibnamefont {Zhang}}, \bibinfo {author} {\bibfnamefont {D.}~\bibnamefont {Yuan}}, \bibinfo {author} {\bibfnamefont {K.}~\bibnamefont {Zhang}}, \bibinfo {author} {\bibfnamefont {B.}~\bibnamefont {Xu}}, \bibinfo {author} {\bibfnamefont {M.}~\bibnamefont {Dalmonte}}, \bibinfo {author} {\bibfnamefont {D.-L.}\ \bibnamefont {Deng}},\ and\ \bibinfo {author} {\bibfnamefont {L.}~\bibnamefont {Li}},\ }\href {https://arxiv.org/abs/2508.06639} {\bibinfo {title} {Real-time scattering and freeze-out dynamics in rydberg-atom lattice gauge theory}} (\bibinfo {year} {2025}),\ \Eprint {https://arxiv.org/abs/2508.06639} {arXiv:2508.06639 [cond-mat.quant-gas]} \BibitemShut {NoStop}%
\bibitem [{\citenamefont {Mark}\ \emph {et~al.}(2025)\citenamefont {Mark}, \citenamefont {Surace}, \citenamefont {Schuster}, \citenamefont {Shaw}, \citenamefont {Gong}, \citenamefont {Choi},\ and\ \citenamefont {Endres}}]{mark2025observation}%
  \BibitemOpen
  \bibfield  {author} {\bibinfo {author} {\bibfnamefont {D.~K.}\ \bibnamefont {Mark}}, \bibinfo {author} {\bibfnamefont {F.~M.}\ \bibnamefont {Surace}}, \bibinfo {author} {\bibfnamefont {T.}~\bibnamefont {Schuster}}, \bibinfo {author} {\bibfnamefont {A.~L.}\ \bibnamefont {Shaw}}, \bibinfo {author} {\bibfnamefont {W.}~\bibnamefont {Gong}}, \bibinfo {author} {\bibfnamefont {S.}~\bibnamefont {Choi}},\ and\ \bibinfo {author} {\bibfnamefont {M.}~\bibnamefont {Endres}},\ }\href {https://arxiv.org/abs/2510.11679} {\bibinfo {title} {Observation of ballistic plasma and memory in high-energy gauge theory dynamics}} (\bibinfo {year} {2025}),\ \Eprint {https://arxiv.org/abs/2510.11679} {arXiv:2510.11679 [quant-ph]} \BibitemShut {NoStop}%
\bibitem [{\citenamefont {Liang}\ \emph {et~al.}(2025)\citenamefont {Liang}, \citenamefont {Yue}, \citenamefont {Chao}, \citenamefont {Hua}, \citenamefont {Lin}, \citenamefont {Tey},\ and\ \citenamefont {You}}]{liang2025observation}%
  \BibitemOpen
  \bibfield  {author} {\bibinfo {author} {\bibfnamefont {X.}~\bibnamefont {Liang}}, \bibinfo {author} {\bibfnamefont {Z.}~\bibnamefont {Yue}}, \bibinfo {author} {\bibfnamefont {Y.-X.}\ \bibnamefont {Chao}}, \bibinfo {author} {\bibfnamefont {Z.-X.}\ \bibnamefont {Hua}}, \bibinfo {author} {\bibfnamefont {Y.}~\bibnamefont {Lin}}, \bibinfo {author} {\bibfnamefont {M.~K.}\ \bibnamefont {Tey}},\ and\ \bibinfo {author} {\bibfnamefont {L.}~\bibnamefont {You}},\ }\bibfield  {title} {\bibinfo {title} {Observation of anomalous information scrambling in a rydberg atom array},\ }\href {http://dx.doi.org/10.1103/w1cp-l5vq} {\bibfield  {journal} {\bibinfo  {journal} {Physical Review Letters}\ }\textbf {\bibinfo {volume} {135}} (\bibinfo {year} {2025})}\BibitemShut {NoStop}%
\bibitem [{\citenamefont {Aidelsburger}\ \emph {et~al.}(2021)\citenamefont {Aidelsburger}, \citenamefont {Barbiero}, \citenamefont {Bermudez}, \citenamefont {Chanda}, \citenamefont {Dauphin}, \citenamefont {González-Cuadra}, \citenamefont {Grzybowski}, \citenamefont {Hands}, \citenamefont {Jendrzejewski}, \citenamefont {J\"{u}nemann}, \citenamefont {Juzeliūnas}, \citenamefont {Kasper}, \citenamefont {Piga}, \citenamefont {Ran}, \citenamefont {Rizzi}, \citenamefont {Sierra}, \citenamefont {Tagliacozzo}, \citenamefont {Tirrito}, \citenamefont {Zache}, \citenamefont {Zakrzewski}, \citenamefont {Zohar},\ and\ \citenamefont {Lewenstein}}]{aidelsburger2021cold}%
  \BibitemOpen
  \bibfield  {author} {\bibinfo {author} {\bibfnamefont {M.}~\bibnamefont {Aidelsburger}}, \bibinfo {author} {\bibfnamefont {L.}~\bibnamefont {Barbiero}}, \bibinfo {author} {\bibfnamefont {A.}~\bibnamefont {Bermudez}}, \bibinfo {author} {\bibfnamefont {T.}~\bibnamefont {Chanda}}, \bibinfo {author} {\bibfnamefont {A.}~\bibnamefont {Dauphin}}, \bibinfo {author} {\bibfnamefont {D.}~\bibnamefont {González-Cuadra}}, \bibinfo {author} {\bibfnamefont {P.~R.}\ \bibnamefont {Grzybowski}}, \bibinfo {author} {\bibfnamefont {S.}~\bibnamefont {Hands}}, \bibinfo {author} {\bibfnamefont {F.}~\bibnamefont {Jendrzejewski}}, \bibinfo {author} {\bibfnamefont {J.}~\bibnamefont {J\"{u}nemann}}, \bibinfo {author} {\bibfnamefont {G.}~\bibnamefont {Juzeliūnas}}, \bibinfo {author} {\bibfnamefont {V.}~\bibnamefont {Kasper}}, \bibinfo {author} {\bibfnamefont {A.}~\bibnamefont {Piga}}, \bibinfo {author} {\bibfnamefont {S.-J.}\ \bibnamefont {Ran}}, \bibinfo {author} {\bibfnamefont {M.}~\bibnamefont {Rizzi}}, \bibinfo {author}
  {\bibfnamefont {G.}~\bibnamefont {Sierra}}, \bibinfo {author} {\bibfnamefont {L.}~\bibnamefont {Tagliacozzo}}, \bibinfo {author} {\bibfnamefont {E.}~\bibnamefont {Tirrito}}, \bibinfo {author} {\bibfnamefont {T.~V.}\ \bibnamefont {Zache}}, \bibinfo {author} {\bibfnamefont {J.}~\bibnamefont {Zakrzewski}}, \bibinfo {author} {\bibfnamefont {E.}~\bibnamefont {Zohar}},\ and\ \bibinfo {author} {\bibfnamefont {M.}~\bibnamefont {Lewenstein}},\ }\bibfield  {title} {\bibinfo {title} {Cold atoms meet lattice gauge theory},\ }\href {http://dx.doi.org/10.1098/rsta.2021.0064} {\bibfield  {journal} {\bibinfo  {journal} {Philosophical Transactions of the Royal Society A: Mathematical, Physical and Engineering Sciences}\ }\textbf {\bibinfo {volume} {380}} (\bibinfo {year} {2021})}\BibitemShut {NoStop}%
\bibitem [{\citenamefont {Mil}\ \emph {et~al.}(2020)\citenamefont {Mil}, \citenamefont {Zache}, \citenamefont {Hegde}, \citenamefont {Xia}, \citenamefont {Bhatt}, \citenamefont {Oberthaler}, \citenamefont {Hauke}, \citenamefont {Berges},\ and\ \citenamefont {Jendrzejewski}}]{mil2020scalable}%
  \BibitemOpen
  \bibfield  {author} {\bibinfo {author} {\bibfnamefont {A.}~\bibnamefont {Mil}}, \bibinfo {author} {\bibfnamefont {T.~V.}\ \bibnamefont {Zache}}, \bibinfo {author} {\bibfnamefont {A.}~\bibnamefont {Hegde}}, \bibinfo {author} {\bibfnamefont {A.}~\bibnamefont {Xia}}, \bibinfo {author} {\bibfnamefont {R.~P.}\ \bibnamefont {Bhatt}}, \bibinfo {author} {\bibfnamefont {M.~K.}\ \bibnamefont {Oberthaler}}, \bibinfo {author} {\bibfnamefont {P.}~\bibnamefont {Hauke}}, \bibinfo {author} {\bibfnamefont {J.}~\bibnamefont {Berges}},\ and\ \bibinfo {author} {\bibfnamefont {F.}~\bibnamefont {Jendrzejewski}},\ }\bibfield  {title} {\bibinfo {title} {A scalable realization of local u(1) gauge invariance in cold atomic mixtures},\ }\href {https://doi.org/10.1126/science.aaz5312} {\bibfield  {journal} {\bibinfo  {journal} {Science}\ }\textbf {\bibinfo {volume} {367}},\ \bibinfo {pages} {1128–1130} (\bibinfo {year} {2020})}\BibitemShut {NoStop}%
\bibitem [{\citenamefont {Zhu}\ \emph {et~al.}(2024)\citenamefont {Zhu}, \citenamefont {Liu}, \citenamefont {Lagnese}, \citenamefont {Surace}, \citenamefont {Zhang}, \citenamefont {He}, \citenamefont {Halimeh}, \citenamefont {Dalmonte}, \citenamefont {Morampudi}, \citenamefont {Wilczek}, \citenamefont {Yuan},\ and\ \citenamefont {Pan}}]{zhu2024probing}%
  \BibitemOpen
  \bibfield  {author} {\bibinfo {author} {\bibfnamefont {Z.-H.}\ \bibnamefont {Zhu}}, \bibinfo {author} {\bibfnamefont {Y.}~\bibnamefont {Liu}}, \bibinfo {author} {\bibfnamefont {G.}~\bibnamefont {Lagnese}}, \bibinfo {author} {\bibfnamefont {F.~M.}\ \bibnamefont {Surace}}, \bibinfo {author} {\bibfnamefont {W.-Y.}\ \bibnamefont {Zhang}}, \bibinfo {author} {\bibfnamefont {M.-G.}\ \bibnamefont {He}}, \bibinfo {author} {\bibfnamefont {J.~C.}\ \bibnamefont {Halimeh}}, \bibinfo {author} {\bibfnamefont {M.}~\bibnamefont {Dalmonte}}, \bibinfo {author} {\bibfnamefont {S.~C.}\ \bibnamefont {Morampudi}}, \bibinfo {author} {\bibfnamefont {F.}~\bibnamefont {Wilczek}}, \bibinfo {author} {\bibfnamefont {Z.-S.}\ \bibnamefont {Yuan}},\ and\ \bibinfo {author} {\bibfnamefont {J.-W.}\ \bibnamefont {Pan}},\ }\href {https://arxiv.org/abs/2411.12565} {\bibinfo {title} {Probing false vacuum decay on a cold-atom gauge-theory quantum simulator}} (\bibinfo {year} {2024}),\ \Eprint {https://arxiv.org/abs/2411.12565} {arXiv:2411.12565
  [cond-mat.quant-gas]} \BibitemShut {NoStop}%
\bibitem [{\citenamefont {Karch}\ \emph {et~al.}(2026)\citenamefont {Karch}, \citenamefont {Will}, \citenamefont {Rodriguez}, \citenamefont {Liebster}, \citenamefont {Huh}, \citenamefont {Knap}, \citenamefont {Pollmann}, \citenamefont {Kuhlenkamp}, \citenamefont {Bloch},\ and\ \citenamefont {Aidelsburger}}]{karch2026dynamical}%
  \BibitemOpen
  \bibfield  {author} {\bibinfo {author} {\bibfnamefont {S.}~\bibnamefont {Karch}}, \bibinfo {author} {\bibfnamefont {M.}~\bibnamefont {Will}}, \bibinfo {author} {\bibfnamefont {I.~P.}\ \bibnamefont {Rodriguez}}, \bibinfo {author} {\bibfnamefont {N.}~\bibnamefont {Liebster}}, \bibinfo {author} {\bibfnamefont {S.}~\bibnamefont {Huh}}, \bibinfo {author} {\bibfnamefont {M.}~\bibnamefont {Knap}}, \bibinfo {author} {\bibfnamefont {F.}~\bibnamefont {Pollmann}}, \bibinfo {author} {\bibfnamefont {C.}~\bibnamefont {Kuhlenkamp}}, \bibinfo {author} {\bibfnamefont {I.}~\bibnamefont {Bloch}},\ and\ \bibinfo {author} {\bibfnamefont {M.}~\bibnamefont {Aidelsburger}},\ }\href {https://arxiv.org/abs/2604.24744} {\bibinfo {title} {Dynamical preparation of u(1) quantum spin liquids in an analogue quantum simulator}} (\bibinfo {year} {2026}),\ \Eprint {https://arxiv.org/abs/2604.24744} {arXiv:2604.24744 [cond-mat.quant-gas]} \BibitemShut {NoStop}%
\bibitem [{\citenamefont {Cochran}\ \emph {et~al.}(2025)\citenamefont {Cochran}, \citenamefont {Jobst}, \citenamefont {Rosenberg}, \citenamefont {Lensky}, \citenamefont {Gyawali}, \citenamefont {Eassa}, \citenamefont {Will}, \citenamefont {Szasz}, \citenamefont {Abanin}, \citenamefont {Acharya},\ and\ \citenamefont {et~al.}}]{Cochran2025_visualizingstring_google}%
  \BibitemOpen
  \bibfield  {author} {\bibinfo {author} {\bibfnamefont {T.~A.}\ \bibnamefont {Cochran}}, \bibinfo {author} {\bibfnamefont {B.}~\bibnamefont {Jobst}}, \bibinfo {author} {\bibfnamefont {E.}~\bibnamefont {Rosenberg}}, \bibinfo {author} {\bibfnamefont {Y.~D.}\ \bibnamefont {Lensky}}, \bibinfo {author} {\bibfnamefont {G.}~\bibnamefont {Gyawali}}, \bibinfo {author} {\bibfnamefont {N.}~\bibnamefont {Eassa}}, \bibinfo {author} {\bibfnamefont {M.}~\bibnamefont {Will}}, \bibinfo {author} {\bibfnamefont {A.}~\bibnamefont {Szasz}}, \bibinfo {author} {\bibfnamefont {D.}~\bibnamefont {Abanin}}, \bibinfo {author} {\bibfnamefont {R.}~\bibnamefont {Acharya}},\ and\ \bibinfo {author} {\bibnamefont {et~al.}},\ }\bibfield  {title} {\bibinfo {title} {Visualizing dynamics of charges and strings in (2 + 1)d lattice gauge theories},\ }\href {http://dx.doi.org/10.1038/s41586-025-08999-9} {\bibfield  {journal} {\bibinfo  {journal} {Nature}\ }\textbf {\bibinfo {volume} {642}},\ \bibinfo {pages} {315} (\bibinfo {year}
  {2025})}\BibitemShut {NoStop}%
\bibitem [{\citenamefont {Schuhmacher}\ \emph {et~al.}(2025)\citenamefont {Schuhmacher}, \citenamefont {Su}, \citenamefont {Osborne}, \citenamefont {Gandon}, \citenamefont {Halimeh},\ and\ \citenamefont {Tavernelli}}]{schuhmacher2025observation}%
  \BibitemOpen
  \bibfield  {author} {\bibinfo {author} {\bibfnamefont {J.}~\bibnamefont {Schuhmacher}}, \bibinfo {author} {\bibfnamefont {G.-X.}\ \bibnamefont {Su}}, \bibinfo {author} {\bibfnamefont {J.~J.}\ \bibnamefont {Osborne}}, \bibinfo {author} {\bibfnamefont {A.}~\bibnamefont {Gandon}}, \bibinfo {author} {\bibfnamefont {J.~C.}\ \bibnamefont {Halimeh}},\ and\ \bibinfo {author} {\bibfnamefont {I.}~\bibnamefont {Tavernelli}},\ }\href {https://arxiv.org/abs/2505.20387} {\bibinfo {title} {Observation of hadron scattering in a lattice gauge theory on a quantum computer}} (\bibinfo {year} {2025}),\ \Eprint {https://arxiv.org/abs/2505.20387} {arXiv:2505.20387 [quant-ph]} \BibitemShut {NoStop}%
\bibitem [{\citenamefont {Cobos}\ \emph {et~al.}(2025)\citenamefont {Cobos}, \citenamefont {Fraxanet}, \citenamefont {Benito}, \citenamefont {di~Marcantonio}, \citenamefont {Rivero}, \citenamefont {Kapás}, \citenamefont {Werner}, \citenamefont {Örs Legeza}, \citenamefont {Bermudez},\ and\ \citenamefont {Rico}}]{cobosRico_2025realtimedynamics21dgauge}%
  \BibitemOpen
  \bibfield  {author} {\bibinfo {author} {\bibfnamefont {J.}~\bibnamefont {Cobos}}, \bibinfo {author} {\bibfnamefont {J.}~\bibnamefont {Fraxanet}}, \bibinfo {author} {\bibfnamefont {C.}~\bibnamefont {Benito}}, \bibinfo {author} {\bibfnamefont {F.}~\bibnamefont {di~Marcantonio}}, \bibinfo {author} {\bibfnamefont {P.}~\bibnamefont {Rivero}}, \bibinfo {author} {\bibfnamefont {K.}~\bibnamefont {Kapás}}, \bibinfo {author} {\bibfnamefont {M.~A.}\ \bibnamefont {Werner}}, \bibinfo {author} {\bibnamefont {Örs Legeza}}, \bibinfo {author} {\bibfnamefont {A.}~\bibnamefont {Bermudez}},\ and\ \bibinfo {author} {\bibfnamefont {E.}~\bibnamefont {Rico}},\ }\href {https://arxiv.org/abs/2507.08088} {\bibinfo {title} {Real-time dynamics in a (2+1)-d gauge theory: The stringy nature on a superconducting quantum simulator}} (\bibinfo {year} {2025}),\ \Eprint {https://arxiv.org/abs/2507.08088} {arXiv:2507.08088 [quant-ph]} \BibitemShut {NoStop}%
\bibitem [{\citenamefont {Montangero}(2018)}]{montangero2018introduction}%
  \BibitemOpen
  \bibfield  {author} {\bibinfo {author} {\bibfnamefont {S.}~\bibnamefont {Montangero}},\ }\href {https://doi.org/10.1007/978-3-030-01409-4} {\emph {\bibinfo {title} {Introduction to Tensor Network Methods: Numerical simulations of low-dimensional many-body quantum systems}}}\ (\bibinfo  {publisher} {Springer International Publishing},\ \bibinfo {year} {2018})\BibitemShut {NoStop}%
\bibitem [{\citenamefont {Silvi}\ \emph {et~al.}(2019)\citenamefont {Silvi}, \citenamefont {Tschirsich}, \citenamefont {Gerster}, \citenamefont {J{\"u}nemann}, \citenamefont {Jaschke}, \citenamefont {Rizzi},\ and\ \citenamefont {Montangero}}]{Silvi2019tn}%
  \BibitemOpen
  \bibfield  {author} {\bibinfo {author} {\bibfnamefont {P.}~\bibnamefont {Silvi}}, \bibinfo {author} {\bibfnamefont {F.}~\bibnamefont {Tschirsich}}, \bibinfo {author} {\bibfnamefont {M.}~\bibnamefont {Gerster}}, \bibinfo {author} {\bibfnamefont {J.}~\bibnamefont {J{\"u}nemann}}, \bibinfo {author} {\bibfnamefont {D.}~\bibnamefont {Jaschke}}, \bibinfo {author} {\bibfnamefont {M.}~\bibnamefont {Rizzi}},\ and\ \bibinfo {author} {\bibfnamefont {S.}~\bibnamefont {Montangero}},\ }\bibfield  {title} {\bibinfo {title} {{The Tensor Networks Anthology: Simulation techniques for many-body quantum lattice systems}},\ }\href {https://doi.org/10.21468/SciPostPhysLectNotes.8} {\bibfield  {journal} {\bibinfo  {journal} {SciPost Phys. Lect. Notes}\ ,\ \bibinfo {pages} {8}} (\bibinfo {year} {2019})}\BibitemShut {NoStop}%
\bibitem [{\citenamefont {Schollw{\"o}ck}(2011)}]{schollwock2011density}%
  \BibitemOpen
  \bibfield  {author} {\bibinfo {author} {\bibfnamefont {U.}~\bibnamefont {Schollw{\"o}ck}},\ }\bibfield  {title} {\bibinfo {title} {The density-matrix renormalization group in the age of matrix product states},\ }\href {https://doi.org/10.1016/j.aop.2010.09.012} {\bibfield  {journal} {\bibinfo  {journal} {Annals of physics}\ }\textbf {\bibinfo {volume} {326}},\ \bibinfo {pages} {96} (\bibinfo {year} {2011})}\BibitemShut {NoStop}%
\bibitem [{\citenamefont {Verstraete}\ \emph {et~al.}(2008)\citenamefont {Verstraete}, \citenamefont {Murg},\ and\ \citenamefont {Cirac}}]{verstraete2008matrix}%
  \BibitemOpen
  \bibfield  {author} {\bibinfo {author} {\bibfnamefont {F.}~\bibnamefont {Verstraete}}, \bibinfo {author} {\bibfnamefont {V.}~\bibnamefont {Murg}},\ and\ \bibinfo {author} {\bibfnamefont {J.~I.}\ \bibnamefont {Cirac}},\ }\bibfield  {title} {\bibinfo {title} {Matrix product states, projected entangled pair states, and variational renormalization group methods for quantum spin systems},\ }\href {https://doi.org/10.1080/14789940801912366} {\bibfield  {journal} {\bibinfo  {journal} {Advances in physics}\ }\textbf {\bibinfo {volume} {57}},\ \bibinfo {pages} {143} (\bibinfo {year} {2008})}\BibitemShut {NoStop}%
\bibitem [{\citenamefont {Carmen~Bañuls}\ and\ \citenamefont {Cichy}(2020)}]{carmen2020review}%
  \BibitemOpen
  \bibfield  {author} {\bibinfo {author} {\bibfnamefont {M.}~\bibnamefont {Carmen~Bañuls}}\ and\ \bibinfo {author} {\bibfnamefont {K.}~\bibnamefont {Cichy}},\ }\bibfield  {title} {\bibinfo {title} {Review on novel methods for lattice gauge theories},\ }\href {https://doi.org/10.1088/1361-6633/ab6311} {\bibfield  {journal} {\bibinfo  {journal} {Reports on Progress in Physics}\ }\textbf {\bibinfo {volume} {83}},\ \bibinfo {pages} {024401} (\bibinfo {year} {2020})}\BibitemShut {NoStop}%
\bibitem [{\citenamefont {Emonts}\ and\ \citenamefont {Zohar}(2020)}]{emonts2020gauss}%
  \BibitemOpen
  \bibfield  {author} {\bibinfo {author} {\bibfnamefont {P.}~\bibnamefont {Emonts}}\ and\ \bibinfo {author} {\bibfnamefont {E.}~\bibnamefont {Zohar}},\ }\bibfield  {title} {\bibinfo {title} {Gauss law, minimal coupling and fermionic peps for lattice gauge theories},\ }\href {http://dx.doi.org/10.21468/SciPostPhysLectNotes.12} {\bibfield  {journal} {\bibinfo  {journal} {SciPost Physics Lecture Notes}\ } (\bibinfo {year} {2020})}\BibitemShut {NoStop}%
\bibitem [{\citenamefont {Montangero}\ \emph {et~al.}(2021)\citenamefont {Montangero}, \citenamefont {Rico},\ and\ \citenamefont {Silvi}}]{montangero2022loop}%
  \BibitemOpen
  \bibfield  {author} {\bibinfo {author} {\bibfnamefont {S.}~\bibnamefont {Montangero}}, \bibinfo {author} {\bibfnamefont {E.}~\bibnamefont {Rico}},\ and\ \bibinfo {author} {\bibfnamefont {P.}~\bibnamefont {Silvi}},\ }\bibfield  {title} {\bibinfo {title} {Loop-free tensor networks for high-energy physics},\ }\bibfield  {journal} {\bibinfo  {journal} {Philosophical Transactions of the Royal Society A: Mathematical, Physical and Engineering Sciences}\ }\textbf {\bibinfo {volume} {380}},\ \href {https://doi.org/10.1098/rsta.2021.0065} {10.1098/rsta.2021.0065} (\bibinfo {year} {2021})\BibitemShut {NoStop}%
\bibitem [{\citenamefont {Zhang}\ \emph {et~al.}(2024)\citenamefont {Zhang}, \citenamefont {Liu}, \citenamefont {Cheng}, \citenamefont {He}, \citenamefont {Wang}, \citenamefont {Wang}, \citenamefont {Zhu}, \citenamefont {Su}, \citenamefont {Zhou}, \citenamefont {Zheng}, \citenamefont {Sun}, \citenamefont {Yang}, \citenamefont {Hauke}, \citenamefont {Zheng}, \citenamefont {Halimeh}, \citenamefont {Yuan},\ and\ \citenamefont {Pan}}]{ZhangPan_natphys2024_1Dlgtexp}%
  \BibitemOpen
  \bibfield  {author} {\bibinfo {author} {\bibfnamefont {W.-Y.}\ \bibnamefont {Zhang}}, \bibinfo {author} {\bibfnamefont {Y.}~\bibnamefont {Liu}}, \bibinfo {author} {\bibfnamefont {Y.}~\bibnamefont {Cheng}}, \bibinfo {author} {\bibfnamefont {M.-G.}\ \bibnamefont {He}}, \bibinfo {author} {\bibfnamefont {H.-Y.}\ \bibnamefont {Wang}}, \bibinfo {author} {\bibfnamefont {T.-Y.}\ \bibnamefont {Wang}}, \bibinfo {author} {\bibfnamefont {Z.-H.}\ \bibnamefont {Zhu}}, \bibinfo {author} {\bibfnamefont {G.-X.}\ \bibnamefont {Su}}, \bibinfo {author} {\bibfnamefont {Z.-Y.}\ \bibnamefont {Zhou}}, \bibinfo {author} {\bibfnamefont {Y.-G.}\ \bibnamefont {Zheng}}, \bibinfo {author} {\bibfnamefont {H.}~\bibnamefont {Sun}}, \bibinfo {author} {\bibfnamefont {B.}~\bibnamefont {Yang}}, \bibinfo {author} {\bibfnamefont {P.}~\bibnamefont {Hauke}}, \bibinfo {author} {\bibfnamefont {W.}~\bibnamefont {Zheng}}, \bibinfo {author} {\bibfnamefont {J.~C.}\ \bibnamefont {Halimeh}}, \bibinfo {author} {\bibfnamefont {Z.-S.}\ \bibnamefont {Yuan}},\
  and\ \bibinfo {author} {\bibfnamefont {J.-W.}\ \bibnamefont {Pan}},\ }\bibfield  {title} {\bibinfo {title} {Observation of microscopic confinement dynamics by a tunable topological $\theta$-angle},\ }\href {https://doi.org/10.1038/s41567-024-02702-x} {\bibfield  {journal} {\bibinfo  {journal} {Nature Physics}\ }\textbf {\bibinfo {volume} {21}},\ \bibinfo {pages} {155–160} (\bibinfo {year} {2024})}\BibitemShut {NoStop}%
\bibitem [{\citenamefont {Mildenberger}\ \emph {et~al.}(2025)\citenamefont {Mildenberger}, \citenamefont {Mruczkiewicz}, \citenamefont {Halimeh}, \citenamefont {Jiang},\ and\ \citenamefont {Hauke}}]{MildenbergerHauke_natphys2025_1Dlgtexp}%
  \BibitemOpen
  \bibfield  {author} {\bibinfo {author} {\bibfnamefont {J.}~\bibnamefont {Mildenberger}}, \bibinfo {author} {\bibfnamefont {W.}~\bibnamefont {Mruczkiewicz}}, \bibinfo {author} {\bibfnamefont {J.~C.}\ \bibnamefont {Halimeh}}, \bibinfo {author} {\bibfnamefont {Z.}~\bibnamefont {Jiang}},\ and\ \bibinfo {author} {\bibfnamefont {P.}~\bibnamefont {Hauke}},\ }\bibfield  {title} {\bibinfo {title} {Confinement in a ${{\mathbb{Z}}}_{2}$ lattice gauge theory on a quantum computer},\ }\href {https://doi.org/10.1038/s41567-024-02723-6} {\bibfield  {journal} {\bibinfo  {journal} {Nature Physics}\ }\textbf {\bibinfo {volume} {21}},\ \bibinfo {pages} {312–317} (\bibinfo {year} {2025})}\BibitemShut {NoStop}%
\bibitem [{\citenamefont {Luo}\ \emph {et~al.}(2025)\citenamefont {Luo}, \citenamefont {Surace}, \citenamefont {De}, \citenamefont {Lerose}, \citenamefont {Bennewitz}, \citenamefont {Ware}, \citenamefont {Schuckert}, \citenamefont {Davoudi}, \citenamefont {Gorshkov}, \citenamefont {Katz},\ and\ \citenamefont {Monroe}}]{luo2025quantumsimulationbubblenucleation}%
  \BibitemOpen
  \bibfield  {author} {\bibinfo {author} {\bibfnamefont {D.}~\bibnamefont {Luo}}, \bibinfo {author} {\bibfnamefont {F.~M.}\ \bibnamefont {Surace}}, \bibinfo {author} {\bibfnamefont {A.}~\bibnamefont {De}}, \bibinfo {author} {\bibfnamefont {A.}~\bibnamefont {Lerose}}, \bibinfo {author} {\bibfnamefont {E.~R.}\ \bibnamefont {Bennewitz}}, \bibinfo {author} {\bibfnamefont {B.}~\bibnamefont {Ware}}, \bibinfo {author} {\bibfnamefont {A.}~\bibnamefont {Schuckert}}, \bibinfo {author} {\bibfnamefont {Z.}~\bibnamefont {Davoudi}}, \bibinfo {author} {\bibfnamefont {A.~V.}\ \bibnamefont {Gorshkov}}, \bibinfo {author} {\bibfnamefont {O.}~\bibnamefont {Katz}},\ and\ \bibinfo {author} {\bibfnamefont {C.}~\bibnamefont {Monroe}},\ }\href {https://arxiv.org/abs/2505.09607} {\bibinfo {title} {Quantum simulation of bubble nucleation across a quantum phase transition}} (\bibinfo {year} {2025}),\ \Eprint {https://arxiv.org/abs/2505.09607} {arXiv:2505.09607 [quant-ph]} \BibitemShut {NoStop}%
\bibitem [{\citenamefont {Liu}\ \emph {et~al.}(2025)\citenamefont {Liu}, \citenamefont {Zhang}, \citenamefont {Zhu}, \citenamefont {He}, \citenamefont {Yuan},\ and\ \citenamefont {Pan}}]{LiuPan_prl2025_stringbreaking}%
  \BibitemOpen
  \bibfield  {author} {\bibinfo {author} {\bibfnamefont {Y.}~\bibnamefont {Liu}}, \bibinfo {author} {\bibfnamefont {W.-Y.}\ \bibnamefont {Zhang}}, \bibinfo {author} {\bibfnamefont {Z.-H.}\ \bibnamefont {Zhu}}, \bibinfo {author} {\bibfnamefont {M.-G.}\ \bibnamefont {He}}, \bibinfo {author} {\bibfnamefont {Z.-S.}\ \bibnamefont {Yuan}},\ and\ \bibinfo {author} {\bibfnamefont {J.-W.}\ \bibnamefont {Pan}},\ }\bibfield  {title} {\bibinfo {title} {String-breaking mechanism in a lattice schwinger model simulator},\ }\href {https://doi.org/10.1103/mwy1-v9hk} {\bibfield  {journal} {\bibinfo  {journal} {Phys. Rev. Lett.}\ }\textbf {\bibinfo {volume} {135}},\ \bibinfo {pages} {101902} (\bibinfo {year} {2025})}\BibitemShut {NoStop}%
\bibitem [{\citenamefont {Tan}\ \emph {et~al.}(2021)\citenamefont {Tan}, \citenamefont {Becker}, \citenamefont {Liu}, \citenamefont {Pagano}, \citenamefont {Collins}, \citenamefont {De}, \citenamefont {Feng}, \citenamefont {Kaplan}, \citenamefont {Kyprianidis}, \citenamefont {Lundgren}, \citenamefont {Morong}, \citenamefont {Whitsitt}, \citenamefont {Gorshkov},\ and\ \citenamefont {Monroe}}]{Tan2021_dwconf}%
  \BibitemOpen
  \bibfield  {author} {\bibinfo {author} {\bibfnamefont {W.~L.}\ \bibnamefont {Tan}}, \bibinfo {author} {\bibfnamefont {P.}~\bibnamefont {Becker}}, \bibinfo {author} {\bibfnamefont {F.}~\bibnamefont {Liu}}, \bibinfo {author} {\bibfnamefont {G.}~\bibnamefont {Pagano}}, \bibinfo {author} {\bibfnamefont {K.~S.}\ \bibnamefont {Collins}}, \bibinfo {author} {\bibfnamefont {A.}~\bibnamefont {De}}, \bibinfo {author} {\bibfnamefont {L.}~\bibnamefont {Feng}}, \bibinfo {author} {\bibfnamefont {H.~B.}\ \bibnamefont {Kaplan}}, \bibinfo {author} {\bibfnamefont {A.}~\bibnamefont {Kyprianidis}}, \bibinfo {author} {\bibfnamefont {R.}~\bibnamefont {Lundgren}}, \bibinfo {author} {\bibfnamefont {W.}~\bibnamefont {Morong}}, \bibinfo {author} {\bibfnamefont {S.}~\bibnamefont {Whitsitt}}, \bibinfo {author} {\bibfnamefont {A.~V.}\ \bibnamefont {Gorshkov}},\ and\ \bibinfo {author} {\bibfnamefont {C.}~\bibnamefont {Monroe}},\ }\bibfield  {title} {\bibinfo {title} {Domain-wall confinement and dynamics in a quantum simulator},\ }\href
  {https://doi.org/10.1038/s41567-021-01194-3} {\bibfield  {journal} {\bibinfo  {journal} {Nature Physics}\ }\textbf {\bibinfo {volume} {17}},\ \bibinfo {pages} {742–747} (\bibinfo {year} {2021})}\BibitemShut {NoStop}%
\bibitem [{\citenamefont {Buyens}\ \emph {et~al.}(2017)\citenamefont {Buyens}, \citenamefont {Haegeman}, \citenamefont {Hebenstreit}, \citenamefont {Verstraete},\ and\ \citenamefont {Van~Acoleyen}}]{buyens2017real}%
  \BibitemOpen
  \bibfield  {author} {\bibinfo {author} {\bibfnamefont {B.}~\bibnamefont {Buyens}}, \bibinfo {author} {\bibfnamefont {J.}~\bibnamefont {Haegeman}}, \bibinfo {author} {\bibfnamefont {F.}~\bibnamefont {Hebenstreit}}, \bibinfo {author} {\bibfnamefont {F.}~\bibnamefont {Verstraete}},\ and\ \bibinfo {author} {\bibfnamefont {K.}~\bibnamefont {Van~Acoleyen}},\ }\bibfield  {title} {\bibinfo {title} {Real-time simulation of the schwinger effect with matrix product states},\ }\href {http://dx.doi.org/10.1103/PhysRevD.96.114501} {\bibfield  {journal} {\bibinfo  {journal} {Physical Review D}\ }\textbf {\bibinfo {volume} {96}} (\bibinfo {year} {2017})}\BibitemShut {NoStop}%
\bibitem [{\citenamefont {Gupta}\ \emph {et~al.}(2026)\citenamefont {Gupta}, \citenamefont {Mathew}, \citenamefont {Kadam}, \citenamefont {Stryker}, \citenamefont {Bapat}, \citenamefont {Mueller}, \citenamefont {Davoudi},\ and\ \citenamefont {Raychowdhury}}]{gupta2026stringbreakingstaticsdynamics11d}%
  \BibitemOpen
  \bibfield  {author} {\bibinfo {author} {\bibfnamefont {N.}~\bibnamefont {Gupta}}, \bibinfo {author} {\bibfnamefont {E.}~\bibnamefont {Mathew}}, \bibinfo {author} {\bibfnamefont {S.~V.}\ \bibnamefont {Kadam}}, \bibinfo {author} {\bibfnamefont {J.~R.}\ \bibnamefont {Stryker}}, \bibinfo {author} {\bibfnamefont {A.}~\bibnamefont {Bapat}}, \bibinfo {author} {\bibfnamefont {N.}~\bibnamefont {Mueller}}, \bibinfo {author} {\bibfnamefont {Z.}~\bibnamefont {Davoudi}},\ and\ \bibinfo {author} {\bibfnamefont {I.}~\bibnamefont {Raychowdhury}},\ }\href {https://arxiv.org/abs/2603.24698} {\bibinfo {title} {String-breaking statics and dynamics in a (1+1)d su(2) lattice gauge theory}} (\bibinfo {year} {2026}),\ \Eprint {https://arxiv.org/abs/2603.24698} {arXiv:2603.24698 [hep-lat]} \BibitemShut {NoStop}%
\bibitem [{\citenamefont {Pichler}\ \emph {et~al.}(2016)\citenamefont {Pichler}, \citenamefont {Dalmonte}, \citenamefont {Rico}, \citenamefont {Zoller},\ and\ \citenamefont {Montangero}}]{pichlermontangero_prx2016_U1TNdynamics}%
  \BibitemOpen
  \bibfield  {author} {\bibinfo {author} {\bibfnamefont {T.}~\bibnamefont {Pichler}}, \bibinfo {author} {\bibfnamefont {M.}~\bibnamefont {Dalmonte}}, \bibinfo {author} {\bibfnamefont {E.}~\bibnamefont {Rico}}, \bibinfo {author} {\bibfnamefont {P.}~\bibnamefont {Zoller}},\ and\ \bibinfo {author} {\bibfnamefont {S.}~\bibnamefont {Montangero}},\ }\bibfield  {title} {\bibinfo {title} {Real-time dynamics in u(1) lattice gauge theories with tensor networks},\ }\href {https://doi.org/10.1103/PhysRevX.6.011023} {\bibfield  {journal} {\bibinfo  {journal} {Phys. Rev. X}\ }\textbf {\bibinfo {volume} {6}},\ \bibinfo {pages} {011023} (\bibinfo {year} {2016})}\BibitemShut {NoStop}%
\bibitem [{\citenamefont {Verdel}\ \emph {et~al.}(2020)\citenamefont {Verdel}, \citenamefont {Liu}, \citenamefont {Whitsitt}, \citenamefont {Gorshkov},\ and\ \citenamefont {Heyl}}]{VerdelHeyl_prb2020_stringbreaking}%
  \BibitemOpen
  \bibfield  {author} {\bibinfo {author} {\bibfnamefont {R.}~\bibnamefont {Verdel}}, \bibinfo {author} {\bibfnamefont {F.}~\bibnamefont {Liu}}, \bibinfo {author} {\bibfnamefont {S.}~\bibnamefont {Whitsitt}}, \bibinfo {author} {\bibfnamefont {A.~V.}\ \bibnamefont {Gorshkov}},\ and\ \bibinfo {author} {\bibfnamefont {M.}~\bibnamefont {Heyl}},\ }\bibfield  {title} {\bibinfo {title} {Real-time dynamics of string breaking in quantum spin chains},\ }\href {https://doi.org/10.1103/PhysRevB.102.014308} {\bibfield  {journal} {\bibinfo  {journal} {Phys. Rev. B}\ }\textbf {\bibinfo {volume} {102}},\ \bibinfo {pages} {014308} (\bibinfo {year} {2020})}\BibitemShut {NoStop}%
\bibitem [{\citenamefont {Surace}\ \emph {et~al.}(2026)\citenamefont {Surace}, \citenamefont {Lerose}, \citenamefont {Katz}, \citenamefont {Bennewitz}, \citenamefont {Schuckert}, \citenamefont {Luo}, \citenamefont {De}, \citenamefont {Ware}, \citenamefont {Morong}, \citenamefont {Collins}, \citenamefont {Monroe}, \citenamefont {Davoudi},\ and\ \citenamefont {Gorshkov}}]{surace_prxquantum2026_stringbreaking1d}%
  \BibitemOpen
  \bibfield  {author} {\bibinfo {author} {\bibfnamefont {F.~M.}\ \bibnamefont {Surace}}, \bibinfo {author} {\bibfnamefont {A.}~\bibnamefont {Lerose}}, \bibinfo {author} {\bibfnamefont {O.}~\bibnamefont {Katz}}, \bibinfo {author} {\bibfnamefont {E.~R.}\ \bibnamefont {Bennewitz}}, \bibinfo {author} {\bibfnamefont {A.}~\bibnamefont {Schuckert}}, \bibinfo {author} {\bibfnamefont {D.}~\bibnamefont {Luo}}, \bibinfo {author} {\bibfnamefont {A.}~\bibnamefont {De}}, \bibinfo {author} {\bibfnamefont {B.}~\bibnamefont {Ware}}, \bibinfo {author} {\bibfnamefont {W.}~\bibnamefont {Morong}}, \bibinfo {author} {\bibfnamefont {K.}~\bibnamefont {Collins}}, \bibinfo {author} {\bibfnamefont {C.}~\bibnamefont {Monroe}}, \bibinfo {author} {\bibfnamefont {Z.}~\bibnamefont {Davoudi}},\ and\ \bibinfo {author} {\bibfnamefont {A.~V.}\ \bibnamefont {Gorshkov}},\ }\bibfield  {title} {\bibinfo {title} {String-breaking dynamics in quantum adiabatic and diabatic processes},\ }\href {https://doi.org/10.1103/c4zd-lbyq} {\bibfield  {journal}
  {\bibinfo  {journal} {PRX Quantum}\ }\textbf {\bibinfo {volume} {7}},\ \bibinfo {pages} {020331} (\bibinfo {year} {2026})}\BibitemShut {NoStop}%
\bibitem [{\citenamefont {Shirali}\ \emph {et~al.}(2025)\citenamefont {Shirali}, \citenamefont {Sherbert}, \citenamefont {Chen}, \citenamefont {Florio}, \citenamefont {Weichselbaum}, \citenamefont {Pisarski},\ and\ \citenamefont {Economou}}]{shirali2025break}%
  \BibitemOpen
  \bibfield  {author} {\bibinfo {author} {\bibfnamefont {K.~S.}\ \bibnamefont {Shirali}}, \bibinfo {author} {\bibfnamefont {K.}~\bibnamefont {Sherbert}}, \bibinfo {author} {\bibfnamefont {Y.}~\bibnamefont {Chen}}, \bibinfo {author} {\bibfnamefont {A.}~\bibnamefont {Florio}}, \bibinfo {author} {\bibfnamefont {A.}~\bibnamefont {Weichselbaum}}, \bibinfo {author} {\bibfnamefont {R.~D.}\ \bibnamefont {Pisarski}},\ and\ \bibinfo {author} {\bibfnamefont {S.~E.}\ \bibnamefont {Economou}},\ }\href {https://arxiv.org/abs/2510.03083} {\bibinfo {title} {To break, or not to break: Symmetries in adaptive quantum simulations, a case study on the schwinger model}} (\bibinfo {year} {2025}),\ \Eprint {https://arxiv.org/abs/2510.03083} {arXiv:2510.03083 [quant-ph]} \BibitemShut {NoStop}%
\bibitem [{\citenamefont {Florio}\ \emph {et~al.}(2024)\citenamefont {Florio}, \citenamefont {Frenklakh}, \citenamefont {Ikeda}, \citenamefont {Kharzeev}, \citenamefont {Korepin}, \citenamefont {Shi},\ and\ \citenamefont {Yu}}]{florio2024quantum}%
  \BibitemOpen
  \bibfield  {author} {\bibinfo {author} {\bibfnamefont {A.}~\bibnamefont {Florio}}, \bibinfo {author} {\bibfnamefont {D.}~\bibnamefont {Frenklakh}}, \bibinfo {author} {\bibfnamefont {K.}~\bibnamefont {Ikeda}}, \bibinfo {author} {\bibfnamefont {D.}~\bibnamefont {Kharzeev}}, \bibinfo {author} {\bibfnamefont {V.}~\bibnamefont {Korepin}}, \bibinfo {author} {\bibfnamefont {S.}~\bibnamefont {Shi}},\ and\ \bibinfo {author} {\bibfnamefont {K.}~\bibnamefont {Yu}},\ }\bibfield  {title} {\bibinfo {title} {Quantum real-time evolution of entanglement and hadronization in jet production: Lessons from the massive schwinger model},\ }\href {http://dx.doi.org/10.1103/PhysRevD.110.094029} {\bibfield  {journal} {\bibinfo  {journal} {Physical Review D}\ }\textbf {\bibinfo {volume} {110}} (\bibinfo {year} {2024})}\BibitemShut {NoStop}%
\bibitem [{\citenamefont {Florio}\ \emph {et~al.}(2023)\citenamefont {Florio}, \citenamefont {Frenklakh}, \citenamefont {Ikeda}, \citenamefont {Kharzeev}, \citenamefont {Korepin}, \citenamefont {Shi},\ and\ \citenamefont {Yu}}]{Florio_prl2024_jetproduction}%
  \BibitemOpen
  \bibfield  {author} {\bibinfo {author} {\bibfnamefont {A.}~\bibnamefont {Florio}}, \bibinfo {author} {\bibfnamefont {D.}~\bibnamefont {Frenklakh}}, \bibinfo {author} {\bibfnamefont {K.}~\bibnamefont {Ikeda}}, \bibinfo {author} {\bibfnamefont {D.}~\bibnamefont {Kharzeev}}, \bibinfo {author} {\bibfnamefont {V.}~\bibnamefont {Korepin}}, \bibinfo {author} {\bibfnamefont {S.}~\bibnamefont {Shi}},\ and\ \bibinfo {author} {\bibfnamefont {K.}~\bibnamefont {Yu}},\ }\bibfield  {title} {\bibinfo {title} {Real-time nonperturbative dynamics of jet production in schwinger model: Quantum entanglement and vacuum modification},\ }\href {https://doi.org/10.1103/PhysRevLett.131.021902} {\bibfield  {journal} {\bibinfo  {journal} {Phys. Rev. Lett.}\ }\textbf {\bibinfo {volume} {131}},\ \bibinfo {pages} {021902} (\bibinfo {year} {2023})}\BibitemShut {NoStop}%
\bibitem [{\citenamefont {Grieninger}\ \emph {et~al.}(2026)\citenamefont {Grieninger}, \citenamefont {Savage},\ and\ \citenamefont {Zemlevskiy}}]{grieninger2026quantumcomplexitystringbreaking}%
  \BibitemOpen
  \bibfield  {author} {\bibinfo {author} {\bibfnamefont {S.}~\bibnamefont {Grieninger}}, \bibinfo {author} {\bibfnamefont {M.~J.}\ \bibnamefont {Savage}},\ and\ \bibinfo {author} {\bibfnamefont {N.~A.}\ \bibnamefont {Zemlevskiy}},\ }\href {https://arxiv.org/abs/2601.08825} {\bibinfo {title} {The quantum complexity of string breaking in the schwinger model}} (\bibinfo {year} {2026}),\ \Eprint {https://arxiv.org/abs/2601.08825} {arXiv:2601.08825 [hep-ph]} \BibitemShut {NoStop}%
\bibitem [{\citenamefont {Magnifico}\ \emph {et~al.}(2020)\citenamefont {Magnifico}, \citenamefont {Dalmonte}, \citenamefont {Facchi}, \citenamefont {Pascazio}, \citenamefont {Pepe},\ and\ \citenamefont {Ercolessi}}]{magnifico2020real}%
  \BibitemOpen
  \bibfield  {author} {\bibinfo {author} {\bibfnamefont {G.}~\bibnamefont {Magnifico}}, \bibinfo {author} {\bibfnamefont {M.}~\bibnamefont {Dalmonte}}, \bibinfo {author} {\bibfnamefont {P.}~\bibnamefont {Facchi}}, \bibinfo {author} {\bibfnamefont {S.}~\bibnamefont {Pascazio}}, \bibinfo {author} {\bibfnamefont {F.~V.}\ \bibnamefont {Pepe}},\ and\ \bibinfo {author} {\bibfnamefont {E.}~\bibnamefont {Ercolessi}},\ }\href@noop {} {\bibfield  {journal} {\bibinfo  {journal} {Quantum}\ }\textbf {\bibinfo {volume} {4}},\ \bibinfo {pages} {281} (\bibinfo {year} {2020})}\BibitemShut {NoStop}%
\bibitem [{\citenamefont {Xu}\ \emph {et~al.}(2026)\citenamefont {Xu}, \citenamefont {Borla}, \citenamefont {Hemery}, \citenamefont {Joshi}, \citenamefont {Dreyer}, \citenamefont {Rinaldi},\ and\ \citenamefont {Halimeh}}]{xuHalimeh_2026observationglueballexcitationsstring}%
  \BibitemOpen
  \bibfield  {author} {\bibinfo {author} {\bibfnamefont {K.}~\bibnamefont {Xu}}, \bibinfo {author} {\bibfnamefont {U.}~\bibnamefont {Borla}}, \bibinfo {author} {\bibfnamefont {K.}~\bibnamefont {Hemery}}, \bibinfo {author} {\bibfnamefont {R.}~\bibnamefont {Joshi}}, \bibinfo {author} {\bibfnamefont {H.}~\bibnamefont {Dreyer}}, \bibinfo {author} {\bibfnamefont {E.}~\bibnamefont {Rinaldi}},\ and\ \bibinfo {author} {\bibfnamefont {J.~C.}\ \bibnamefont {Halimeh}},\ }\href {https://arxiv.org/abs/2604.07435} {\bibinfo {title} {Observation of glueball excitations and string breaking in a $2+1$d $\mathbb{Z}_2$ lattice gauge theory on a trapped-ion quantum computer}} (\bibinfo {year} {2026}),\ \Eprint {https://arxiv.org/abs/2604.07435} {arXiv:2604.07435 [hep-lat]} \BibitemShut {NoStop}%
\bibitem [{\citenamefont {Xu}\ \emph {et~al.}(2025{\natexlab{a}})\citenamefont {Xu}, \citenamefont {Knap},\ and\ \citenamefont {Pollmann}}]{Pollman2025roughening}%
  \BibitemOpen
  \bibfield  {author} {\bibinfo {author} {\bibfnamefont {W.-T.}\ \bibnamefont {Xu}}, \bibinfo {author} {\bibfnamefont {M.}~\bibnamefont {Knap}},\ and\ \bibinfo {author} {\bibfnamefont {F.}~\bibnamefont {Pollmann}},\ }\bibfield  {title} {\bibinfo {title} {Tensor-network study of the roughening transition in a $(2+1)\mathrm{D}$ ${\mathbb{z}}_{2}$ lattice gauge theory with matter},\ }\href {https://doi.org/10.1103/3lxj-dx76} {\bibfield  {journal} {\bibinfo  {journal} {Phys. Rev. Lett.}\ }\textbf {\bibinfo {volume} {135}},\ \bibinfo {pages} {036503} (\bibinfo {year} {2025}{\natexlab{a}})}\BibitemShut {NoStop}%
\bibitem [{\citenamefont {Ueda}\ \emph {et~al.}(2026)\citenamefont {Ueda}, \citenamefont {Burgelman}, \citenamefont {Tagliacozzo},\ and\ \citenamefont {Vanderstraeten}}]{uedatagliacozzo_2026interfaceroughening3dising}%
  \BibitemOpen
  \bibfield  {author} {\bibinfo {author} {\bibfnamefont {A.}~\bibnamefont {Ueda}}, \bibinfo {author} {\bibfnamefont {L.}~\bibnamefont {Burgelman}}, \bibinfo {author} {\bibfnamefont {L.}~\bibnamefont {Tagliacozzo}},\ and\ \bibinfo {author} {\bibfnamefont {L.}~\bibnamefont {Vanderstraeten}},\ }\href {https://arxiv.org/abs/2601.07829} {\bibinfo {title} {Interface roughening in the 3-d ising model with tensor networks}} (\bibinfo {year} {2026}),\ \Eprint {https://arxiv.org/abs/2601.07829} {arXiv:2601.07829 [cond-mat.str-el]} \BibitemShut {NoStop}%
\bibitem [{\citenamefont {Negro}\ \emph {et~al.}(2026)\citenamefont {Negro}, \citenamefont {Rosanowski}, \citenamefont {Funcke}, \citenamefont {Jakobs}, \citenamefont {Jansen}, \citenamefont {Ludwig},\ and\ \citenamefont {Urbach}}]{negro2026finitevolumeschemecontinuumextrapolation}%
  \BibitemOpen
  \bibfield  {author} {\bibinfo {author} {\bibfnamefont {A.}~\bibnamefont {Negro}}, \bibinfo {author} {\bibfnamefont {E.~O.}\ \bibnamefont {Rosanowski}}, \bibinfo {author} {\bibfnamefont {L.}~\bibnamefont {Funcke}}, \bibinfo {author} {\bibfnamefont {T.}~\bibnamefont {Jakobs}}, \bibinfo {author} {\bibfnamefont {K.}~\bibnamefont {Jansen}}, \bibinfo {author} {\bibfnamefont {P.}~\bibnamefont {Ludwig}},\ and\ \bibinfo {author} {\bibfnamefont {C.}~\bibnamefont {Urbach}},\ }\href {https://arxiv.org/abs/2606.20029} {\bibinfo {title} {A finite-volume scheme for the continuum extrapolation of lattice step-scaling in (2+1)d hamiltonian u(1) gauge theory}} (\bibinfo {year} {2026}),\ \Eprint {https://arxiv.org/abs/2606.20029} {arXiv:2606.20029 [hep-lat]} \BibitemShut {NoStop}%
\bibitem [{\citenamefont {Wu}\ and\ \citenamefont {Liu}(2025)}]{wu2025_pepsdynamics_purelgt}%
  \BibitemOpen
  \bibfield  {author} {\bibinfo {author} {\bibfnamefont {Y.}~\bibnamefont {Wu}}\ and\ \bibinfo {author} {\bibfnamefont {W.-Y.}\ \bibnamefont {Liu}},\ }\href {https://arxiv.org/abs/2503.20566} {\bibinfo {title} {Accurate gauge-invariant tensor network simulations for abelian lattice gauge theory in (2+1)d: ground state and real-time dynamics}} (\bibinfo {year} {2025}),\ \Eprint {https://arxiv.org/abs/2503.20566} {arXiv:2503.20566 [cond-mat.str-el]} \BibitemShut {NoStop}%
\bibitem [{\citenamefont {Di~Marcantonio}\ \emph {et~al.}(2026)\citenamefont {Di~Marcantonio}, \citenamefont {Pradhan}, \citenamefont {Vallecorsa}, \citenamefont {Bañuls},\ and\ \citenamefont {Rico~Ortega}}]{DiMarcantonioRico_2026commphys_roughening}%
  \BibitemOpen
  \bibfield  {author} {\bibinfo {author} {\bibfnamefont {F.}~\bibnamefont {Di~Marcantonio}}, \bibinfo {author} {\bibfnamefont {S.}~\bibnamefont {Pradhan}}, \bibinfo {author} {\bibfnamefont {S.}~\bibnamefont {Vallecorsa}}, \bibinfo {author} {\bibfnamefont {M.~C.}\ \bibnamefont {Bañuls}},\ and\ \bibinfo {author} {\bibfnamefont {E.}~\bibnamefont {Rico~Ortega}},\ }\bibfield  {title} {\bibinfo {title} {Roughening and dynamics of an electric flux string in a (2+1)d lattice gauge theory},\ }\href {http://dx.doi.org/10.1038/s42005-026-02659-8} {\bibfield  {journal} {\bibinfo  {journal} {Communications Physics}\ }\textbf {\bibinfo {volume} {9}} (\bibinfo {year} {2026})}\BibitemShut {NoStop}%
\bibitem [{\citenamefont {Bombieri}\ \emph {et~al.}(2026)\citenamefont {Bombieri}, \citenamefont {Zache}, \citenamefont {Pichler},\ and\ \citenamefont {González-Cuadra}}]{bombieri2026u1latticegaugetheory}%
  \BibitemOpen
  \bibfield  {author} {\bibinfo {author} {\bibfnamefont {L.}~\bibnamefont {Bombieri}}, \bibinfo {author} {\bibfnamefont {T.~V.}\ \bibnamefont {Zache}}, \bibinfo {author} {\bibfnamefont {H.}~\bibnamefont {Pichler}},\ and\ \bibinfo {author} {\bibfnamefont {D.}~\bibnamefont {González-Cuadra}},\ }\href {https://arxiv.org/abs/2602.06123} {\bibinfo {title} {U(1) lattice gauge theory and string roughening on a triangular rydberg array}} (\bibinfo {year} {2026}),\ \Eprint {https://arxiv.org/abs/2602.06123} {arXiv:2602.06123 [quant-ph]} \BibitemShut {NoStop}%
\bibitem [{\citenamefont {Krinitsin}\ \emph {et~al.}(2025)\citenamefont {Krinitsin}, \citenamefont {Tausendpfund}, \citenamefont {Rizzi}, \citenamefont {Heyl},\ and\ \citenamefont {Schmitt}}]{KrinitsinSchmitt_prl2025_rougheningising}%
  \BibitemOpen
  \bibfield  {author} {\bibinfo {author} {\bibfnamefont {W.}~\bibnamefont {Krinitsin}}, \bibinfo {author} {\bibfnamefont {N.}~\bibnamefont {Tausendpfund}}, \bibinfo {author} {\bibfnamefont {M.}~\bibnamefont {Rizzi}}, \bibinfo {author} {\bibfnamefont {M.}~\bibnamefont {Heyl}},\ and\ \bibinfo {author} {\bibfnamefont {M.}~\bibnamefont {Schmitt}},\ }\bibfield  {title} {\bibinfo {title} {Roughening dynamics of interfaces in the two-dimensional quantum ising model},\ }\href {https://doi.org/10.1103/9bsk-x9rw} {\bibfield  {journal} {\bibinfo  {journal} {Phys. Rev. Lett.}\ }\textbf {\bibinfo {volume} {134}},\ \bibinfo {pages} {240402} (\bibinfo {year} {2025})}\BibitemShut {NoStop}%
\bibitem [{\citenamefont {Xu}\ \emph {et~al.}(2025{\natexlab{b}})\citenamefont {Xu}, \citenamefont {Borla}, \citenamefont {Moroz},\ and\ \citenamefont {Halimeh}}]{xu2025string}%
  \BibitemOpen
  \bibfield  {author} {\bibinfo {author} {\bibfnamefont {K.}~\bibnamefont {Xu}}, \bibinfo {author} {\bibfnamefont {U.}~\bibnamefont {Borla}}, \bibinfo {author} {\bibfnamefont {S.}~\bibnamefont {Moroz}},\ and\ \bibinfo {author} {\bibfnamefont {J.~C.}\ \bibnamefont {Halimeh}},\ }\bibfield  {title} {\bibinfo {title} {String breaking dynamics and glueball formation in a $2+ 1$ d lattice gauge theory},\ }\href@noop {} {\bibfield  {journal} {\bibinfo  {journal} {arXiv preprint arXiv:2507.01950}\ } (\bibinfo {year} {2025}{\natexlab{b}})}\BibitemShut {NoStop}%
\bibitem [{\citenamefont {White}(1993)}]{white1993density}%
  \BibitemOpen
  \bibfield  {author} {\bibinfo {author} {\bibfnamefont {S.~R.}\ \bibnamefont {White}},\ }\bibfield  {title} {\bibinfo {title} {Density-matrix algorithms for quantum renormalization groups},\ }\href {https://doi.org/10.1103/PhysRevB.48.10345} {\bibfield  {journal} {\bibinfo  {journal} {Physical review b}\ }\textbf {\bibinfo {volume} {48}},\ \bibinfo {pages} {10345} (\bibinfo {year} {1993})}\BibitemShut {NoStop}%
\bibitem [{\citenamefont {White}(1992)}]{white1992_prl}%
  \BibitemOpen
  \bibfield  {author} {\bibinfo {author} {\bibfnamefont {S.~R.}\ \bibnamefont {White}},\ }\bibfield  {title} {\bibinfo {title} {Density matrix formulation for quantum renormalization groups},\ }\href {https://doi.org/10.1103/PhysRevLett.69.2863} {\bibfield  {journal} {\bibinfo  {journal} {Phys. Rev. Lett.}\ }\textbf {\bibinfo {volume} {69}},\ \bibinfo {pages} {2863} (\bibinfo {year} {1992})}\BibitemShut {NoStop}%
\bibitem [{\citenamefont {Gottesman}(1997)}]{gottesman1997stabilizer}%
  \BibitemOpen
  \bibfield  {author} {\bibinfo {author} {\bibfnamefont {D.}~\bibnamefont {Gottesman}},\ }\href@noop {} {\emph {\bibinfo {title} {Stabilizer codes and quantum error correction}}}\ (\bibinfo  {publisher} {California Institute of Technology},\ \bibinfo {year} {1997})\BibitemShut {NoStop}%
\bibitem [{\citenamefont {Qian}\ \emph {et~al.}(2024{\natexlab{a}})\citenamefont {Qian}, \citenamefont {Huang},\ and\ \citenamefont {Qin}}]{qian2024augmenting}%
  \BibitemOpen
  \bibfield  {author} {\bibinfo {author} {\bibfnamefont {X.}~\bibnamefont {Qian}}, \bibinfo {author} {\bibfnamefont {J.}~\bibnamefont {Huang}},\ and\ \bibinfo {author} {\bibfnamefont {M.}~\bibnamefont {Qin}},\ }\bibfield  {title} {\bibinfo {title} {Augmenting density matrix renormalization group with clifford circuits},\ }\href {https://doi.org/10.1103/PhysRevLett.133.190402} {\bibfield  {journal} {\bibinfo  {journal} {Phys. Rev. Lett.}\ }\textbf {\bibinfo {volume} {133}},\ \bibinfo {pages} {190402} (\bibinfo {year} {2024}{\natexlab{a}})}\BibitemShut {NoStop}%
\bibitem [{\citenamefont {Qian}\ \emph {et~al.}(2024{\natexlab{b}})\citenamefont {Qian}, \citenamefont {Huang},\ and\ \citenamefont {Qin}}]{qian2024clifford}%
  \BibitemOpen
  \bibfield  {author} {\bibinfo {author} {\bibfnamefont {X.}~\bibnamefont {Qian}}, \bibinfo {author} {\bibfnamefont {J.}~\bibnamefont {Huang}},\ and\ \bibinfo {author} {\bibfnamefont {M.}~\bibnamefont {Qin}},\ }\href {https://arxiv.org/abs/2407.03202} {\bibinfo {title} {Clifford circuits augmented time-dependent variational principle}} (\bibinfo {year} {2024}{\natexlab{b}}),\ \Eprint {https://arxiv.org/abs/2407.03202} {arXiv:2407.03202 [cond-mat.str-el]} \BibitemShut {NoStop}%
\bibitem [{\citenamefont {Mello}\ \emph {et~al.}(2025)\citenamefont {Mello}, \citenamefont {Santini}, \citenamefont {Lami}, \citenamefont {De~Nardis},\ and\ \citenamefont {Collura}}]{mello2024clifford}%
  \BibitemOpen
  \bibfield  {author} {\bibinfo {author} {\bibfnamefont {A.~F.}\ \bibnamefont {Mello}}, \bibinfo {author} {\bibfnamefont {A.}~\bibnamefont {Santini}}, \bibinfo {author} {\bibfnamefont {G.}~\bibnamefont {Lami}}, \bibinfo {author} {\bibfnamefont {J.}~\bibnamefont {De~Nardis}},\ and\ \bibinfo {author} {\bibfnamefont {M.}~\bibnamefont {Collura}},\ }\bibfield  {title} {\bibinfo {title} {Clifford dressed time-dependent variational principle},\ }\href {http://dx.doi.org/10.1103/PhysRevLett.134.150403} {\bibfield  {journal} {\bibinfo  {journal} {Physical Review Letters}\ }\textbf {\bibinfo {volume} {134}} (\bibinfo {year} {2025})}\BibitemShut {NoStop}%
\bibitem [{\citenamefont {Masot-Llima}\ and\ \citenamefont {Garcia-Saez}(2024)}]{masotllima2024}%
  \BibitemOpen
  \bibfield  {author} {\bibinfo {author} {\bibfnamefont {S.}~\bibnamefont {Masot-Llima}}\ and\ \bibinfo {author} {\bibfnamefont {A.}~\bibnamefont {Garcia-Saez}},\ }\bibfield  {title} {\bibinfo {title} {Stabilizer tensor networks: Universal quantum simulator on a basis of stabilizer states},\ }\href {https://doi.org/10.1103/PhysRevLett.133.230601} {\bibfield  {journal} {\bibinfo  {journal} {Phys. Rev. Lett.}\ }\textbf {\bibinfo {volume} {133}},\ \bibinfo {pages} {230601} (\bibinfo {year} {2024})}\BibitemShut {NoStop}%
\bibitem [{\citenamefont {Nakhl}\ \emph {et~al.}(2025)\citenamefont {Nakhl}, \citenamefont {Harper}, \citenamefont {West}, \citenamefont {Dowling}, \citenamefont {Sevior}, \citenamefont {Quella},\ and\ \citenamefont {Usman}}]{AzarQuella_prl2025_stabilizer}%
  \BibitemOpen
  \bibfield  {author} {\bibinfo {author} {\bibfnamefont {A.~C.}\ \bibnamefont {Nakhl}}, \bibinfo {author} {\bibfnamefont {B.}~\bibnamefont {Harper}}, \bibinfo {author} {\bibfnamefont {M.}~\bibnamefont {West}}, \bibinfo {author} {\bibfnamefont {N.}~\bibnamefont {Dowling}}, \bibinfo {author} {\bibfnamefont {M.}~\bibnamefont {Sevior}}, \bibinfo {author} {\bibfnamefont {T.}~\bibnamefont {Quella}},\ and\ \bibinfo {author} {\bibfnamefont {M.}~\bibnamefont {Usman}},\ }\bibfield  {title} {\bibinfo {title} {Stabilizer tensor networks with magic state injection},\ }\href {https://doi.org/10.1103/PhysRevLett.134.190602} {\bibfield  {journal} {\bibinfo  {journal} {Phys. Rev. Lett.}\ }\textbf {\bibinfo {volume} {134}},\ \bibinfo {pages} {190602} (\bibinfo {year} {2025})}\BibitemShut {NoStop}%
\bibitem [{\citenamefont {Haegeman}\ \emph {et~al.}(2011)\citenamefont {Haegeman}, \citenamefont {Cirac}, \citenamefont {Osborne}, \citenamefont {Pi\ifmmode~\check{z}\else \v{z}\fi{}orn}, \citenamefont {Verschelde},\ and\ \citenamefont {Verstraete}}]{Haegeman_prl2011_tdvp}%
  \BibitemOpen
  \bibfield  {author} {\bibinfo {author} {\bibfnamefont {J.}~\bibnamefont {Haegeman}}, \bibinfo {author} {\bibfnamefont {J.~I.}\ \bibnamefont {Cirac}}, \bibinfo {author} {\bibfnamefont {T.~J.}\ \bibnamefont {Osborne}}, \bibinfo {author} {\bibfnamefont {I.}~\bibnamefont {Pi\ifmmode~\check{z}\else \v{z}\fi{}orn}}, \bibinfo {author} {\bibfnamefont {H.}~\bibnamefont {Verschelde}},\ and\ \bibinfo {author} {\bibfnamefont {F.}~\bibnamefont {Verstraete}},\ }\bibfield  {title} {\bibinfo {title} {Time-dependent variational principle for quantum lattices},\ }\href {https://doi.org/10.1103/PhysRevLett.107.070601} {\bibfield  {journal} {\bibinfo  {journal} {Phys. Rev. Lett.}\ }\textbf {\bibinfo {volume} {107}},\ \bibinfo {pages} {070601} (\bibinfo {year} {2011})}\BibitemShut {NoStop}%
\bibitem [{\citenamefont {L{\"u}scher}(1981)}]{Luscher:1980ac}%
  \BibitemOpen
  \bibfield  {author} {\bibinfo {author} {\bibfnamefont {M.}~\bibnamefont {L{\"u}scher}},\ }\bibfield  {title} {\bibinfo {title} {{Symmetry Breaking Aspects of the Roughening Transition in Gauge Theories}},\ }\href {https://doi.org/10.1016/0550-3213(81)90423-5} {\bibfield  {journal} {\bibinfo  {journal} {Nucl. Phys.}\ }\textbf {\bibinfo {volume} {B180}},\ \bibinfo {pages} {317} (\bibinfo {year} {1981})}\BibitemShut {NoStop}%
\bibitem [{\citenamefont {L{\"u}scher}\ \emph {et~al.}(1980)\citenamefont {L{\"u}scher}, \citenamefont {Symanzik},\ and\ \citenamefont {Weisz}}]{Luscher:1980fr}%
  \BibitemOpen
  \bibfield  {author} {\bibinfo {author} {\bibfnamefont {M.}~\bibnamefont {L{\"u}scher}}, \bibinfo {author} {\bibfnamefont {K.}~\bibnamefont {Symanzik}},\ and\ \bibinfo {author} {\bibfnamefont {P.}~\bibnamefont {Weisz}},\ }\bibfield  {title} {\bibinfo {title} {{Anomalies of the Free Loop Wave Equation in the WKB Approximation}},\ }\href {https://doi.org/10.1016/0550-3213(80)90009-7} {\bibfield  {journal} {\bibinfo  {journal} {Nucl. Phys.}\ }\textbf {\bibinfo {volume} {B173}},\ \bibinfo {pages} {365} (\bibinfo {year} {1980})}\BibitemShut {NoStop}%
\bibitem [{\citenamefont {Tupitsyn}\ \emph {et~al.}(2010)\citenamefont {Tupitsyn}, \citenamefont {Kitaev}, \citenamefont {Prokof'ev},\ and\ \citenamefont {Stamp}}]{TupitsynKitaevProkofevStamp2010}%
  \BibitemOpen
  \bibfield  {author} {\bibinfo {author} {\bibfnamefont {I.~S.}\ \bibnamefont {Tupitsyn}}, \bibinfo {author} {\bibfnamefont {A.}~\bibnamefont {Kitaev}}, \bibinfo {author} {\bibfnamefont {N.~V.}\ \bibnamefont {Prokof'ev}},\ and\ \bibinfo {author} {\bibfnamefont {P.~C.~E.}\ \bibnamefont {Stamp}},\ }\bibfield  {title} {\bibinfo {title} {Topological multicritical point in the phase diagram of the toric code model and three-dimensional lattice gauge higgs model},\ }\href {https://doi.org/10.1103/PhysRevB.82.085114} {\bibfield  {journal} {\bibinfo  {journal} {Physical Review B}\ }\textbf {\bibinfo {volume} {82}},\ \bibinfo {pages} {085114} (\bibinfo {year} {2010})}\BibitemShut {NoStop}%
\bibitem [{\citenamefont {Wu}\ \emph {et~al.}(2012)\citenamefont {Wu}, \citenamefont {Deng},\ and\ \citenamefont {Prokof'ev}}]{DengProkofev2012}%
  \BibitemOpen
  \bibfield  {author} {\bibinfo {author} {\bibfnamefont {F.}~\bibnamefont {Wu}}, \bibinfo {author} {\bibfnamefont {Y.}~\bibnamefont {Deng}},\ and\ \bibinfo {author} {\bibfnamefont {N.}~\bibnamefont {Prokof'ev}},\ }\bibfield  {title} {\bibinfo {title} {Phase diagram of the toric code model in a parallel magnetic field},\ }\href {https://doi.org/10.1103/PhysRevB.85.195104} {\bibfield  {journal} {\bibinfo  {journal} {Physical Review B}\ }\textbf {\bibinfo {volume} {85}},\ \bibinfo {pages} {195104} (\bibinfo {year} {2012})},\ \Eprint {https://arxiv.org/abs/1201.6409} {arXiv:1201.6409 [cond-mat.stat-mech]} \BibitemShut {NoStop}%
\bibitem [{\citenamefont {Somoza}\ \emph {et~al.}(2021)\citenamefont {Somoza}, \citenamefont {Serna},\ and\ \citenamefont {Nahum}}]{SomozaSernaNahum2021}%
  \BibitemOpen
  \bibfield  {author} {\bibinfo {author} {\bibfnamefont {A.~M.}\ \bibnamefont {Somoza}}, \bibinfo {author} {\bibfnamefont {P.}~\bibnamefont {Serna}},\ and\ \bibinfo {author} {\bibfnamefont {A.}~\bibnamefont {Nahum}},\ }\bibfield  {title} {\bibinfo {title} {Self-dual criticality in three-dimensional {$\mathbb{Z}_2$} gauge theory with matter},\ }\href {https://doi.org/10.1103/PhysRevX.11.041008} {\bibfield  {journal} {\bibinfo  {journal} {Physical Review X}\ }\textbf {\bibinfo {volume} {11}},\ \bibinfo {pages} {041008} (\bibinfo {year} {2021})},\ \Eprint {https://arxiv.org/abs/2012.15845} {arXiv:2012.15845 [cond-mat.stat-mech]} \BibitemShut {NoStop}%
\bibitem [{\citenamefont {Fradkin}\ and\ \citenamefont {Shenker}(1979)}]{Fradkin:1978dv}%
  \BibitemOpen
  \bibfield  {author} {\bibinfo {author} {\bibfnamefont {E.~H.}\ \bibnamefont {Fradkin}}\ and\ \bibinfo {author} {\bibfnamefont {S.~H.}\ \bibnamefont {Shenker}},\ }\bibfield  {title} {\bibinfo {title} {{Phase Diagrams of Lattice Gauge Theories with Higgs Fields}},\ }\href {https://doi.org/10.1103/PhysRevD.19.3682} {\bibfield  {journal} {\bibinfo  {journal} {Phys. Rev. D}\ }\textbf {\bibinfo {volume} {19}},\ \bibinfo {pages} {3682} (\bibinfo {year} {1979})}\BibitemShut {NoStop}%
\bibitem [{\citenamefont {Kogut}\ and\ \citenamefont {Susskind}(1975)}]{KogutSusskind1975}%
  \BibitemOpen
  \bibfield  {author} {\bibinfo {author} {\bibfnamefont {J.~B.}\ \bibnamefont {Kogut}}\ and\ \bibinfo {author} {\bibfnamefont {L.}~\bibnamefont {Susskind}},\ }\bibfield  {title} {\bibinfo {title} {Hamiltonian formulation of wilson's lattice gauge theories},\ }\href {https://doi.org/10.1103/PhysRevD.11.395} {\bibfield  {journal} {\bibinfo  {journal} {Physical Review D}\ }\textbf {\bibinfo {volume} {11}},\ \bibinfo {pages} {395} (\bibinfo {year} {1975})}\BibitemShut {NoStop}%
\bibitem [{\citenamefont {Creutz}(1977)}]{Creutz1977}%
  \BibitemOpen
  \bibfield  {author} {\bibinfo {author} {\bibfnamefont {M.}~\bibnamefont {Creutz}},\ }\bibfield  {title} {\bibinfo {title} {Gauge fixing, the transfer matrix, and confinement on a lattice},\ }\href {https://doi.org/10.1103/PhysRevD.15.1128} {\bibfield  {journal} {\bibinfo  {journal} {Physical Review D}\ }\textbf {\bibinfo {volume} {15}},\ \bibinfo {pages} {1128} (\bibinfo {year} {1977})}\BibitemShut {NoStop}%
\bibitem [{\citenamefont {Luscher}\ \emph {et~al.}(1981)\citenamefont {Luscher}, \citenamefont {Munster},\ and\ \citenamefont {Weisz}}]{Luscher:1980iy}%
  \BibitemOpen
  \bibfield  {author} {\bibinfo {author} {\bibfnamefont {M.}~\bibnamefont {Luscher}}, \bibinfo {author} {\bibfnamefont {G.}~\bibnamefont {Munster}},\ and\ \bibinfo {author} {\bibfnamefont {P.}~\bibnamefont {Weisz}},\ }\bibfield  {title} {\bibinfo {title} {{How Thick Are Chromoelectric Flux Tubes?}},\ }\href {https://doi.org/10.1016/0550-3213(81)90151-6} {\bibfield  {journal} {\bibinfo  {journal} {Nucl. Phys. B}\ }\textbf {\bibinfo {volume} {180}},\ \bibinfo {pages} {1} (\bibinfo {year} {1981})}\BibitemShut {NoStop}%
\bibitem [{\citenamefont {Nambu}(1979)}]{Nambu:1978bd}%
  \BibitemOpen
  \bibfield  {author} {\bibinfo {author} {\bibfnamefont {Y.}~\bibnamefont {Nambu}},\ }\bibfield  {title} {\bibinfo {title} {{QCD and the String Model}},\ }\href {https://doi.org/10.1016/0370-2693(79)91193-6} {\bibfield  {journal} {\bibinfo  {journal} {Phys.Lett.}\ }\textbf {\bibinfo {volume} {B80}},\ \bibinfo {pages} {372} (\bibinfo {year} {1979})}\BibitemShut {NoStop}%
\bibitem [{\citenamefont {Goto}(1971)}]{Goto:1971ce}%
  \BibitemOpen
  \bibfield  {author} {\bibinfo {author} {\bibfnamefont {T.}~\bibnamefont {Goto}},\ }\bibfield  {title} {\bibinfo {title} {{Relativistic quantum mechanics of one-dimensional mechanical continuum and subsidiary condition of dual resonance model}},\ }\href {https://doi.org/10.1143/PTP.46.1560} {\bibfield  {journal} {\bibinfo  {journal} {Prog.Theor.Phys.}\ }\textbf {\bibinfo {volume} {46}},\ \bibinfo {pages} {1560} (\bibinfo {year} {1971})}\BibitemShut {NoStop}%
\bibitem [{\citenamefont {Caselle}\ \emph {et~al.}(1996)\citenamefont {Caselle}, \citenamefont {Gliozzi}, \citenamefont {Magnea},\ and\ \citenamefont {Vinti}}]{Caselle:1995fh}%
  \BibitemOpen
  \bibfield  {author} {\bibinfo {author} {\bibfnamefont {M.}~\bibnamefont {Caselle}}, \bibinfo {author} {\bibfnamefont {F.}~\bibnamefont {Gliozzi}}, \bibinfo {author} {\bibfnamefont {U.}~\bibnamefont {Magnea}},\ and\ \bibinfo {author} {\bibfnamefont {S.}~\bibnamefont {Vinti}},\ }\bibfield  {title} {\bibinfo {title} {{Width of long color flux tubes in lattice gauge systems}},\ }\href {https://doi.org/10.1016/0550-3213(95)00639-7} {\bibfield  {journal} {\bibinfo  {journal} {Nucl. Phys. B}\ }\textbf {\bibinfo {volume} {460}},\ \bibinfo {pages} {397} (\bibinfo {year} {1996})}\BibitemShut {NoStop}%
\bibitem [{\citenamefont {Gliozzi}\ \emph {et~al.}(2010)\citenamefont {Gliozzi}, \citenamefont {Pepe},\ and\ \citenamefont {Wiese}}]{Gliozzi:2010zv}%
  \BibitemOpen
  \bibfield  {author} {\bibinfo {author} {\bibfnamefont {F.}~\bibnamefont {Gliozzi}}, \bibinfo {author} {\bibfnamefont {M.}~\bibnamefont {Pepe}},\ and\ \bibinfo {author} {\bibfnamefont {U.~J.}\ \bibnamefont {Wiese}},\ }\bibfield  {title} {\bibinfo {title} {{The Width of the Confining String in Yang-Mills Theory}},\ }\href {https://doi.org/10.1103/PhysRevLett.104.232001} {\bibfield  {journal} {\bibinfo  {journal} {Phys. Rev. Lett.}\ }\textbf {\bibinfo {volume} {104}},\ \bibinfo {pages} {232001} (\bibinfo {year} {2010})},\ \Eprint {https://arxiv.org/abs/1002.4888} {arXiv:1002.4888 [hep-lat]} \BibitemShut {NoStop}%
\bibitem [{\citenamefont {Caselle}\ \emph {et~al.}(2026)\citenamefont {Caselle}, \citenamefont {Cellini}, \citenamefont {Nada}, \citenamefont {Panfalone},\ and\ \citenamefont {Verzichelli}}]{caselle2026intrinsicwidthfluxtube}%
  \BibitemOpen
  \bibfield  {author} {\bibinfo {author} {\bibfnamefont {M.}~\bibnamefont {Caselle}}, \bibinfo {author} {\bibfnamefont {E.}~\bibnamefont {Cellini}}, \bibinfo {author} {\bibfnamefont {A.}~\bibnamefont {Nada}}, \bibinfo {author} {\bibfnamefont {D.}~\bibnamefont {Panfalone}},\ and\ \bibinfo {author} {\bibfnamefont {L.}~\bibnamefont {Verzichelli}},\ }\bibfield  {title} {\bibinfo {title} {{Intrinsic width of the flux tube as a tool to explore confining mechanisms in lattice gauge theories}},\ }\href {https://doi.org/10.1007/JHEP07(2026)027} {\bibfield  {journal} {\bibinfo  {journal} {JHEP}\ }\textbf {\bibinfo {volume} {07}},\ \bibinfo {pages} {027}},\ \Eprint {https://arxiv.org/abs/2601.19520} {arXiv:2601.19520 [hep-lat]} \BibitemShut {NoStop}%
\bibitem [{\citenamefont {Wegner}(1971)}]{Wegner:1971app}%
  \BibitemOpen
  \bibfield  {author} {\bibinfo {author} {\bibfnamefont {F.~J.}\ \bibnamefont {Wegner}},\ }\bibfield  {title} {\bibinfo {title} {{Duality in Generalized Ising Models and Phase Transitions Without Local Order Parameters}},\ }\href {https://doi.org/10.1063/1.1665530} {\bibfield  {journal} {\bibinfo  {journal} {J. Math. Phys.}\ }\textbf {\bibinfo {volume} {12}},\ \bibinfo {pages} {2259} (\bibinfo {year} {1971})}\BibitemShut {NoStop}%
\bibitem [{\citenamefont {Caselle}\ \emph {et~al.}(2003)\citenamefont {Caselle}, \citenamefont {Hasenbusch},\ and\ \citenamefont {Panero}}]{Caselle:2002ah}%
  \BibitemOpen
  \bibfield  {author} {\bibinfo {author} {\bibfnamefont {M.}~\bibnamefont {Caselle}}, \bibinfo {author} {\bibfnamefont {M.}~\bibnamefont {Hasenbusch}},\ and\ \bibinfo {author} {\bibfnamefont {M.}~\bibnamefont {Panero}},\ }\bibfield  {title} {\bibinfo {title} {{String effects in the 3-d gauge Ising model}},\ }\href {https://doi.org/10.1088/1126-6708/2003/01/057} {\bibfield  {journal} {\bibinfo  {journal} {JHEP}\ }\textbf {\bibinfo {volume} {01}},\ \bibinfo {pages} {057}},\ \Eprint {https://arxiv.org/abs/hep-lat/0211012} {arXiv:hep-lat/0211012} \BibitemShut {NoStop}%
\bibitem [{\citenamefont {Giamarchi}(2003)}]{Giamarchi_1d}%
  \BibitemOpen
  \bibfield  {author} {\bibinfo {author} {\bibfnamefont {T.}~\bibnamefont {Giamarchi}},\ }\href {https://doi.org/10.1093/acprof:oso/9780198525004.001.0001} {\emph {\bibinfo {title} {Quantum Physics in One Dimension}}}\ (\bibinfo  {publisher} {Oxford University Press},\ \bibinfo {year} {2003})\BibitemShut {NoStop}%
\bibitem [{\citenamefont {Frau}\ \emph {et~al.}(2025)\citenamefont {Frau}, \citenamefont {Tarabunga}, \citenamefont {Collura}, \citenamefont {Tirrito},\ and\ \citenamefont {Dalmonte}}]{Frau_2025}%
  \BibitemOpen
  \bibfield  {author} {\bibinfo {author} {\bibfnamefont {M.}~\bibnamefont {Frau}}, \bibinfo {author} {\bibfnamefont {P.~S.}\ \bibnamefont {Tarabunga}}, \bibinfo {author} {\bibfnamefont {M.}~\bibnamefont {Collura}}, \bibinfo {author} {\bibfnamefont {E.}~\bibnamefont {Tirrito}},\ and\ \bibinfo {author} {\bibfnamefont {M.}~\bibnamefont {Dalmonte}},\ }\bibfield  {title} {\bibinfo {title} {Stabilizer disentangling of conformal field theories},\ }\href {http://dx.doi.org/10.21468/SciPostPhys.18.5.165} {\bibfield  {journal} {\bibinfo  {journal} {SciPost Physics}\ }\textbf {\bibinfo {volume} {18}} (\bibinfo {year} {2025})}\BibitemShut {NoStop}%
\bibitem [{\citenamefont {Tagliacozzo}\ and\ \citenamefont {Vidal}(2011)}]{entrenandgaugesymmetry}%
  \BibitemOpen
  \bibfield  {author} {\bibinfo {author} {\bibfnamefont {L.}~\bibnamefont {Tagliacozzo}}\ and\ \bibinfo {author} {\bibfnamefont {G.}~\bibnamefont {Vidal}},\ }\bibfield  {title} {\bibinfo {title} {Entanglement renormalization and gauge symmetry},\ }\href {https://doi.org/10.1103/PhysRevB.83.115127} {\bibfield  {journal} {\bibinfo  {journal} {Phys. Rev. B}\ }\textbf {\bibinfo {volume} {83}},\ \bibinfo {pages} {115127} (\bibinfo {year} {2011})}\BibitemShut {NoStop}%
\bibitem [{\citenamefont {Kufel}\ \emph {et~al.}(2025)\citenamefont {Kufel}, \citenamefont {Kemp}, \citenamefont {Vu}, \citenamefont {Linsel}, \citenamefont {Laumann},\ and\ \citenamefont {Yao}}]{KufelYao_prl2025_nqs_toriccode}%
  \BibitemOpen
  \bibfield  {author} {\bibinfo {author} {\bibfnamefont {D.~S.}\ \bibnamefont {Kufel}}, \bibinfo {author} {\bibfnamefont {J.}~\bibnamefont {Kemp}}, \bibinfo {author} {\bibfnamefont {D.}~\bibnamefont {Vu}}, \bibinfo {author} {\bibfnamefont {S.~M.}\ \bibnamefont {Linsel}}, \bibinfo {author} {\bibfnamefont {C.~R.}\ \bibnamefont {Laumann}},\ and\ \bibinfo {author} {\bibfnamefont {N.~Y.}\ \bibnamefont {Yao}},\ }\bibfield  {title} {\bibinfo {title} {Approximately symmetric neural networks for quantum spin liquids},\ }\href {https://doi.org/10.1103/pgnx-11ph} {\bibfield  {journal} {\bibinfo  {journal} {Phys. Rev. Lett.}\ }\textbf {\bibinfo {volume} {135}},\ \bibinfo {pages} {056702} (\bibinfo {year} {2025})}\BibitemShut {NoStop}%
\bibitem [{\citenamefont {Bonati}\ and\ \citenamefont {Morlacchi}(2020)}]{Bonati:2020orj}%
  \BibitemOpen
  \bibfield  {author} {\bibinfo {author} {\bibfnamefont {C.}~\bibnamefont {Bonati}}\ and\ \bibinfo {author} {\bibfnamefont {S.}~\bibnamefont {Morlacchi}},\ }\bibfield  {title} {\bibinfo {title} {{Flux tubes and string breaking in three dimensional SU(2) Yang-Mills theory}},\ }\href {https://doi.org/10.1103/PhysRevD.101.094506} {\bibfield  {journal} {\bibinfo  {journal} {Phys. Rev. D}\ }\textbf {\bibinfo {volume} {101}},\ \bibinfo {pages} {094506} (\bibinfo {year} {2020})},\ \Eprint {https://arxiv.org/abs/2003.07244} {arXiv:2003.07244 [hep-lat]} \BibitemShut {NoStop}%
\bibitem [{\citenamefont {Bonati}\ \emph {et~al.}(2021)\citenamefont {Bonati}, \citenamefont {Caselle},\ and\ \citenamefont {Morlacchi}}]{Bonati:2021vbc}%
  \BibitemOpen
  \bibfield  {author} {\bibinfo {author} {\bibfnamefont {C.}~\bibnamefont {Bonati}}, \bibinfo {author} {\bibfnamefont {M.}~\bibnamefont {Caselle}},\ and\ \bibinfo {author} {\bibfnamefont {S.}~\bibnamefont {Morlacchi}},\ }\bibfield  {title} {\bibinfo {title} {{The Unreasonable effectiveness of effective string theory: The case of the 3D SU(2) Higgs model}},\ }\href {https://doi.org/10.1103/PhysRevD.104.054501} {\bibfield  {journal} {\bibinfo  {journal} {Phys. Rev. D}\ }\textbf {\bibinfo {volume} {104}},\ \bibinfo {pages} {054501} (\bibinfo {year} {2021})},\ \Eprint {https://arxiv.org/abs/2106.08784} {arXiv:2106.08784 [hep-lat]} \BibitemShut {NoStop}%
\bibitem [{\citenamefont {Clem}(1975)}]{Clem:1975ohd}%
  \BibitemOpen
  \bibfield  {author} {\bibinfo {author} {\bibfnamefont {J.~R.}\ \bibnamefont {Clem}},\ }\bibfield  {title} {\bibinfo {title} {{Simple model for the vortex core in a type II superconductor}},\ }\href {https://doi.org/10.1007/BF00116134} {\bibfield  {journal} {\bibinfo  {journal} {J. Low Temp. Phys.}\ }\textbf {\bibinfo {volume} {18}},\ \bibinfo {pages} {427} (\bibinfo {year} {1975})}\BibitemShut {NoStop}%
\bibitem [{\citenamefont {Mandelstam}(1976)}]{Mandelstam:1974pi}%
  \BibitemOpen
  \bibfield  {author} {\bibinfo {author} {\bibfnamefont {S.}~\bibnamefont {Mandelstam}},\ }\bibfield  {title} {\bibinfo {title} {{Vortices and Quark Confinement in Nonabelian Gauge Theories}},\ }\href {https://doi.org/10.1016/0370-1573(76)90043-0} {\bibfield  {journal} {\bibinfo  {journal} {Phys. Rept.}\ }\textbf {\bibinfo {volume} {23}},\ \bibinfo {pages} {245} (\bibinfo {year} {1976})}\BibitemShut {NoStop}%
\bibitem [{\citenamefont {'t~Hooft}(1981)}]{tHooft:1981bkw}%
  \BibitemOpen
  \bibfield  {author} {\bibinfo {author} {\bibfnamefont {G.}~\bibnamefont {'t~Hooft}},\ }\bibfield  {title} {\bibinfo {title} {{Topology of the Gauge Condition and New Confinement Phases in Nonabelian Gauge Theories}},\ }\href {https://doi.org/10.1016/0550-3213(81)90442-9} {\bibfield  {journal} {\bibinfo  {journal} {Nucl. Phys. B}\ }\textbf {\bibinfo {volume} {190}},\ \bibinfo {pages} {455} (\bibinfo {year} {1981})}\BibitemShut {NoStop}%
\bibitem [{\citenamefont {Cea}\ \emph {et~al.}(2012)\citenamefont {Cea}, \citenamefont {Cosmai},\ and\ \citenamefont {Papa}}]{Cea:2012qw}%
  \BibitemOpen
  \bibfield  {author} {\bibinfo {author} {\bibfnamefont {P.}~\bibnamefont {Cea}}, \bibinfo {author} {\bibfnamefont {L.}~\bibnamefont {Cosmai}},\ and\ \bibinfo {author} {\bibfnamefont {A.}~\bibnamefont {Papa}},\ }\bibfield  {title} {\bibinfo {title} {{Chromoelectric flux tubes and coherence length in QCD}},\ }\href {https://doi.org/10.1103/PhysRevD.86.054501} {\bibfield  {journal} {\bibinfo  {journal} {Phys. Rev. D}\ }\textbf {\bibinfo {volume} {86}},\ \bibinfo {pages} {054501} (\bibinfo {year} {2012})},\ \Eprint {https://arxiv.org/abs/1208.1362} {arXiv:1208.1362 [hep-lat]} \BibitemShut {NoStop}%
\bibitem [{\citenamefont {Cea}\ \emph {et~al.}(2017)\citenamefont {Cea}, \citenamefont {Cosmai}, \citenamefont {Cuteri},\ and\ \citenamefont {Papa}}]{Cea:2017ocq}%
  \BibitemOpen
  \bibfield  {author} {\bibinfo {author} {\bibfnamefont {P.}~\bibnamefont {Cea}}, \bibinfo {author} {\bibfnamefont {L.}~\bibnamefont {Cosmai}}, \bibinfo {author} {\bibfnamefont {F.}~\bibnamefont {Cuteri}},\ and\ \bibinfo {author} {\bibfnamefont {A.}~\bibnamefont {Papa}},\ }\bibfield  {title} {\bibinfo {title} {{Flux tubes in the QCD vacuum}},\ }\href {https://doi.org/10.1103/PhysRevD.95.114511} {\bibfield  {journal} {\bibinfo  {journal} {Phys. Rev. D}\ }\textbf {\bibinfo {volume} {95}},\ \bibinfo {pages} {114511} (\bibinfo {year} {2017})},\ \Eprint {https://arxiv.org/abs/1702.06437} {arXiv:1702.06437 [hep-lat]} \BibitemShut {NoStop}%
\bibitem [{\citenamefont {Agostini}\ \emph {et~al.}(1997)\citenamefont {Agostini}, \citenamefont {Carlino}, \citenamefont {Caselle},\ and\ \citenamefont {Hasenbusch}}]{Agostini:1996xy}%
  \BibitemOpen
  \bibfield  {author} {\bibinfo {author} {\bibfnamefont {V.}~\bibnamefont {Agostini}}, \bibinfo {author} {\bibfnamefont {G.}~\bibnamefont {Carlino}}, \bibinfo {author} {\bibfnamefont {M.}~\bibnamefont {Caselle}},\ and\ \bibinfo {author} {\bibfnamefont {M.}~\bibnamefont {Hasenbusch}},\ }\bibfield  {title} {\bibinfo {title} {{The Spectrum of the (2+1)-dimensional gauge Ising model}},\ }\href {https://doi.org/10.1016/S0550-3213(96)00539-1} {\bibfield  {journal} {\bibinfo  {journal} {Nucl. Phys. B}\ }\textbf {\bibinfo {volume} {484}},\ \bibinfo {pages} {331} (\bibinfo {year} {1997})},\ \Eprint {https://arxiv.org/abs/hep-lat/9607029} {arXiv:hep-lat/9607029} \BibitemShut {NoStop}%
\bibitem [{\citenamefont {Felser}\ \emph {et~al.}(2021)\citenamefont {Felser}, \citenamefont {Notarnicola},\ and\ \citenamefont {Montangero}}]{felser2021efficient}%
  \BibitemOpen
  \bibfield  {author} {\bibinfo {author} {\bibfnamefont {T.}~\bibnamefont {Felser}}, \bibinfo {author} {\bibfnamefont {S.}~\bibnamefont {Notarnicola}},\ and\ \bibinfo {author} {\bibfnamefont {S.}~\bibnamefont {Montangero}},\ }\bibfield  {title} {\bibinfo {title} {Efficient tensor network ansatz for high-dimensional quantum many-body problems},\ }\href {https://doi.org/10.1103/physrevlett.126.170603} {\bibfield  {journal} {\bibinfo  {journal} {Physical Review Letters}\ }\textbf {\bibinfo {volume} {126}},\ \bibinfo {pages} {170603} (\bibinfo {year} {2021})}\BibitemShut {NoStop}%
\bibitem [{\citenamefont {Yosprakob}\ \emph {et~al.}(2026)\citenamefont {Yosprakob}, \citenamefont {Tu}, \citenamefont {Okubo}, \citenamefont {Okunishi},\ and\ \citenamefont {Kim}}]{yosprakob2026cliffordcircuitsaugmentedgrassmann}%
  \BibitemOpen
  \bibfield  {author} {\bibinfo {author} {\bibfnamefont {A.}~\bibnamefont {Yosprakob}}, \bibinfo {author} {\bibfnamefont {W.-L.}\ \bibnamefont {Tu}}, \bibinfo {author} {\bibfnamefont {T.}~\bibnamefont {Okubo}}, \bibinfo {author} {\bibfnamefont {K.}~\bibnamefont {Okunishi}},\ and\ \bibinfo {author} {\bibfnamefont {D.}~\bibnamefont {Kim}},\ }\href {https://arxiv.org/abs/2510.04164} {\bibinfo {title} {Clifford circuits augmented grassmann matrix product states}} (\bibinfo {year} {2026}),\ \Eprint {https://arxiv.org/abs/2510.04164} {arXiv:2510.04164 [quant-ph]} \BibitemShut {NoStop}%
\bibitem [{\citenamefont {Banerjee}\ \emph {et~al.}(2022)\citenamefont {Banerjee}, \citenamefont {Caspar}, \citenamefont {Jiang}, \citenamefont {Peng},\ and\ \citenamefont {Wiese}}]{banerjee2022nematic}%
  \BibitemOpen
  \bibfield  {author} {\bibinfo {author} {\bibfnamefont {D.}~\bibnamefont {Banerjee}}, \bibinfo {author} {\bibfnamefont {S.}~\bibnamefont {Caspar}}, \bibinfo {author} {\bibfnamefont {F.-J.}\ \bibnamefont {Jiang}}, \bibinfo {author} {\bibfnamefont {J.-H.}\ \bibnamefont {Peng}},\ and\ \bibinfo {author} {\bibfnamefont {U.-J.}\ \bibnamefont {Wiese}},\ }\bibfield  {title} {\bibinfo {title} {Nematic confined phases in the u(1) quantum link model on a triangular lattice: Near-term quantum computations of string dynamics on a chip},\ }\bibfield  {journal} {\bibinfo  {journal} {Physical Review Research}\ }\textbf {\bibinfo {volume} {4}},\ \href {https://doi.org/10.1103/physrevresearch.4.023176} {10.1103/physrevresearch.4.023176} (\bibinfo {year} {2022})\BibitemShut {NoStop}%
\bibitem [{\citenamefont {Klebanov}\ \emph {et~al.}(2008)\citenamefont {Klebanov}, \citenamefont {Kutasov},\ and\ \citenamefont {Murugan}}]{Klebanov_2008}%
  \BibitemOpen
  \bibfield  {author} {\bibinfo {author} {\bibfnamefont {I.~R.}\ \bibnamefont {Klebanov}}, \bibinfo {author} {\bibfnamefont {D.}~\bibnamefont {Kutasov}},\ and\ \bibinfo {author} {\bibfnamefont {A.}~\bibnamefont {Murugan}},\ }\bibfield  {title} {\bibinfo {title} {Entanglement as a probe of confinement},\ }\href {https://doi.org/10.1016/j.nuclphysb.2007.12.017} {\bibfield  {journal} {\bibinfo  {journal} {Nuclear Physics B}\ }\textbf {\bibinfo {volume} {796}},\ \bibinfo {pages} {274–293} (\bibinfo {year} {2008})}\BibitemShut {NoStop}%
\bibitem [{\citenamefont {Tu}\ \emph {et~al.}(2020)\citenamefont {Tu}, \citenamefont {Kharzeev},\ and\ \citenamefont {Ullrich}}]{tu2020einstein}%
  \BibitemOpen
  \bibfield  {author} {\bibinfo {author} {\bibfnamefont {Z.}~\bibnamefont {Tu}}, \bibinfo {author} {\bibfnamefont {D.~E.}\ \bibnamefont {Kharzeev}},\ and\ \bibinfo {author} {\bibfnamefont {T.}~\bibnamefont {Ullrich}},\ }\bibfield  {title} {\bibinfo {title} {Einstein-podolsky-rosen paradox and quantum entanglement at subnucleonic scales},\ }\href {http://dx.doi.org/10.1103/PhysRevLett.124.062001} {\bibfield  {journal} {\bibinfo  {journal} {Physical Review Letters}\ }\textbf {\bibinfo {volume} {124}} (\bibinfo {year} {2020})}\BibitemShut {NoStop}%
\bibitem [{\citenamefont {Amorosso}\ \emph {et~al.}(2024)\citenamefont {Amorosso}, \citenamefont {Syritsyn},\ and\ \citenamefont {Venugopalan}}]{Amorosso:2024leg}%
  \BibitemOpen
  \bibfield  {author} {\bibinfo {author} {\bibfnamefont {R.}~\bibnamefont {Amorosso}}, \bibinfo {author} {\bibfnamefont {S.}~\bibnamefont {Syritsyn}},\ and\ \bibinfo {author} {\bibfnamefont {R.}~\bibnamefont {Venugopalan}},\ }\bibfield  {title} {\bibinfo {title} {{Entanglement entropy of a color flux tube in (2+1)D Yang-Mills theory}},\ }\href {https://doi.org/10.1007/JHEP12(2024)177} {\bibfield  {journal} {\bibinfo  {journal} {JHEP}\ }\textbf {\bibinfo {volume} {12}},\ \bibinfo {pages} {177}},\ \Eprint {https://arxiv.org/abs/2410.00112} {arXiv:2410.00112 [hep-lat]} \BibitemShut {NoStop}%
\bibitem [{\citenamefont {Amorosso}\ \emph {et~al.}(2026)\citenamefont {Amorosso}, \citenamefont {Syritsyn},\ and\ \citenamefont {Venugopalan}}]{Amorosso:2026mdo}%
  \BibitemOpen
  \bibfield  {author} {\bibinfo {author} {\bibfnamefont {R.}~\bibnamefont {Amorosso}}, \bibinfo {author} {\bibfnamefont {S.}~\bibnamefont {Syritsyn}},\ and\ \bibinfo {author} {\bibfnamefont {R.}~\bibnamefont {Venugopalan}},\ }\href {https://arxiv.org/abs/2601.17199} {\bibinfo {title} {Entanglement enabled tomography of flux tubes in (2+1)d yang-mills theory}} (\bibinfo {year} {2026}),\ \Eprint {https://arxiv.org/abs/2601.17199} {arXiv:2601.17199 [hep-th]} \BibitemShut {NoStop}%
\bibitem [{\citenamefont {Fishman}\ \emph {et~al.}(2022)\citenamefont {Fishman}, \citenamefont {White},\ and\ \citenamefont {Stoudenmire}}]{ITensor}%
  \BibitemOpen
  \bibfield  {author} {\bibinfo {author} {\bibfnamefont {M.}~\bibnamefont {Fishman}}, \bibinfo {author} {\bibfnamefont {S.~R.}\ \bibnamefont {White}},\ and\ \bibinfo {author} {\bibfnamefont {E.~M.}\ \bibnamefont {Stoudenmire}},\ }\bibfield  {title} {\bibinfo {title} {{The ITensor Software Library for Tensor Network Calculations}},\ }\href {https://doi.org/10.21468/SciPostPhysCodeb.4} {\bibfield  {journal} {\bibinfo  {journal} {SciPost Phys. Codebases}\ ,\ \bibinfo {pages} {4}} (\bibinfo {year} {2022})}\BibitemShut {NoStop}%
\end{thebibliography}%

\end{document}